\documentclass[usenatbib]{mnras}

\newcommand{\gsim}{\raisebox{-0.3ex}{\mbox{$\stackrel{>}{_\sim} \,$}}}

\usepackage[T1]{fontenc}
\usepackage{ae,aecompl}
\usepackage{graphicx}
\usepackage{amsmath,amssymb}
\usepackage{color,ulem}
\usepackage[compatibility=false]{caption }
\usepackage{natbib}

\usepackage{verbatim}

\definecolor{webgreen}{rgb}{0,.5,0}
\definecolor{webbrown}{rgb}{.6,0,0}

\hypersetup{%
   colorlinks=true,%
   breaklinks=true,%
   plainpages=false, bookmarksnumbered, bookmarksopen=true,
   bookmarksopenlevel=1,%
   urlcolor=webbrown, linkcolor=blue, citecolor=webgreen,
   }

\begin{document}
\title [] 
{Thermodynamic properties, multiphase gas and AGN feedback in a large sample of giant ellipticals}
\author[]{K. Lakhchaura$^{1,2}$\thanks{lakhchaura.k@gmail.com}, N. Werner$^{1,3,4}$, M. Sun$^{5}$, R. E. A. Canning$^{6}$, 
\newauthor M. Gaspari$^{7}$\thanks{{\it Einstein} and {\it Spitzer} Fellow}, S. W. Allen$^{6}$, T. Connor$^{8}$, M.~Donahue$^{9}$, C. Sarazin$^{10}$\\
$^1$MTA-E\"otv\"os University Lend\"ulet Hot Universe Research Group, P\'azm\'any P\'eter s\'et\'any 1/A, Budapest, 1117, Hungary \\
$^2$MTA-ELTE Astrophysics Research Group, P\'azm\'any P\'eter s\'et\'any 1/A, Budapest, 1117, Hungary \\
$^3$Department of Theoretical Physics and Astrophysics, Faculty of Science, Masaryk University, Kotl\'a\v{r}sk\'a 2, Brno, 611 37, Czech Republic \\
$^4$School of Science, Hiroshima University, 1-3-1 Kagamiyama, Higashi-Hiroshima 739-8526, Japan \\
$^5$Department of Physics, University of Alabama in Huntsville, Huntsville, AL 35899, USA\\
$^6$Kavli Institute for Particle Astrophysics and Cosmology, Stanford University, 452 Lomita Mall, Stanford, CA 94305-4085, USA \\
$^7$Department of Astrophysical Sciences, Princeton University, 4 Ivy Lane, Princeton, NJ 08544-1001, USA \\
$^8$The Observatories of the Carnegie Institution for Science, 813 Santa Barbara Street, Pasadena, CA 91101, USA\\
$^9$Physics \& Astronomy Department, Michigan State University, East Lansing, MI 48824-2320, USA \\
$^{10}$Department of Astronomy, University of Virginia, 530 McCormick Rd., Charlottesville, VA 22904, USA\\
}
\maketitle

\begin{abstract}

We present a study of the thermal structure of the hot X-ray 
emitting atmospheres for a sample of 49 nearby X-ray and optically bright elliptical galaxies using {\it Chandra} X-ray data. We focus on the connection 
between the properties of the hot X-ray emitting gas and the cooler H$\alpha$+[NII] emitting phase, 
and the possible role of the latter in the AGN (Active Galactic Nuclei) feedback cycle. 
We do not find evident correlations between the H$\alpha$+[NII] emission and global properties such as X-ray luminosity, mass of hot gas, and gas mass fraction. 
We find that the presence of H$\alpha$+[NII] emission 
is more likely in systems with higher densities, lower entropies, shorter cooling times, shallower entropy profiles, lower values of min($t_{\rm cool}/t_{\rm ff}$), and 
disturbed X-ray morphologies (linked to turbulent motions). However, we see no clear separations in the observables obtained for galaxies with and without 
optical emission line nebulae. 
%We detected X-ray emitting central AGNs in 11 of the 25 H$\alpha$+[NII] emitting galaxies ($\sim$44\%), and in only three of the 20 galaxies without any optical emission line nebulae. 
The AGN jet powers of the galaxies with X-ray cavities show hint of a possible weak positive correlation with their H$\alpha$+[NII] luminosities. 
This correlation and the observed trends in the thermodynamic 
properties may result from chaotic cold accretion (CCA) powering AGN jets, as seen in some high-resolution hydrodynamic simulations.
\end{abstract}


\section{Introduction} 
\label{sec:intro}
Until the 1980s, elliptical galaxies were thought to be gas-less dormant systems containing 
mostly old stars, a picture that was drastically changed with the advent of sensitive instruments 
in the X-ray, infrared and mm-bands. Many elliptical galaxies are now known to 
host a complex multiphase interstellar medium, ranging from the cold $\lesssim$30 K
molecular clouds traced by sub-mm CO lines 
\citep[e.g.,][]{Edge2001,Edge2003,Salome2003,McDonald2012,Temi2018}; the cool $\sim$100 K 
gas detected through the FIR cooling lines of [CII], [NII] and [OI] 
\citep[e.g.,][]{Edge2010,Mittal2011,Mittal2012,Werner2013}; the warm $\sim$1000 K H$_2$ 
molecular gas seen in the NIR \citep[e.g.,][]{Jaffe1997,Falcke1998,Donahue2000,Edge2002,Hatch2005,Jaffe2005,Johnstone2007,Oonk2010,Lim2012};
 the ionised $\sim$10000 K nebulae seen in the optical H$\alpha+$[NII] emission
\citep[e.g.,][]{Cowie1983,Johnstone1987,Heckman1989,Donahue1992,Crawford1999,McDonald2010}; 
the moderately hot $\sim$100000 K gas detected in the FUV \citep[e.g.,][]{Sparks2012};
and the very hot $\sim$10$^7$ K X-ray gas. 

The role of the cool gas in feeding the active galactic nuclei (AGN) in these systems has remained an open question. 
The correlation between the jet powers, calculated from the radio-filled X-ray cavities, and the 
Bondi accretion rate of hot gas found by \cite{Allen2006} initially suggested ongoing hot accretion in giant ellipticals 
although \citet[][]{Russell2013} later on did not find a clear correlation in a larger sample. 

Using high-resolution 3D hydrodynamic simulations of massive galaxies, 
\cite{Gaspari2013,Gaspari2015,Gaspari2018} found that \textquoteleft{}chaotic cold (gas) accretion\textquoteright{} (CCA) plays an important role in the 
evolution of the central supermassive black hole (SMBH) and the host galaxy; this view has also been supported in other 
similar studies \citep[e.g., ][]{Prasad2015}. However, the exact nature of the material feeding and powering the AGN is still 
a subject of debate. 

The cool gas in giant ellipticals has most likely an internal origin and 
formed through the radiative cooling of the hot X-ray emitting gas and through stellar mass loss. \cite{Werner2014} analysed a sample of ten
optically and X-ray bright giant ellipticals, and found that the galaxies with extended cool gas nebulae 
have significantly lower entropies than the galaxies without cool gas, with a clear separation in the entropy profiles of the two groups. This 
indicates that the cool gas resulted from the radiative cooling of the hot phase. The cool gas develops through the formation of cooling instabilities 
from the hot gas, and feeds the central AGN; the radio mode feedback from the central AGN then heats the surrounding hot medium preventing it from 
cooling catastrophically, thus completing what is known as the \textquoteleft{}AGN feedback cycle\textquoteright{} \citep[see][for reviews]{Fabian2012,McNamaraNulsen2012,Soker2016}.

This scenario would lead to a correlation between the properties of the hot and cool phases. 
The ratio $t_{\rm cool}/t_{\rm ff}$, where $t_{\rm cool}$ is the local cooling time and $t_{\rm ff}$ is the free-fall 
time of a cooling blob, was found to be an important parameter for the formation of cooling 
instabilities \citep[see][]{Gaspari2012a,McCourt2012,Sharma2012}. Based on hydrodynamic simulations of massive ellipticals and clusters of galaxies, 
it has been found that $t_{\rm cool}$/$t_{\rm ff}\lesssim 10$ is the critical condition for the cooling instabilities to form in the cores of these systems. 
This result was also found to be supported observationally \citep[see][]{Cavagnolo2009,Lakhchaura2016}, although recently there have been some disagreements on 
the robustness of the cooling instability threshold \citep[e.g.,][]{Hogan2017,Pulido2018,Babyk2018b}.

\citet{McNamara2016} and \cite{Voit2017} found that the 
formation of cooling instabilities is 
also promoted by the adiabatic uplift of the hot gas by rising AGN jet inflated bubbles. Based on results obtained from both hydrodynamic simulations and 
observations, \cite{Gaspari2018} found that condensations are also promoted by subsonic turbulence and suggested the 
criterion $t_{\rm cool}$/$t_{\rm eddy}\approx 1$, where $t_{\rm eddy}$ is the turbulent eddy time, 
to be the best tracer of multiphase gas. Thus, in addition to entropy profiles and the 
$t_{\rm cool}/t_{\rm ff}$ ratio, gas motions should also be investigated in order to understand the formation of cooling instabilities in massive halos.

An alternative explanation for the presence/absence of multiphase gas in giant elliptical galaxies was given by \cite{Voit2015}.  Based on the results obtained for 
the small sample of \cite{Werner2014}, \cite{Voit2015} found that all but one (NGC 4261) of the five single phase galaxies in the sample were found to have 
$t_{\rm cool}/t_{\rm ff}$ $\gtrsim$20 while all five multiphase galaxies had 5 $<t_{\rm cool}/t_{\rm ff}\lesssim$ 20, in the 1--10 kpc radial range. They suggest that 
the single-phase and multi-phase ellipticals are two intrinsically different categories of massive ellipticals. While in the single-phase ellipticals, the feedback from supernova 
explosions prevents the stellar ejecta from forming stars by sweeping it out of the galaxy, in multiphase ellipticals, supernova feedback is not sufficient and there 
the central AGN feedback maintains $t_{\rm cool}/t_{\rm ff}$ $\approx$10. Although the study was based on a small sample of galaxies, similar results were also 
obtained in the hydrodynamic simulations of \cite{Wang2018}.

So far, most of the studies related to the non-gravitational processes (gas cooling/heating and AGN feedback) have focused
on bright massive clusters of galaxies. However, to understand the details of these processes better, it is crucial 
to also study the giant elliptical galaxies, where we can resolve the central regions (where most of the non-gravitational 
processes take place) in a greater detail than in clusters.

In this work, we have analysed the X-ray and H$\alpha$+[NII] observations of a sample of 49 nearby 
X-ray and optically bright elliptical galaxies, in order to understand the cool-hot gas connection, their 
interplay and their role in the AGN feedback cycle. About 19 of the 49 galaxies are the central galaxies of 
their respective groups; four are central galaxies of clusters; 23 are non-central galaxies of groups/clusters 
and 3 are isolated/fossil galaxies. Our sample has a high degree of completeness above certain 
X-ray and optical luminosity thresholds (see \S\ref{sec:sampl_sel}). The sample selection is described in \S\ref{sec:sampl_sel}, and the 
data reduction and analysis are detailed in \S\ref{sec:xray_data_red_analys}. The results are presented in \S\ref{sec:results}, discussed in \S\ref{sec:discussion}, and 
the conclusions are summarised in \S\ref{sec:summary}. A lambda cold dark
matter cosmology with H$_{0}=$ 70 km s$^{-1}$ Mpc$^{-1}$ and $\Omega_{\rm M}=$ 0.3
($\Omega_{\Lambda}=$ 0.7) has been assumed throughout.

\begin{table*}
 \caption{The names, redshifts ($z$), mean redshift-independent distances ($D$) \citep[from NED;][]{NED}, positions (RA, Dec), X-ray temperatures and 0.5--7.0 keV 
 intrinsic X-ray luminosities 
 estimated from a 10 kpc radius circular region around the X-ray peak (see \S\ref{sec:global_properties}), 2--10 keV intrinsic luminosities of the central point 
 sources and their ratio with the Eddington luminosities (see \S\ref{sec:cen_PS}), morphologies and luminosities of the H$\alpha$+[NII] emission, 
 the 1.4 GHz flux densities and 
 absolute visible band  magnitudes (B$_T$) of the galaxies. 
 The H$\alpha$+[NII] morphology/extent classification (column 10) is as follows: N: no 
cool gas emission, NE: H$\alpha$+[NII] extent $<$ 2 kpc, E: H$\alpha$+[NII] extent $\geq$ 2 kpc and U: galaxies for which the presence/absence of H$\alpha$+[NII] 
could not be confirmed. References for columns 8, 10 ,11, 12 and 13, given as superscripts, are described at the bottom.}
\label{tab:sample_info}
\vskip 0.5cm
\centering
{\scriptsize
\setlength\tabcolsep{1.5pt} 
\begin{tabular}{c c c c c c c c c c c c c}
\hline
Target & $z$ & $D$ & RA & Dec & kT &  L$_{X_{\rm Halo}}$ (0.5--7 keV) & \multicolumn{2}{c}{L$_{X_{\rm AGN}}$ (2--10 keV)} & H${\alpha}$+[NII] & L$_{H{\alpha}+[NII]}$ & $S_{1.4}$ & B$_T$\\
Name & & (Mpc) & (J2000) & (J2000) & (keV) & (10$^{42}$ erg/s) & (10$^{41}$ erg/s) & (10$^{-7}$ L$_{Edd}$) & morph. & (10$^{40}$ erg/s) & (Jy) & Magnitude\\
\hline
3C 449    & 0.01711 & 72.5$^*$ & 22 31 20.63 &  39 21 29.81 & 1.01$\pm$0.03 & 0.11$\pm$0.01 &  -   &          -           &  U            &       -     & 3.674$\pm$0.123$^{27}$ & -20.24$\pm$0.14$^{23}$\\ 
IC 1860   & 0.0229 & 95.75     & 02 49 33.88 & -31 11 21.94 & 1.03$\pm$0.01 & 0.68$\pm$0.05 &   -  &          -           & NE$^{1,2}$        &       -     & - & -21.21$\pm$0.03$^{24}$\\ 
IC 4296   & 0.0124 & 47.31   & 13 36 39.05 & -33 57 57.30 & 0.81$\pm$0.01 & 0.139$\pm$0.003 & 0.87$\pm$0.03 &  7.09$\pm$0.24     & E$^{1,2,3}$  & 0.55$^{12}$ & 18$\pm$1$^{28}$ & -21.76$\pm$0.09$^{24}$\\ 
IC 4765   & 0.0150 & 59.52     & 18 47 18.15 & -63 19 52.14 & 0.91$\pm$0.02 & 0.49$\pm$0.05 &    -  &          -           & NE$^{1,2}$        &       -     & 0.0056$^{29}$ & -21.54$\pm$0.16$^{25}$\\ 
NGC 57    & 0.0181 & 77.15     & 00 15 30.87 &  17 19 42.22 & 0.90$\pm$0.05 & 0.22$\pm$0.05 &    - &          -           &  N$^{1,2}$        &       -     & 0.0009$\pm$0.0005$^{28}$ & -21.77$\pm$0.26$^{24}$\\ 
NGC 315   & 0.0164 & 56.01     & 00 57 48.88 &  30 21 08.81 & 0.72$\pm$0.01 & 0.12$\pm$0.01 & 3.69$\pm$0.09  &    46.14$\pm$1.12     &  U$^{1,2}$        & 0.26$^{14}$ & 1.8$\pm$0.1$^{28}$ & -21.54$\pm$0.35$^{24}$\\ 
NGC 410   & 0.0176 & 66.0      & 01 10 58.87 &  33 09 07.30 & 0.83$\pm$0.03 & 0.35$\pm$0.14 &   -  &          -           & NE$^{1,2}$        &       -     & 0.0058$\pm$0.0005$^{28}$ & -21.58$\pm$0.22$^{24}$\\ 
NGC 499   & 0.0147 & 60.74     & 01 23 11.46 &  33 27 36.30 & 1.08$\pm$0.01 & 0.41$\pm$0.03 &    -  &          -           & NE$^{1,2}$        &       -     & 0.0007$\pm$0.0005$^{28}$ & -21.75$\pm$0.20$^{24}$\\ 
NGC 507   & 0.0164 & 59.83     & 01 23 39.95 &  33 15 22.22 & 1.04$\pm$0.01 & 0.23$\pm$0.02 &   - &          -           &  N$^{1,2}$        &       -     & 0.062$\pm$0.002$^{28}$ & -21.68$\pm$0.36$^{24}$\\ 
NGC 533   & 0.0184 & 61.58     & 01 25 31.43 &  01 45 33.57 & 0.91$\pm$0.01 & 0.51$\pm$0.03 &   -  &          -           & E$^{1,2}$        &   3.24$^4$  & 0.029$\pm$0.001$^{28}$ & -21.54$\pm$0.26$^{24}$\\ 
NGC 708   & 0.0162 & 64.19     & 01 52 46.48 &  36 09 06.53 & 1.25$\pm$0.01 & 0.88$\pm$0.02 &  -   &          -           &  E$^{1,2,5,16}$ &   3.40$^5$  & 0.067$\pm$0.002$^{30}$ & -20.74$\pm$0.23$^{26}$\\
NGC 741   & 0.0186 & 64.39     & 01 56 20.96 &  05 37 43.77 & 0.81$\pm$0.01 & 0.21$\pm$0.01 &   -  &          -           &  N$^{1,2}$        &   1.22$^4$  & 0.94$\pm$0.06$^{28}$ & -21.84$\pm$0.36$^{24}$\\  
NGC 777   & 0.0167 & 58.08     & 02 00 14.93 &  31 25 45.78 & 0.62$\pm$0.13 & 0.60$\pm$0.12 &  -   &          -           &  N$^{1,2}$        & 0.03$^{15}$ &  0.007$\pm$0.0005$^{28}$ & -21.33$\pm$0.24$^{24}$\\ 
NGC 1132  & 0.0232 & 87.9      & 02 52 51.82 & -01 16 29.0  & 0.95$\pm$0.02 & 0.17$\pm$0.02 &   -  &          -           &  N$^{1,2}$        &       -     & 0.0054$^{31}$ & -21.47$\pm$0.26$^{24}$\\ 
NGC 1316  & 0.0059 & 19.25   & 03 22 41.79 & -37 12 29.52 & 0.71$\pm$0.01 & 0.035$\pm$0.002 & 0.021$\pm$0.003 &   0.80$\pm$0.11    &  E$^{1,2,6}$    &   0.36$^6$  & 150$\pm$10$^{28}$ & -22.00$\pm$0.19$^{24}$\\ 
NGC 1399   & 0.0048 & 17.75  & 03 38 29.08 & -35 27 02.67 & 1.01$\pm$0.01 & 0.156$\pm$0.004 &  -   &          -           &  N$^{1,2,7}$    &   0.01$^4$  & 2.2$\pm$0.1$^{28}$ & -20.70$\pm$0.20$^{24}$\\ 
NGC 1404  & 0.0065 & 19.18   & 03 38 51.92 & -35 35 39.81 & 0.62$\pm$0.00 & 0.119$\pm$0.001 &   -  &          -           &  N$^{1,2}$        &       -     & 0.0039$\pm$0.0006$^{28}$ & -20.44$\pm$0.24$^{24}$\\ 
NGC 1407  & 0.0060 & 23.27   & 03 40 11.90 & -18 34 49.36 & 0.82$\pm$0.01 & 0.067$\pm$0.003 & 0.016$\pm$0.002 &   0.57$\pm$0.07    &  N$^{1,2}$    &   0.31$^4$  & 0.088$\pm$0.004$^{28}$  & -21.13$\pm$0.40$^{24}$\\ 
NGC 1521  & 0.0140 & 50.93     & 04 08 18.94 & -21 03 06.98 & 0.58$\pm$0.01 & 0.07$\pm$0.01 &    -  &          -           & NE$^{1,2}$        & 0.04$^{15}$ & 0.0042$\pm$0.0005$^{28}$ & -21.14$\pm$0.10$^{24}$\\ 
NGC 1550  & 0.0123 & 67.30     & 04 19 37.92 &  02 24 35.58 & 1.16$\pm$0.03 & 1.64$\pm$0.10 &   -  &          -           &  N$^{1,2}$        &       -     & 0.017$\pm$0.002$^{28}$ & -21.07$\pm$0.24$^{24}$\\ 
NGC 1600  & 0.0158 & 45.77     & 04 31 39.86 & -05 05 09.97 & 1.01$\pm$0.02 & 0.07$\pm$0.01 &   -  &          -           &  N$^{1,2}$        & 0.40$^{10}$ & 0.062$\pm$0.003$^{28}$ & -21.37$\pm$0.14$^{24}$\\ 
NGC 2300  & 0.0064 & 41.45     & 07 32 20.49 &  85 42 31.90 & 0.66$\pm$0.02 & 0.10$\pm$0.01 &   -  &          -           &  N$^{1,2}$        &      -      & 0.0029$\pm$0.0005$^{28}$ & -20.68$\pm$0.22$^{24}$\\ 
NGC 2305  & 0.0113 & 47.88     & 06 48 37.30 & -64 16 24.05 & 0.60$\pm$0.02 & 0.19$\pm$0.02 &  -   &          -           & NE$^{1,2}$        &       -     & - & -20.35$\pm$0.21$^{24}$\\ 
NGC 3091  & 0.0122 & 48.32     & 10 00 14.13 & -19 38 11.32 & 0.80$\pm$0.01 & 0.20$\pm$0.02 &   -  &          -           &  N$^{1,2}$        &       -     & 0.0025$\pm$0.0005$^{28}$ & -21.29$\pm$0.18$^{24}$\\ 
NGC 3923  & 0.0058 & 20.97   & 11 51 01.78 & -28 48 22.36 & 0.81$\pm$0.05 & 0.037$\pm$0.001 &  -   &          -           &  N$^{1,2}$        &       -     & 0.0010$\pm$0.0005$^{28}$ & -20.81$\pm$0.77$^{24}$\\ 
NGC 4073  & 0.0197 & 60.08     & 12 04 27.06 &  01 53 45.65 & 1.63$\pm$0.01 & 1.05$\pm$0.05 &   -  &          -           &  N$^{1,2}$        &       -     & 0.0012$\pm$0.0005$^{28}$ & -21.48$\pm$0.26$^{24}$\\ 
NGC 4125  & 0.0045 & 21.41   & 12 08 06.02 &  65 10 26.88 & 0.47$\pm$0.01 & 0.023$\pm$0.001 & 0.006$\pm$0.001 &   0.17$\pm$0.03    &  U$^{1,2}$   & 1.84$^{17}$ & 0.025$\pm$0.001$^{28}$ & -21.00$\pm$0.26$^{24}$\\ 
NGC 4261  & 0.0073 & 29.58   & 12 19 23.22 &  05 49 29.69 & 0.70$\pm$0.01 & 0.064$\pm$0.003 & 0.74$\pm$0.02 &   8.88$\pm$0.24  & NE$^{1,2,8}$ & 0.05$^{18}$ & 22$\pm$1$^{28}$  & -20.95$\pm$0.11$^{24}$\\ 
NGC 4374  & 0.0033 & 16.68   & 12 25 03.74 &  12 53 13.14 & 0.68$\pm$0.01 & 0.049$\pm$0.002 & 0.044$\pm$0.003 &  0.69$\pm$0.05  & NE$^{1,2,9}$  & 0.43$^9$  & 7.0$\pm$0.6$^{28}$ & -21.02$\pm$0.10$^{24}$\\ 
NGC 4406  & 0.0006 & 16.08   & 12 26 11.81 &  12 56 45.49 & 0.79$\pm$0.01 & 0.097$\pm$0.004 & 0.007$\pm$0.001 &   0.23$\pm$0.03    &  E$^{10}$     &  2.97$^4$  & 5.0$\pm$1.5$^{32}$ & -21.20$\pm$0.11$^{24}$\\ 
NGC 4472  & 0.0032 & 15.82   & 12 28 46.80 &  08 00 01.48 & 0.94$\pm$0.00 & 0.158$\pm$0.001 &  -   &          -           &  N$^{1,2}$        &   0.52$^4$  & 0.22$\pm$0.01$^{28}$ & -21.63$\pm$0.14$^{24}$\\ 
NGC 4486  & 0.0042 & 16.56   & 12 30 49.42 &  12 23 28.04 & 1.64$\pm$0.00 & 2.157$\pm$0.004 & 0.63$\pm$0.14$^{19}$ & 5.40$\pm$1.20 &  E$^4$  & 3.87$^{22}$  & 210$\pm$10$^{28}$ & -21.51$\pm$0.09$^{24}$\\ 
NGC 4552  & 0.0009 & 15.97   & 12 35 39.80 &  12 33 23.00 & 0.61$\pm$0.00 & 0.029$\pm$0.001 & 0.16$\pm$0.01 & 3.84$\pm$0.24 &  N$^{1,2}$      &   0.24$^4$  & 0.100$\pm$0.003$^{28}$ & -20.29$\pm$0.09$^{24}$\\ 
NGC 4636  & 0.0031 & 15.96   & 12 42 49.87 &  02 41 16.01 & 0.68$\pm$0.00 & 0.198$\pm$0.002 &    -  &          -           & NE$^{4,7}$    &   0.65$^{21}$  & 0.078$\pm$0.003$^{28}$ & -20.58$\pm$0.20$^{24}$\\ 
NGC 4649  & 0.0034 & 16.55   & 12 43 40.01 &  11 33 09.40 & 0.86$\pm$0.00 & 0.106$\pm$0.002 &   -  &          -           &  N$^{1,2}$        & 0.91$^{10}$ & 0.029$\pm$0.001$^{28}$ & -21.28$\pm$0.11$^{24}$\\
NGC 4696  & 0.0098 & 37.48     & 12 48 49.28 & -41 18 39.92 & 1.40$\pm$0.00 & 2.49$\pm$0.01 &   -  &          -           &  E$^{11}$     & 4.67$^{11}$ & 3.98$\pm$0.11$^{33}$ & -21.66$\pm$0.17$^{25}$\\
NGC 4778 & 0.0137 & 59.29$^*$ & 12 53 05.6  & -09 12 21    & 0.81$\pm$0.00 & 0.59$\pm$0.01 &   -  &          -           & NE$^{1,2}$        & 0.157$^{13}$ & 0.0049$\pm$0.0005$^{28}$ & -20.39$\pm$0.23$^{24}$\\
NGC 4782  & 0.0133 & 48.63     & 12 54 35.70 & -12 34 06.92 & 0.72$\pm$0.03 & 0.05$\pm$0.01 & 0.048$\pm$0.007 &   0.48$\pm$0.07 & NE$^{1,2,4}$    &   1.02$^4$  & 7.0$\pm$0.6$^{28}$  & -20.74$\pm$0.31$^{24}$\\ 
NGC 4936  & 0.0103 & 31.36     & 13 04 17.09 & -30 31 34.71 & 0.89$\pm$0.04 & 0.06$\pm$0.01 &   -   &          -           &  E$^{1,2}$        &   1.54$^4$  & 0.040$\pm$0.002$^{28}$  & -20.72$\pm$0.16$^{25}$\\ 
NGC 5044  & 0.009  & 35.75     & 13 15 23.97 & -16 23 08.00 & 0.85$\pm$0.00 & 1.29$\pm$0.01 & 0.041$\pm$0.002 &   1.50$\pm$0.07  &  E$^{1,2}$    &  6.0$^{21}$  & 0.035$\pm$0.001$^{28}$  & -20.94$\pm$0.41$^{24}$\\ 
NGC 5129  & 0.0230 & 86.85     & 13 24 10.00 &  13 58 35.19 & 0.79$\pm$0.02 & 0.23$\pm$0.02 &   -  &          -           & NE$^{1,2}$        &       -     & 0.0072$^{31}$  & -21.67$\pm$0.22$^{24}$\\ 
NGC 5419  & 0.0139 & 50.87     & 14 03 38.77 & -33 58 42.20 & 1.19$\pm$0.06 & 0.24$\pm$0.02 & 0.71$\pm$0.07 &   4.72$\pm$0.47    -   &  N$^{1,2}$  &   1.28$^4$  & 0.79$\pm$0.06$^{28}$ & -21.63$\pm$0.36$^{24}$\\ 
NGC 5813  & 0.0064 & 29.23   & 15 01 11.27 &  01 42 07.09 & 0.67$\pm$0.00 & 0.497$\pm$0.003 & 0.008$\pm$0.001 &   0.24$\pm$0.03 &  E$^{4,7,19}$ & 1.16$^{21}$  & 0.015$\pm$0.001$^{28}$ & -20.88$\pm$0.24$^{24}$\\ 
NGC 5846  & 0.0057 & 27.13     & 15 06 29.25 &  01 36 20.29 & 0.66$\pm$0.00 & 0.29$\pm$0.01 &   -  &          -           & E$^{1,2}$        &   2.47$^{21}$  & 0.021$\pm$0.001$^{28}$ & -21.12$\pm$0.25$^{24}$\\ 
NGC 6407  & 0.0154 & 64.93     & 17 44 57.66 & -60 44 23.28 & 0.86$\pm$0.05 & 0.54$\pm$0.09 &  -   &          -           &  U$^{1,2}$        & $<$0.03$^4$ & - & -21.18$\pm$0.15$^{25}$\\ 
NGC 6861  & 0.0094 & 30.09   & 20 07 19.48 & -48 22 12.94 & 0.97$\pm$0.02 & 0.051$\pm$0.004 & 0.076$\pm$0.007 & 0.26$\pm$0.02    & E$^{1,2}$    &   2.15$^4$  & - & -20.27$\pm$0.22$^{24}$\\ 
NGC 6868  & 0.0094 & 32.32   & 20 09 54.08 & -48 22 46.25 & 0.71$\pm$0.01 & 0.046$\pm$0.003 & 0.14$\pm$0.01 &    3.41$\pm$0.24    & E$^{1,2}$  &   3.46$^4$  & - & -20.89$\pm$0.23$^{24}$\\ 
NGC 7619  & 0.0132 & 50.53     & 23 20 14.52 &  08 12 22.63 & 0.85$\pm$0.01 & 0.14$\pm$0.01 &  -  &          -           &  N$^{1,2}$        &   1.86$^4$  & 0.02$\pm$0.001$^{28}$ & -21.42$\pm$0.23$^{24}$\\ 
NGC 7796  & 0.0113 & 50.06   & 23 58 59.81 & -55 27 30.12 & 0.60$\pm$0.01 & 0.035$\pm$0.004 &   -   &          -           &  N$^{1,2}$        & $<$0.01$^4$ & - & -21.04$\pm$0.22$^{24}$\\
\hline
\end{tabular}}
\footnotesize{* redshift-independent distances were not available for these sources.\\
References for column 8, 10, 11, 12 and 13: 1. \cite{Connor2018} (in prep.) 2. \cite{Sun2018} (in prep.)  3. \cite{Grossova2018} (in prep.) 4. \cite{Macchetto1996} 5. \cite{Plana1998} 
6. \cite{Mackie1998} 
7. \cite{Werner2014} 8. \cite{Ferrarese1996} 9. \cite{Bower1997} 10. \cite{Trinchieri1991} 11. \cite{Fabian2016}
12. \cite{Phillips1986} 13. \cite{Valluri1996} 14. \cite{Ho1997} 15. \cite{Annibali2010} 16. \cite{Blanton2004} 17. \cite{Kulkarni2014} 18. \cite{Ferrarese1996} 
19. \cite{Randall2011} 20. \cite{GonzalezMartin2009} 21. \cite{Caon2000} 22. \cite{Gavazzi2000} 23. \cite{Smith1989}  24. \cite{deVaucouleurs1991} 25.\cite{Lauberts1989} 
26. \cite{Gavazzi1996} 27. \cite{LaingPeacock1980} 28. \cite{Brown2011} 29. \cite{Oosterloo2007} 30. \cite{Condon1998} 31. \cite{Condon2002} 32. \cite{Vollmer2004} 33. \cite{Kuehr1981} \\}
\end{table*}

\section{Sample and Data}
\subsection{Sample Selection}
\label{sec:sampl_sel}
We started with the parent sample of \cite{Dunn2010} and selected 52 galaxies within 100 Mpc for which archival \textit{Chandra} X-ray observations were available. 
We also included an additional 16 X-ray and optically bright galaxies which were missing from the original selection 
(e.g., NGC5813). To make our sample represent the actual population of nearby bright 
ellipticals, we selected our final sample based on the intrinsic properties of the galaxies (the X-ray luminosity of the hot gas 
and the absolute visible band magnitude). 

The 0.5--7.0 keV X-ray luminosities at $r<10$~kpc for all the galaxies (see \S\ref{sec:xray_data_red_analys} and \S\ref{sec:global_properties}) 
are given in Table \ref{tab:sample_info}.  We obtained the visible band magnitudes for the entire sample 
from the NASA/IPAC Extragalactic Database \citep[NED;][]{NED}, which were then converted to absolute magnitudes (B$_T$) based on the mean 
redshift-independent distances given in NED (see Table \ref{tab:sample_info}). 
We applied a lower limit of 10$^{40}$ erg/s to the 0.5--7.0 keV X-ray halo luminosity and an upper limit of -20 to B$_T$. 
These selection criteria led to a sample size of 54. Due to short exposure times, the \textit{Chandra} observations of four of the 54 systems 
provided too few counts for the temperature within 10 kpc to be determined with a sufficient accuracy, rendering them unsuitable for detailed analysis. 
Also, the X-ray emission from the galaxy IC310 was 
found to be strongly dominated by the central point source. After excluding these five systems, our final sample was reduced to 49 galaxies.

For the H$\alpha$+[NII] information presented in this work, we have mainly used the results 
from an analysis of the H$\alpha$+[NII] observations 
carried out using the SOAR optical Imager (SOI) and Goodman High Throughput Spectrograph of the 4.1 m SOuthern Astrophysical Research 
(SOAR) telescope \citep[][in prep.]{Connor2018}, as well as the Apache Point Observatory (APO) Astrophysics Research Consortium (ARC) 3.5 m telescope \citep[][in prep.]{Sun2018}. 
Note that, the H$\alpha$+[NII] morphology information presented in this paper comes from the APO and SOAR data, while the luminosities are taken from the literature.

Significant H$\alpha$+[NII] emission was detected in about half (24/49) 
of the galaxies. Based on the SOAR/APO results, 
we classified the H$\alpha$+[NII] morphologies of our sample into 4 categories. These include no 
cool gas emission (N; total 20 galaxies), nuclear emission (NE: 12 galaxies with H$\alpha$+[NII] emission extent $<$ 2 kpc), extended filamentary emission (E:  13 galaxies with H$\alpha$+[NII] emission extent $\geq$ 2 kpc), 
and unsure (U: 4 galaxies for which presence/absence of H$\alpha$+[NII] emission 
could not be confirmed). The 10$^{th}$ and 11$^{th}$ columns of 
table \ref{tab:sample_info} show the morphology class \citep[based on the SOAR/APO results][]{Connor2018,Sun2018} and the luminosities (from literature) of the H${\alpha}$+[NII] emission for the sample.

The detailed discussion on the imaging and spectroscopic data from SOAR and APO will be presented in two papers \citep{Connor2018,Sun2018}. The depth of the SOAR data is comparable with the APO data, both for imaging and spectroscopy. Both telescopes have similar mirror sizes and similar optical instruments for imaging/spectroscopy. Within 9" diameter of the nucleus, the 5-$\sigma$ limit reached by our data is $5\times 10^{-15}$ erg s$^{-1}$ cm$^{-2}$-- $2\times 10^{-14}$ erg s$^{-1}$ cm$^{-2}$, depending on the continuum brightness. Since the limit is essentially an 
equivalent width (EW) limit, our final constraint on the emission-line luminosity is better than $5\times 10^{39}$ erg s$^{-1}$ \citep[see][for comparison]{Werner2014}. 

Some of the H${\alpha}$+[NII] flux estimates in Table \ref{tab:sample_info}, are taken from  \cite{Macchetto1996} and are based on narrow band H$\alpha$+[NII] images. 
In these observations, the stellar continuum is 
removed by subtracting a scaled broad R-band
image from the narrow band H$\alpha$+[NII] image. The scaling factor is a critical parameter in such an analysis, and
various factors (e.g., a non-uniform colour across the field)
may lead to wrong stellar continuum subtraction leading to spurious detections, especially when the 
H$\alpha$+[NII] emission is uniform (non-filamentary). 
As an example, for the galaxies NGC~1399 and NGC~4472, \cite{Macchetto1996} detected significant 
emission with disk like morphologies, although no significant H$\alpha$+[NII] emission was detected in the SOAR images and spectra. 
Therefore, we caution our readers that some of the disk-like emission detected in \cite{Macchetto1996} might be an artefact, and hence the accuracy of the 
flux estimates is limited by that of the stellar subtraction.

\subsection{X-ray Data Reduction and Analysis}
\label{sec:xray_data_red_analys}

\subsubsection{Data Reduction} 
We obtained the publicly available {\it Chandra} observations for our sample 
from the High Energy Astrophysics Science Archive Research Centre (HEASARC). The observation log 
for all the data used in the analysis is given in Table \ref{tab:obsn_log}. We used 
CIAO version 4.9 \citep{Fruscione2006} and CALDB version 4.7.3 for the data reduction, and the X-ray spectral 
fitting package XSPEC version 12.9.1 (AtomDB version 3.0.7) \citep{Arnaud1996} for the spectral analyses. 
Throughout the paper, the metallicities are given with respect to the Solar abundances of \cite{Grevesse1998}. All the data were reprocessed 
using the standard \textit{chandra\_repro} tool. Periods of strong background flares 
were filtered using the \textit{lc\_clean} script, and the threshold was set to match the blanksky background
maps. Point sources were detected using the CIAO task \textit{wavdetect} with a false-positive probability threshold 
of 10$^{-6}$, they were verified by visual inspection of the X-ray images and finally 
filtered (except for the central point sources, see \S\ref{sec:cen_PS}) from the event files. 
Note that, the point source detection is prone to be affected by the quality of data and S/N ratio, especially for faint sources.

\subsubsection{Central X-ray Point Sources}
\label{sec:cen_PS}
For the galaxies for which central point sources (coinciding with the galaxy's X-ray emission peak) were detected, a visual inspection 
was not sufficient for verification. For these central sources, X-ray spectra were extracted from the central regions of radius 3 pixels 
(1.476$"$). The spectra were first modeled with a \textit{wabs*apec} model and then with a \textit{wabs(apec+pow)} model in XSPEC; 
the power-law index was frozen to 1.5\footnote{To avoid the degeneracy between the \textit{apec} and \textit{pow} components, 
it was required to freeze the power law index. The value of 1.5 is consistent with the values typically seen for such sources 
\citep[e.g.][]{David2009}}. For some of the sources an additional 
absorption \textit{zwabs} model was required with the power-law to account for the intrinsic absorption of the AGN. The sources 
were confirmed if the addition of the powerlaw component lead to a significant improvement in the fit. 
In the end, central point sources were confirmed in 16 of the 49 galaxies of our sample. Interestingly, 11 of the 16 galaxies 
were found in systems containing cool gas (NE and E) while only three were in systems with no detectable optical emission line nebulae (N); 
the remaining two galaxies were in the unsure (U) systems. Note however that based on the radio flux densities given in the literature 
\citep[e.g.][]{Brown2011,Dunn2010}, practically all galaxies in our sample harbour central radio sources (except NGC 2305 for which we did not find a reported detection).

The intrinsic 2--10 keV central AGN luminosities estimated from the power-law components of the spectral models and their ratio with the 
Eddington luminosities are given in the eighth and ninth columns 
of Table \ref{tab:sample_info}, respectively. The Eddington luminosities were calculated using the relation 
$L_{\rm Edd}=1.26\times10^{47} (M_{\rm BH}/10^9 M_{\odot})$ erg/s \citep{Russell2013}. 
The Black Hole masses ($M_{\rm BH}$) were estimated from the empirical correlation of $M_{BH}$-$\sigma_*$ relation 
\citep[][]{Gebhardt2000,Tremaine2002}, $M_{\rm BH}=10^{8.13} ( \sigma_* /200\;{\rm km\;s^{-1}})^{4.02}$ $M_{\odot}$. 
The velocity dispersions ($\sigma_*$) were obtained from the Hyperleda database \citep{Hyperleda}. 
For NGC 4486 (M87), the central source was heavily affected by pile-up, therefore 
the AGN luminosity of \cite{GonzalezMartin2009} was used. All the AGNs are found to have very low Eddington ratios ($<$10$^{-5}$)
and seem to be operating in the gentle radio mechanical feedback mode. For the remaining analyses, the central point sources for all the 
16 galaxies were removed by excluding the central regions of radius 3 pixels (1.476").

\subsubsection{Spectral Extraction} For this study we determined emission weighted average properties within a radius of 10 kpc 
as well as the deprojected radial profiles of the thermodynamic properties of the hot diffuse halos. For the average properties, 
we restricted the spectral analyses to $r<10$ kpc \citep[$\sim R_{\rm e}$ for most of the galaxies, see][]{Goulding2016} since the hot gas 
properties within this region are dominated by the galaxy scale physics. For this, spectra were extracted from a circular region within 
$r<10$ kpc centred on the galaxy's X-ray peak, using the CIAO task \textit{specextract}. For the radial 
profiles, spectra were extracted from a number of circular annuli centred on the X-ray peak. 
The radial ranges of the annuli were chosen based on the requirement that each annulus should have at least 100 counts in the 0.5--5 keV energy range. 
The total number of annuli was limited to be $\leq$25. For the radial analyses, the spectra from the outermost annuli in some of the galaxies may be 
contaminated by the emission from the surrounding group or cluster. However, the values obtained for these annuli do not affect our main results.

\subsubsection{Background Spectra} For each source spectrum, corresponding background spectra were extracted  
from the standard Chandra blanksky background event files matching the source observations, 
obtained from Maxim Markevitch's blank-sky background database. The event files were reprojected to match 
the source observations. To match the time-dependent particle background levels in the source and 
blanksky observations, all the blanksky spectra were scaled by the ratio of the 9.5--12 keV count rates of the source 
and blanksky observations.

We also checked for contamination by soft Galactic 
foreground (most significant in the outermost annuli) and for differences in the Galactic 
foreground level in the scaled blanksky and source spectra. For this, we obtained the {\it ROSAT} All Sky Survey 0.47--1.21 keV 
(RASS 45 band) count rates from the outer 0.7--1.0 degree annular regions around each galaxy 
using the HEASOFT X-ray background tool. These count-rates were compared with the 
source count-rates in the outermost annuli. For most of our galaxies (39/49), the R45 
count rates were $<$10\% of the total 0.47--1.21 keV count-rates. 

We chose the two worst affected sources, NGC~4552 and NGC~4778 (HCG~62), for which the RASS R45 count rate was $\gsim$20\% of the total 0.47--1.21~keV count-rate. The 
annuli spectra for these sources were analysed with two additional model components: a 0.25 keV \textit{APEC} component with one-third 
solar metallicity for the Galactic halo and a 0.3 keV \textit{APEC} component with solar metallicity for the 
Local Hot Bubble; both non-deprojected. The normalizations of these components were allowed to be negative and their values for all outer annuli 
were tied to that of the innermost annulus with a multiplicative factor equal to the 
ratio of the respective areas. We found that the addition of the Galactic background components in the model for the two most affected galaxies 
lead to no significant changes in the final results. %Therefore, we believe that our results are not significantly affected by differences in the Galactic background between the source and the scaled blanksky observations.

\begin{figure*}
\centering
  \includegraphics[width=.48\linewidth]{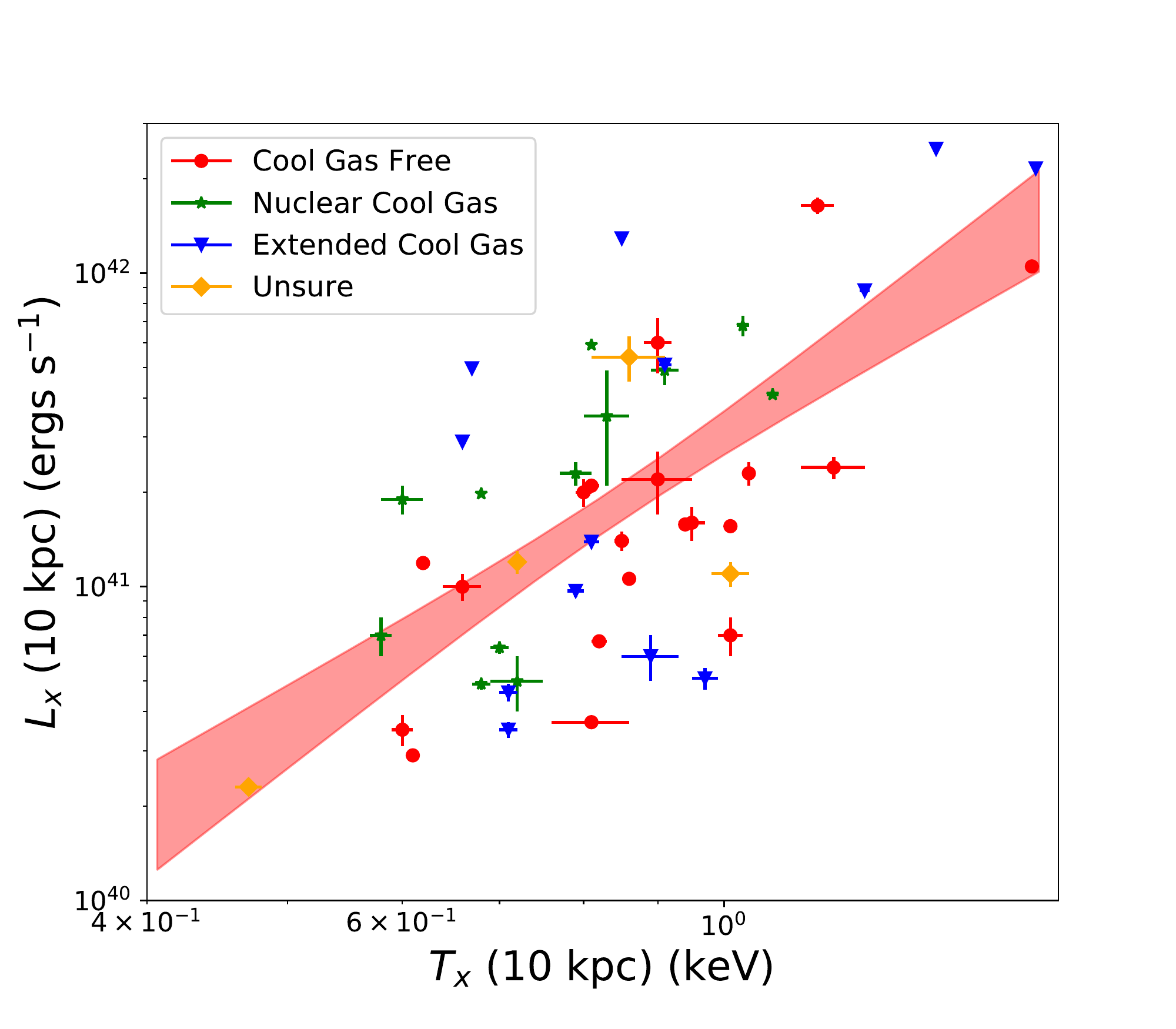}
  \includegraphics[width=.48\linewidth]{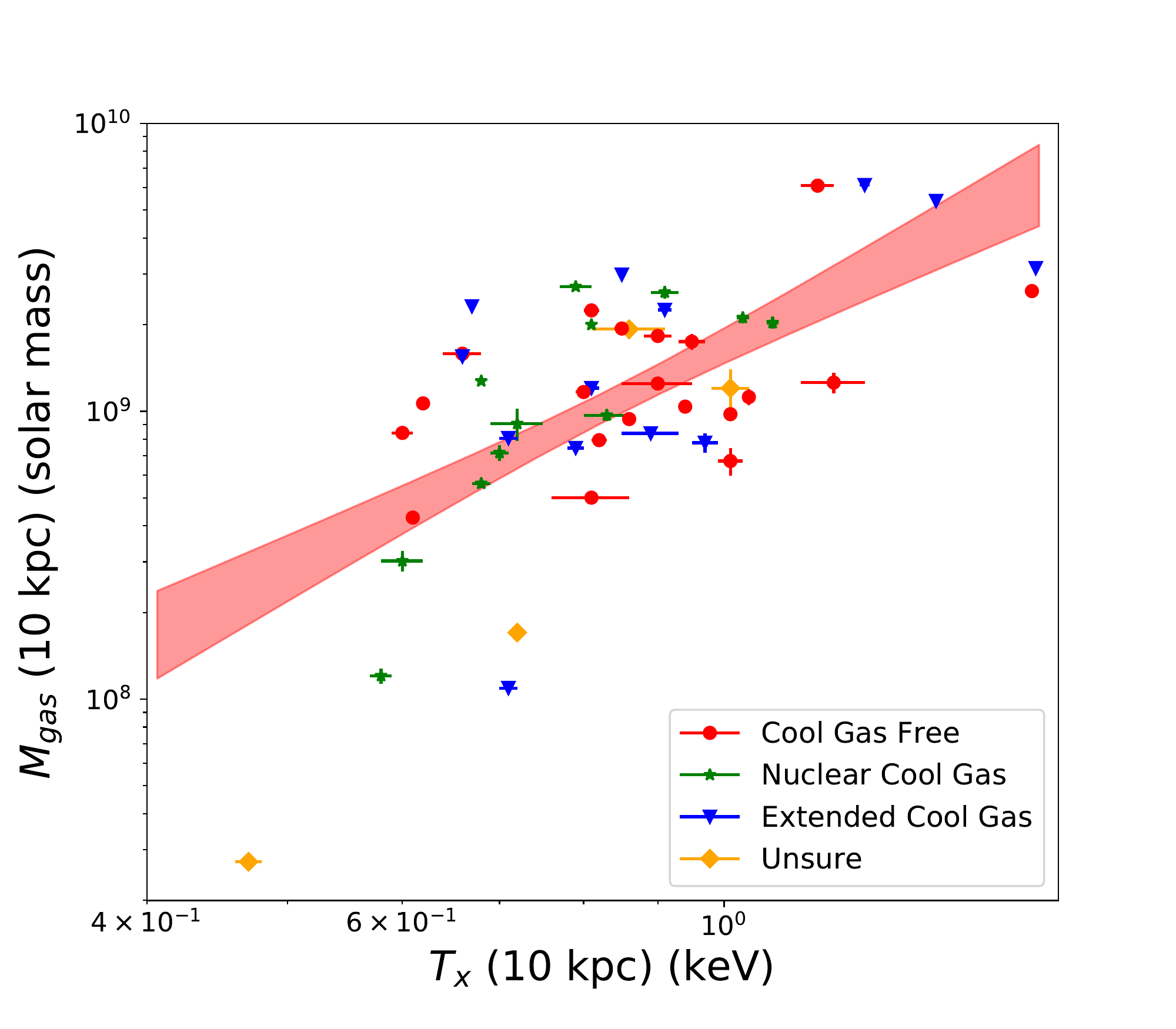}
  \includegraphics[width=.48\linewidth]{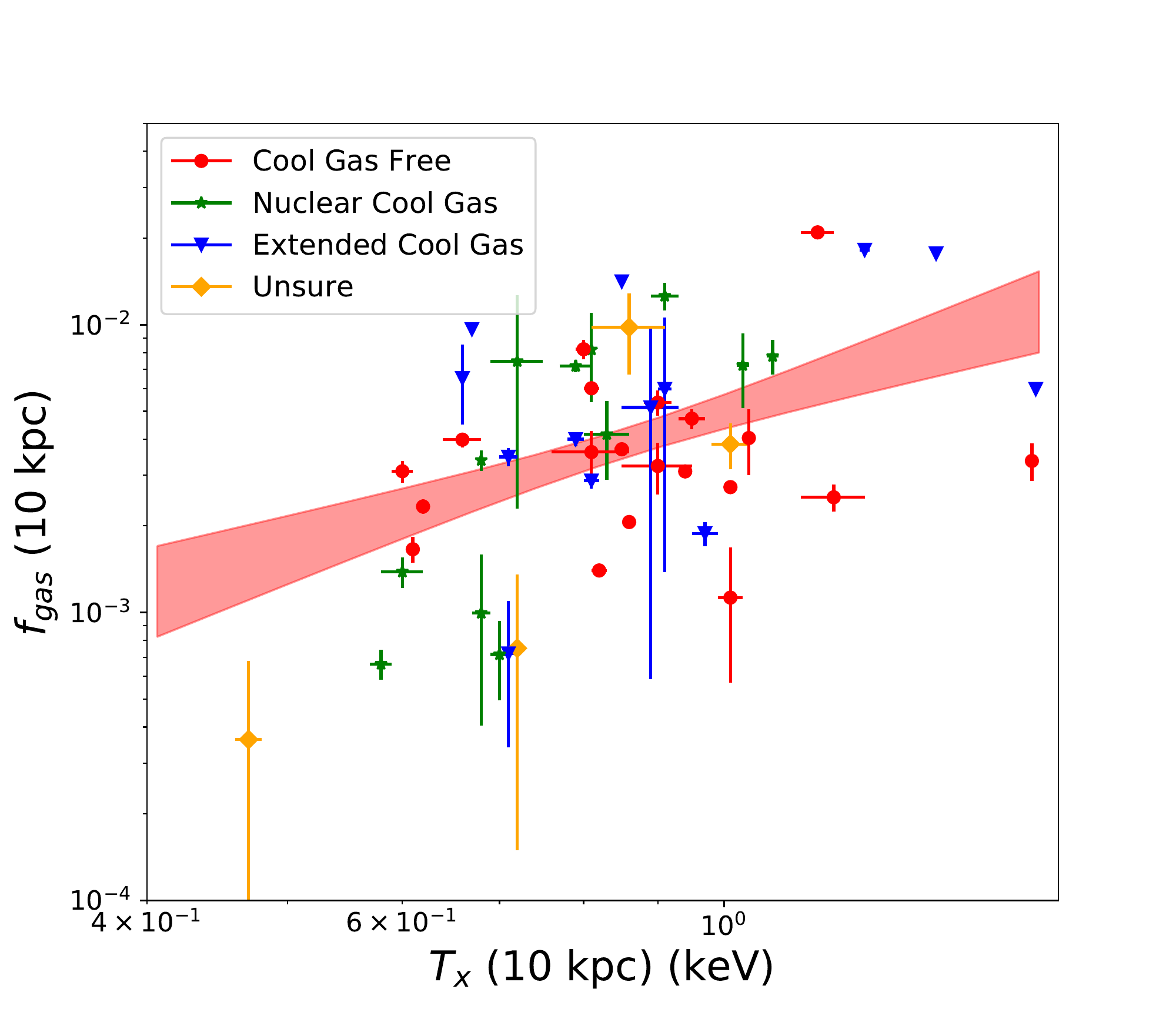}  
  \includegraphics[width=.48\linewidth]{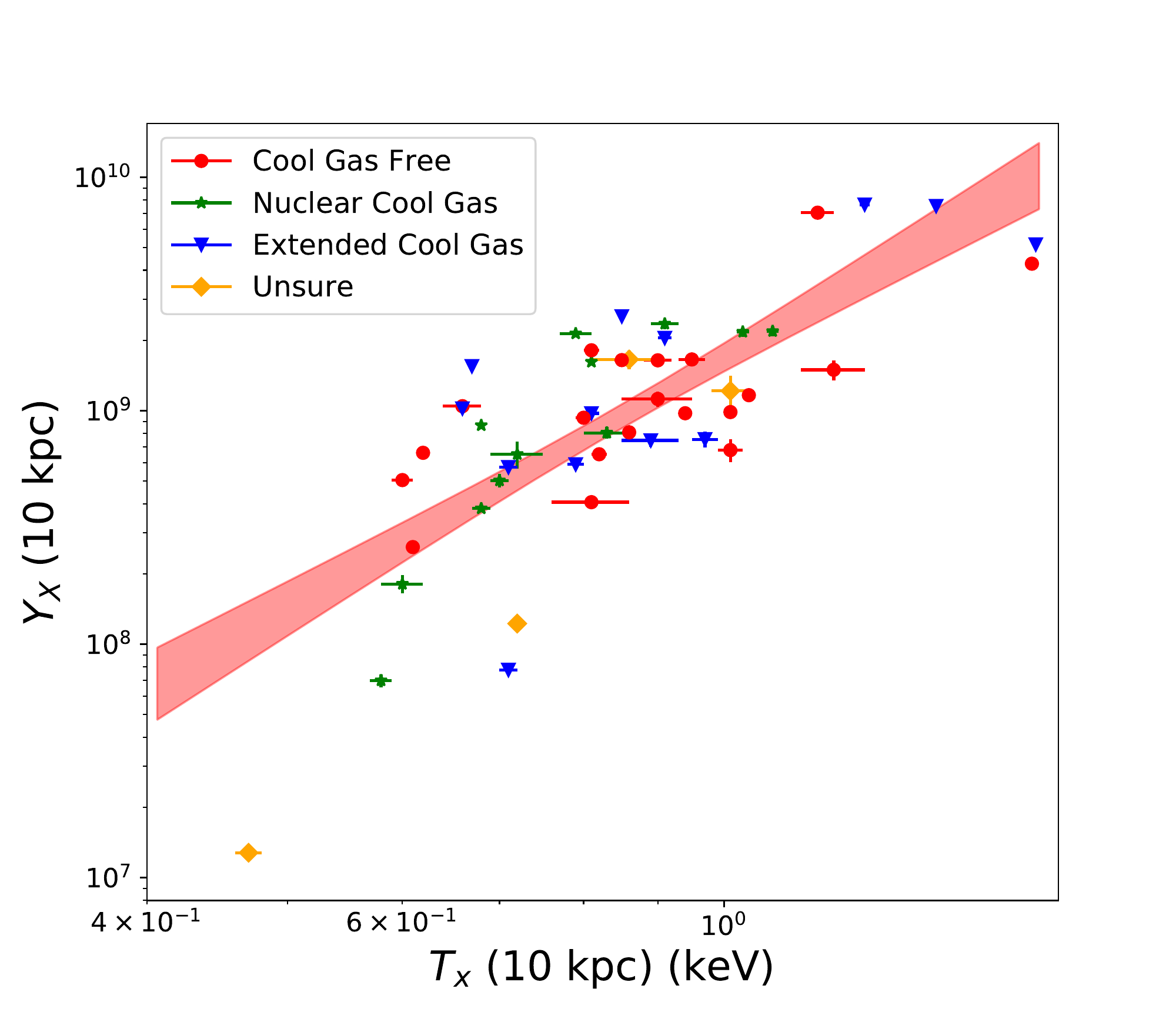}  
\caption{The 0.5--7.0 keV X-ray luminosities (upper left), the total gas masses (upper right), the gas mass fractions (lower left) and 
the $Y_{\rm X}=M_{\rm gas} T_{\rm X}$ (lower right) values, estimated from 
within a radius of 10 kpc plotted as a function of the gas temperatures determined from the same region. The red, green, blue and orange colors denote the cool gas free, 
nuclear cool gas, extended cool gas and unsure systems, respectively. The shaded regions show the best fit relations ($Y=A X^B$) of the y-axis variables with the x-axis, see Table \ref{tab:correlns}.}
\label{fig:comb_Lx_Mgas_fgas_10kpc_with_cg_morphs}
\end{figure*}

\subsubsection{Spectral Analysis} 
The spectra for the central $r=$10 kpc were fitted with a single-temperature absorbed 
\textit{VAPEC} model in XSPEC, using the C-statistics. A thermal bremsstrahlung component  with $kT=7.3$~keV was added 
to account for the unresolved point sources \citep[see][]{Irwin2003}. The neutral hydrogen column density was obtained 
from the \textit{Swift} Galactic $N_{\rm H}$ tool which gives the total (atomic$+$molecular) X-ray absorbing Hydrogen column 
density, using the method of \cite{Willingale2013}. The redshift was fixed to the value obtained from the 
SIMBAD database \citep{SIMBAD}. 
%For the global fits, the metallicities of all elements were initially kept free, and were later frozen to their best-fit values 
%(except for Mg, Si and Fe), restricted in the range of 0.3-3.0 times the solar values. 
The abundances of Mg, Si and Fe were kept free for this analysis. However, for the 
galaxies for which the abundances could not be constrained, they were frozen to one-third of the solar value, which 
is the value obtained for most of the galaxies in the sample.

\begin{figure*}
\centering
  \includegraphics[width=\linewidth]{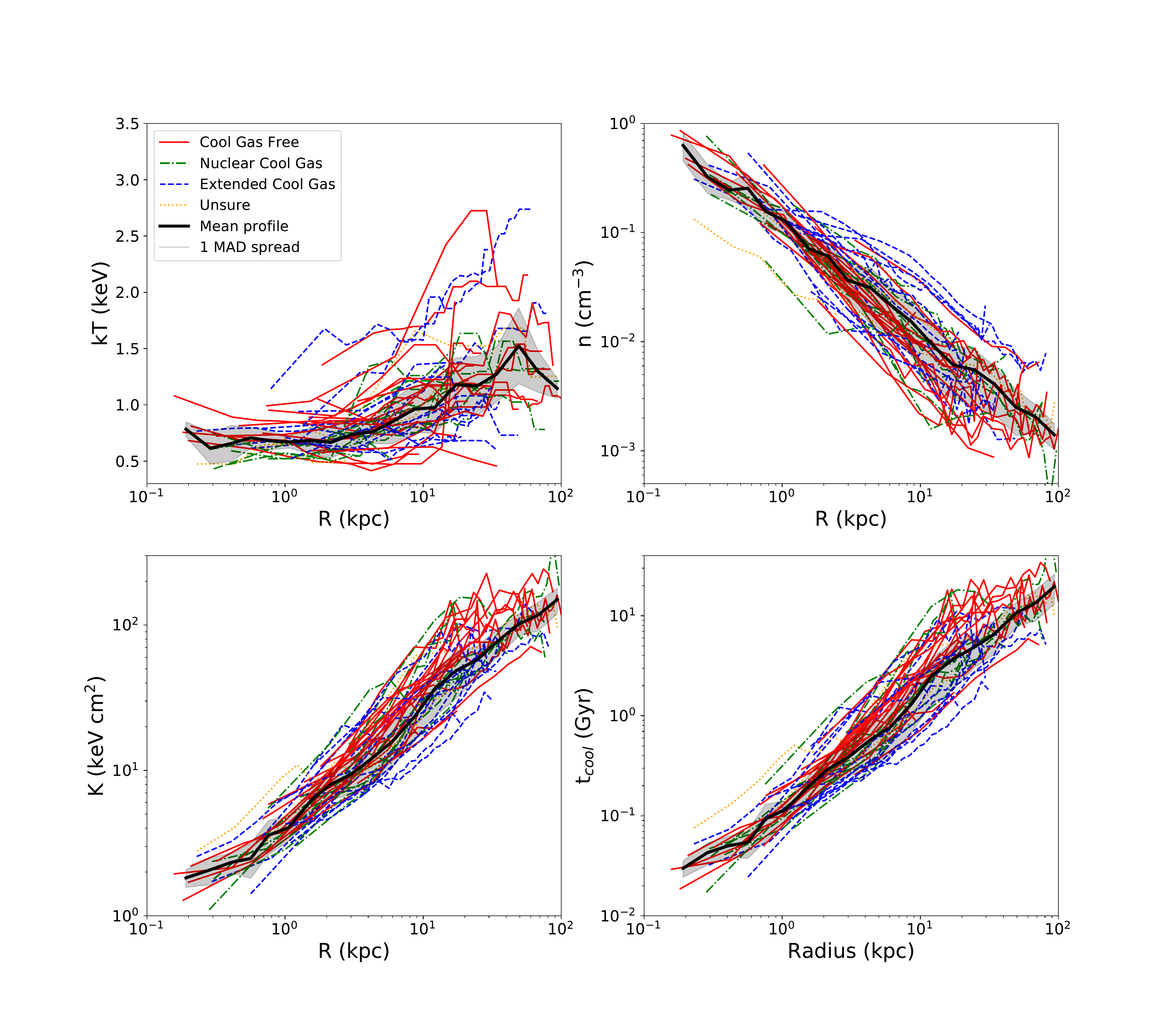}
  \vspace{-35pt}
\caption{The profiles of temperature (top left), density (top right), entropy (bottom left) and cooling time (bottom right) 
for the full sample (see \S\ref{sec:depr_profs}). 
The red (solid), green (dashed-dotted), blue (dashed) and orange (dotted) lines denote the cool gas free, 
nuclear cool gas, extended cool gas and unsure systems, respectively. The black line shows the median profile for the full sample and the grey 
shaded regions show the median absolute deviation (MAD) spreads about the medians.}
\label{fig:comb_profs_with_cg_morphs}
\end{figure*}

\begin{figure*}
\centering
  \includegraphics[width=\linewidth]{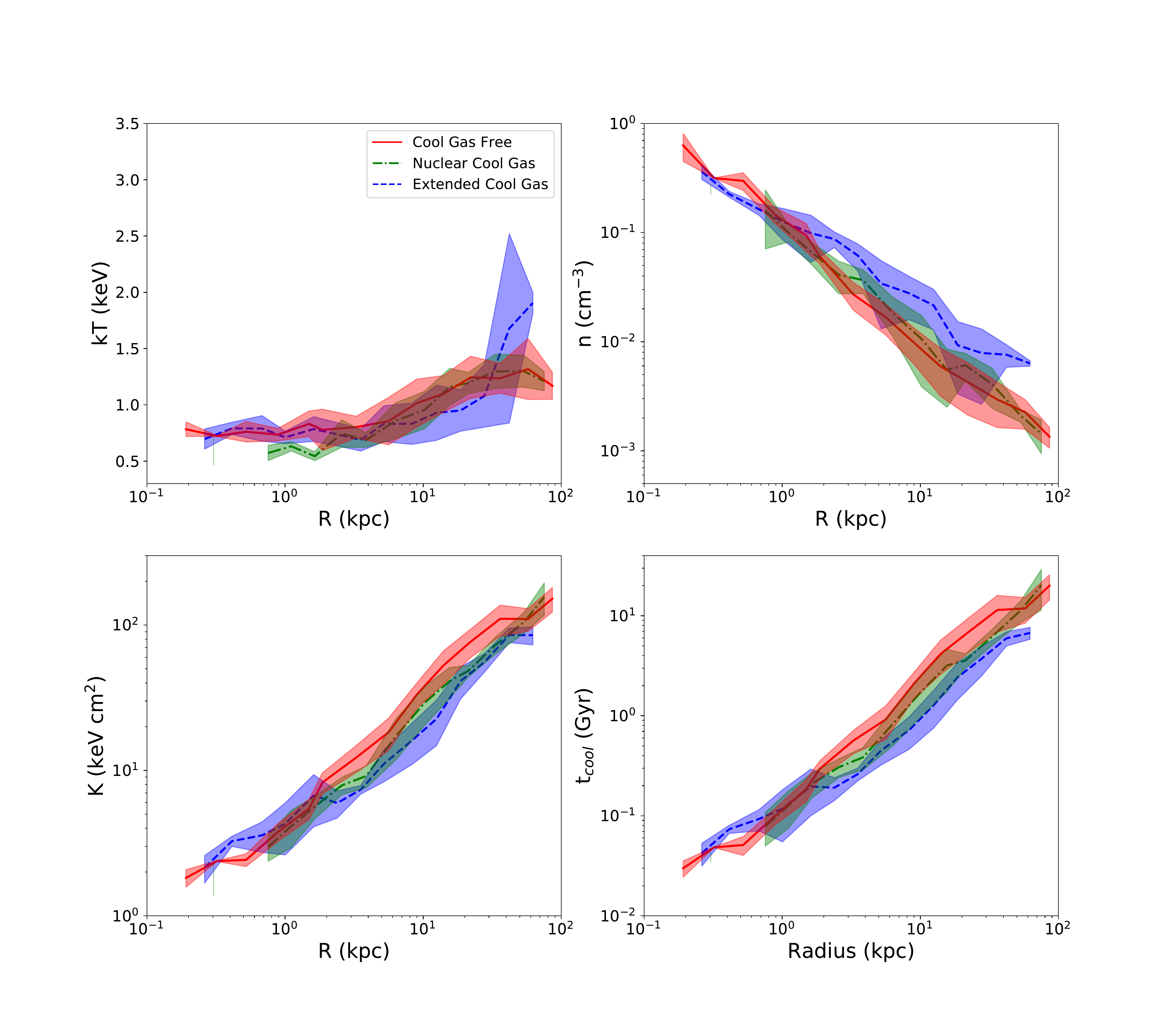}
  \vspace{-35pt}
\caption{The combined radially binned profiles of temperature (top left), density (top right), entropy (bottom left) and cooling time (bottom right) 
for the full sample (see \S\ref{sec:depr_profs}).  The red (solid), green (dashed-dotted) and blue (dashed) lines show median profiles for the cool gas 
free, nuclear cool gas and extended cool gas systems, respectively and the shaded regions 
show the median absolute deviation (MAD) spreads about the medians. The figure shows higher densities and lower entropies and cooling times for the extended cool gas galaxies 
than the rest of the sample, outside the innermost regions ($\sim$2 kpc).}
\label{fig:comb_band_profs_with_cg_morphs}
\end{figure*}

For the radial profiles, all the metallicities were frozen 
to the values obtained from the above analysis of the central 10 kpc radius regions. For low-temperature systems ($kT\sim$0.5--1.0 keV), this assumption might 
lead to an underestimation of metallicities and overestimation of densities 
\citep[see][]{Buote2000,Werner2008a}, particularly in the inner regions where the gas is expected to 
be multiphase. However, for many of the galaxies, the data quality did not allow us to resolve the multi-temperature structure and the metallicities could not be 
constrained for the individual annuli. Therefore, to analyse the entire sample in a uniform way, we assumed the central 10 kpc region 
metallicities when fitting the radial profiles. Note that underestimating the metallicity by a factor of 2, will result in 
overestimating the density by a factor of $\sim$1.35\footnote{We also checked for the effect of using projected metallicity profiles 
(with 2T apec models for the inner shells) instead of fixed metallicities, for two (a cool gas free and a cool gas rich) galaxies with high data quality. This lead to 
very similar changes ($<$10\% decrease in the densities, $<$20\% increase in the entropy and $\sim$8\% decrease in the slopes of the entropy profiles) in both the galaxies. We think that 
for all our galaxies, a free metallicity will shift all the density profiles slightly upwards and the entropy and cooling times profiles downwards, however, the general trends 
in Figs. \ref{fig:comb_profs_with_cg_morphs} and \ref{fig:comb_band_profs_with_cg_morphs} will remain the same.} \citep{Werner2012}.

The deprojection analysis to determine the radial profiles of thermodynamic quantities was performed using the \textit{projct} model in
XSPEC. The free parameters in the fit were the temperature and normalization of the 
\textit{APEC} component and the normalization of the bremsstrahlung component (not deprojected). We assumed a constant density and temperature 
in each 3D shell. The \textit{APEC} normalizations ($\eta$) were converted to the individual shell gas densities ($n=n_e+n_i$) using the relation 
\begin{equation}
\eta = \frac{10^{-14} \int{n_e n_p dV} }{4 \pi {D_{A}}^{2} (1+{\rm z})^2}.
\end{equation}
Here $D_A$ is the angular diameter 
distance to the source, $n_e$ and $n_p$ are the electron and proton number densities, where for a fully 
ionised gas with one-third solar abundance $n_e=0.53n$ and $n_p=n_e/1.2$. The densities and temperatures were used to calculate the gas entropy 
($K=kTn_e^{-2/3}$), pressure ($P=nkT$), and cooling time ($t_{\rm cool}=1.5nkT/(n_{\rm e} n_{\rm i} \Lambda(T,Z))$, where $\Lambda(T,Z)$ is the cooling 
function and Z is the metallicity; note that we are using here the metallicities obtained from the central 10 kpc region spectral analysis).

\begin{figure*}
\centering
\includegraphics[width=.49\linewidth]{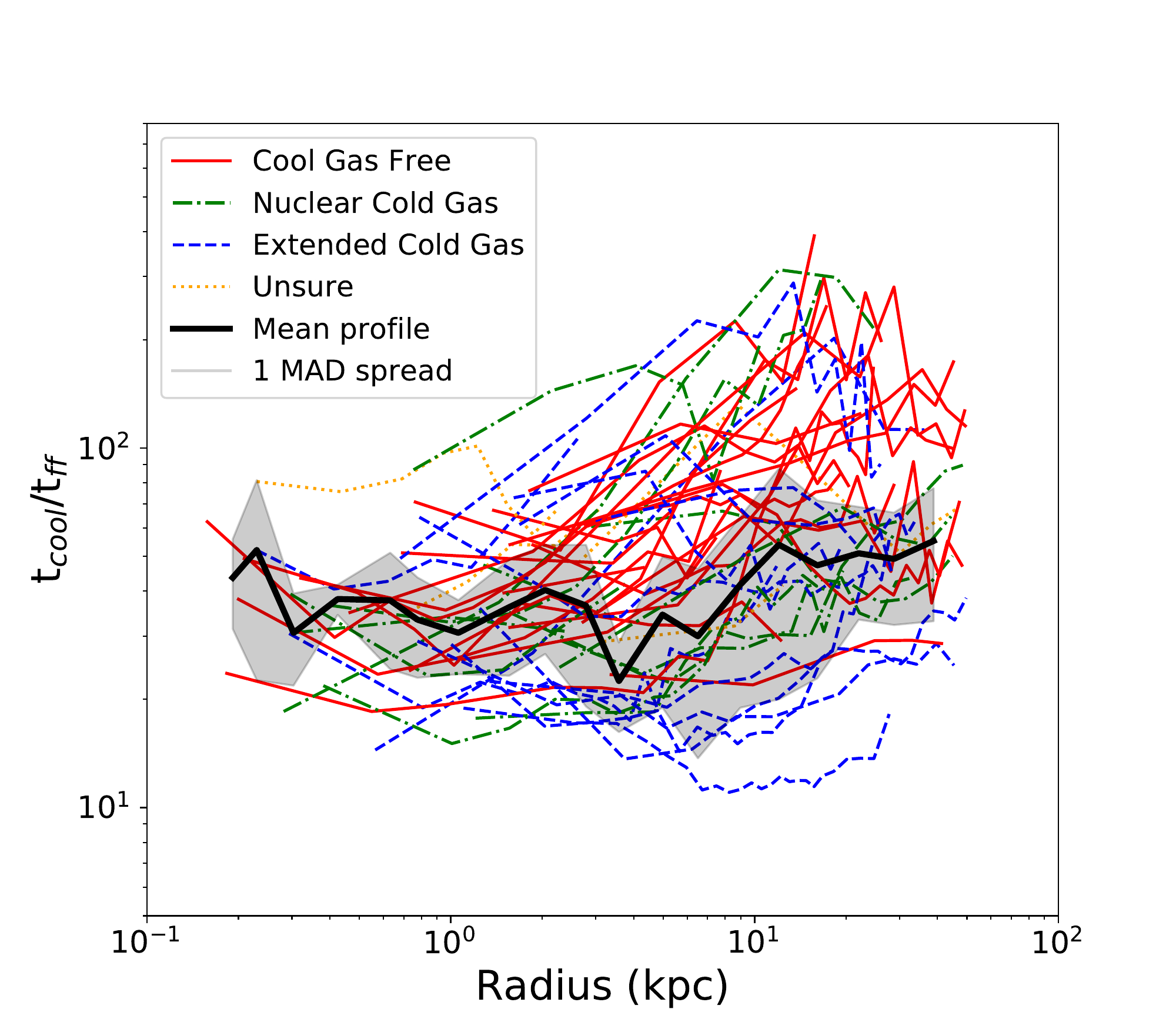}
\includegraphics[width=.49\linewidth]{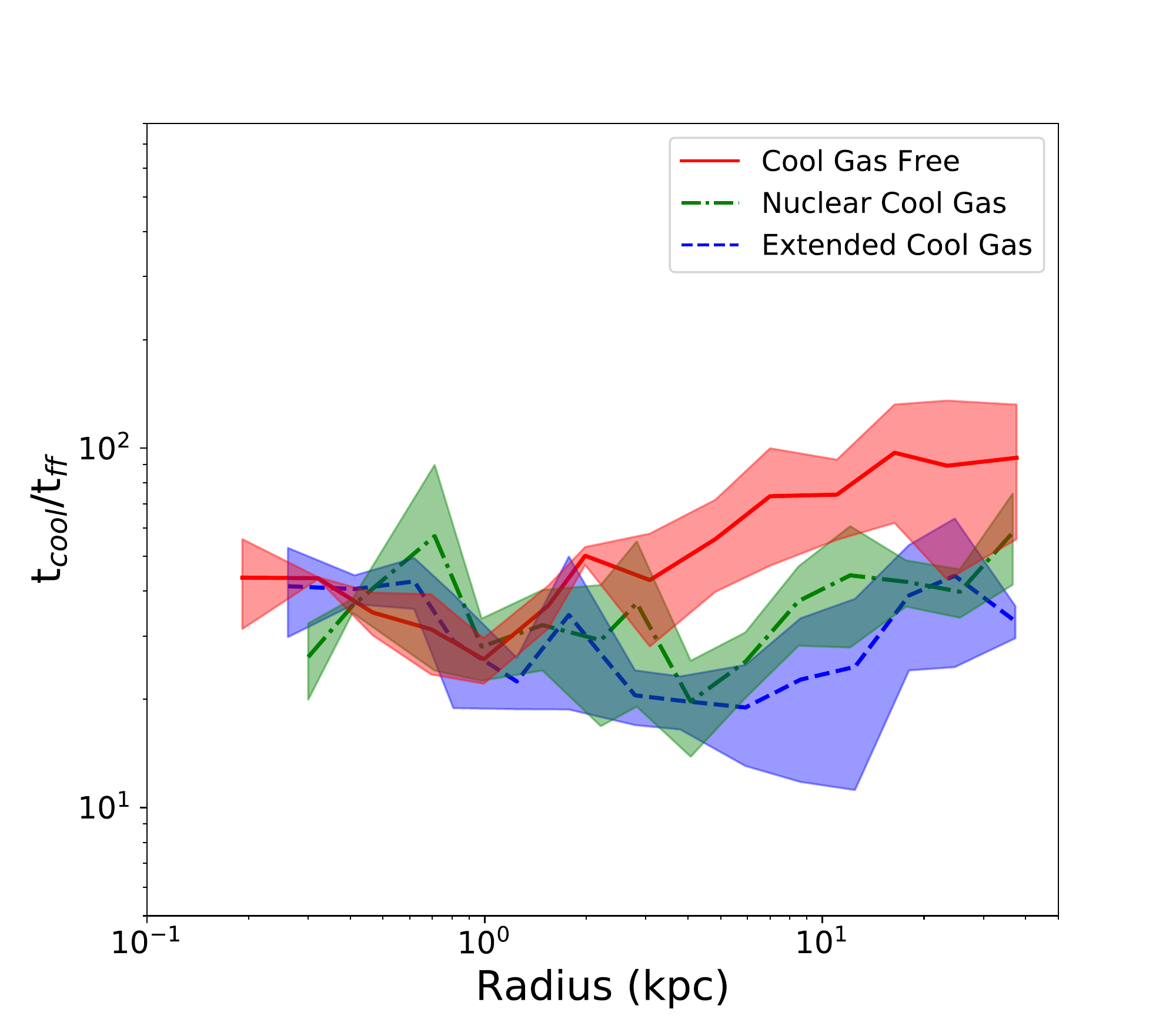}
\includegraphics[width=.49\linewidth]{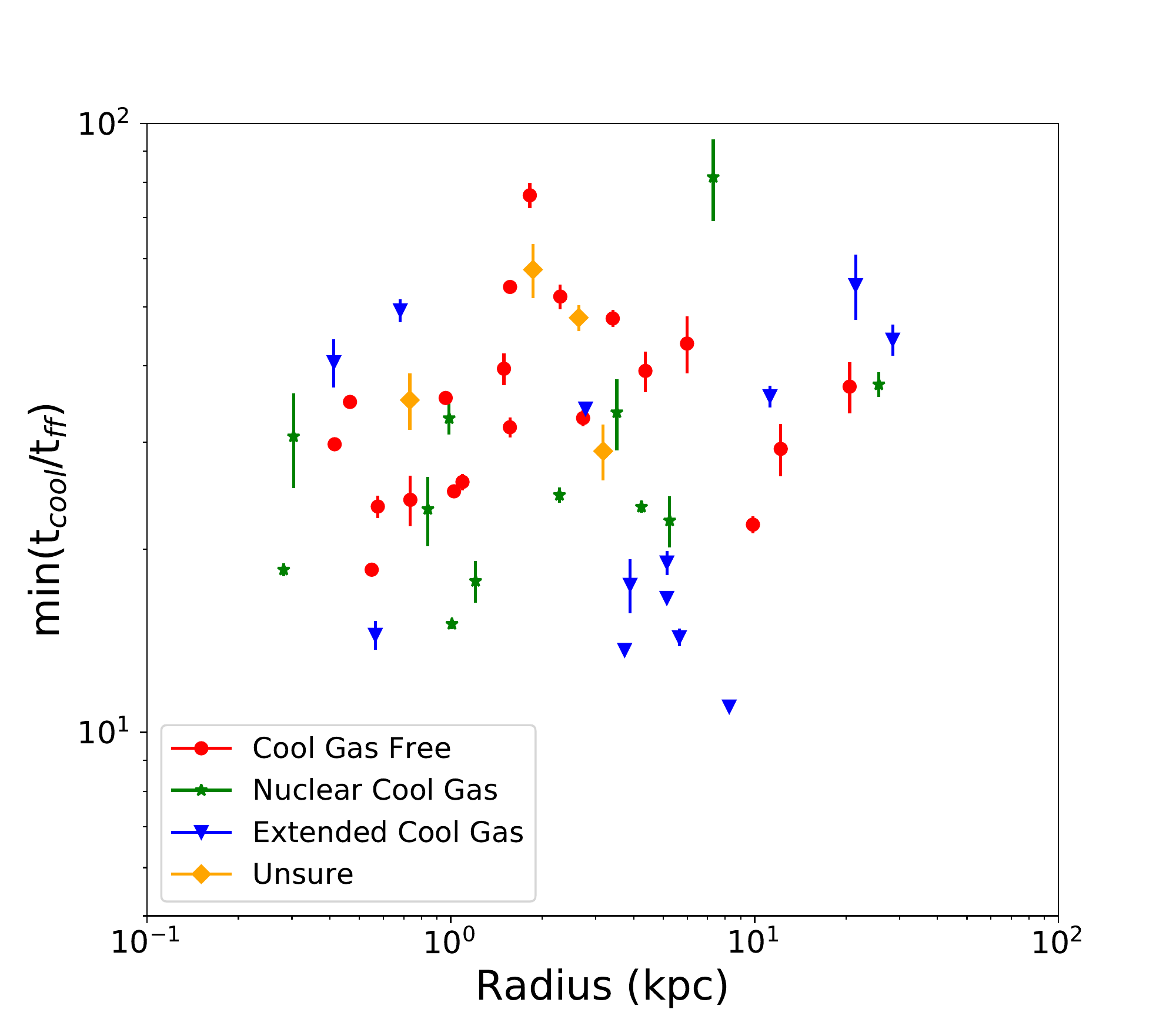}
\caption{The $t_{\rm cool}/t_{\rm ff}$ profiles of the full sample (top left), the radially binned combined $t_{\rm cool}/t_{\rm ff}$ profiles of the 
different cold gas morphology groups (top right) and the min($t_{\rm cool}/t_{\rm ff}$) values of the full sample (bottom panel) (see \S\ref{sec:bimodality}). 
The red (solid lines/circles), green (dashed-dotted lines/stars), blue (dashed lines/triangles) and orange (dotted lines/diamonds) symbols denote the cool gas free, 
nuclear cool gas, extended cool gas and unsure systems, respectively. The black solid lines in the top left panel, and the red (solid), green (dashed-dotted) and blue (dashed) lines in the top 
right panel show the median profiles for the full sample, and the 
cool gas free, nuclear cool gas and extended cool gas galaxies, respectively.  
The shaded regions show the median absolute deviation (MAD) spreads about the medians. The presence 
of cool gas seems to be preferred in systems with lower values of 
min($t_{\rm cool}/t_{\rm ff}$).}
\label{fig:comb_tcbtf_profs_mintcbtf_with_cg_morphs}
\end{figure*}
%\begin{figure*}
%\centering
%  \includegraphics[width=.63\linewidth,angle=90]{Figures/combined_Mgas_10kpc_with_CG_morphologies_new_classfn.pdf}
%  \includegraphics[width=.63\linewidth,angle=90]{Figures/combined_fgas_10kpc_with_CG_morphologies_new_classfn.pdf}
%\caption{The total gas masses (left) and the gas mass fractions within a radius of 10 kpc, for the entire sample 
%(with cold gas morphology classification; color scheme same as 
%Fig.~\ref{fig:comb_profs_with_cg_morphs}). The three fast rotators (NGC1316, NGC6861 and NGC6868) have been 
%circled.}
%\label{fig:comb_Mgas_fgas_10kpc_with_cg_morphs}
%\end{figure*}

\section{Results}
\label{sec:results}

\subsection{X-ray Properties within r=10 kpc}
\label{sec:global_properties}
The spectra for all the 49 galaxies extracted from circular regions of $r=10$~kpc around the X-ray peaks 
were analyzed as described in \S\ref{sec:xray_data_red_analys}. 
The resulting best-fit values of the gas temperatures span a range of values from 0.47~keV to 1.64 keV. 
The 0.5--7.0 keV X-ray luminosities ($L_{\rm X}$) determined within $r=10$~kpc span two orders of magnitude from $2.3\times10^{40}$~erg~s$^{-1}$ to $2.5\times10^{42}$~erg~s$^{-1}$. 
The temperatures and X-ray halo luminosities obtained for the entire sample are listed in Table~\ref{tab:sample_info}. The X-ray halo luminosities 
plotted vs. the average X-ray temperatures of the galaxies are shown in the top left panel of Fig. \ref{fig:comb_Lx_Mgas_fgas_10kpc_with_cg_morphs}.

\subsection{Deprojected Profiles}
\label{sec:depr_profs}
The deprojected temperature ($kT$), density ($n$), entropy ($K$), pressure ($P$) and cooling time ($t_{\rm cool}$) profiles of the individual galaxies 
are given in the appendix (Figs.~\ref{fig:kT_profs}, 
\ref{fig:n_profs}, \ref{fig:K_profs}, \ref{fig:P_profs} and \ref{fig:tcool_profs}, respectively). 
The $kT$, $n$, $K$ and cooling time ($t_{\rm cool}$) profiles of the full sample, classified based on the cool gas extents (see \S\ref{sec:sampl_sel}) 
are shown in Fig.~\ref{fig:comb_profs_with_cg_morphs}. The temperature profiles in the top left panel, do not show any distinction between the different 
cool gas morphology/extent groups.  
The density profiles show higher densities for the extended cool gas (blue)
galaxies than the rest of the galaxies. However, at least 3 (NGC 4936, NGC 6868 and IC 4296) of the 13 extended cool gas galaxies seem to 
have low densities. Note that, NGC 6868 was found to have  
indications for a rotating cold gas disk in the velocity  distribution maps of [CII] emission \citep[see][]{Werner2014}. It is possible that the cool gas in 
the low density galaxies is supported by rotation. 
In general, the profiles of entropy and cooling time show lower values for the extended cool gas galaxies 
than the cool gas free galaxies. 
The three outliers with low density and extended cool gas, also have higher 
entropies and cooling times than the rest of the extended cool gas galaxies.

To see the trends in the thermodynamic profiles and their scatter more clearly, in 
Fig. \ref{fig:comb_band_profs_with_cg_morphs} we show the median temperature, density, entropy and cooling time profiles of the cool gas free (red), 
nuclear cool gas (green) and extended cool gas (blue) groups. The profiles were obtained by finding the median values in 15 radial bins. The shaded regions 
show the median absolute deviations (MAD) about the medians for each group. The trends seen in Fig. \ref{fig:comb_profs_with_cg_morphs} 
are much more clearly visible in Fig. \ref{fig:comb_band_profs_with_cg_morphs}. The median temperature  profiles of the three 
groups are found to be very similar. The density profiles show higher 
values for the extended cool gas galaxies than the cool gas free galaxies. In the entropy and cooling time profiles also, the extended cool gas galaxies seem to 
have lower values than the cool gas free galaxies, especially outside the innermost regions ($\sim$2 kpc), but with significant spread. The nuclear cool gas galaxies 
are found to have densities, entropies and cooling times in between the extended cool gas and cool gas free galaxies.

\begin{table}
 \caption{Results of the linear correlation analysis, discussed in \S\ref{sec:bimodality}, for the 0.5--7.0 keV X-ray luminosities ($L_{\rm X}$; in erg s$^{-1}$), 
 the total gas masses ($M_{\rm gas}$; in $M_{\odot}$), the 
gas mass fractions ($f_{\rm gas}$) and the $Y_{\rm X}$ ($=M_{\rm gas} T_{\rm X}$; in $M_{\odot}$ keV) values, estimated from 
within a radius of 10 kpc with the gas temperatures ($T_{\rm X}$; in keV) determined from the same region (results also shown in Fig. \ref{fig:comb_Lx_Mgas_fgas_10kpc_with_cg_morphs}).}
\label{tab:correlns}
\vskip 0.5cm
\centering
{\scriptsize
\setlength\tabcolsep{4pt} 
\begin{tabular}{c c c c}
\hline
\hline
Relation                                    & Intercept    &  Slope  & Corr. Coeff.  \\
                                                  &                    &              &                      \\    
\hline
 $L_{\rm X}$-$T_{\rm X}$       & (3.1$\pm$0.5)$\times 10^{41}$ &  3.1$\pm$0.5 &  0.67$\pm$0.09  \\
 $M_{\rm gas}$-$T_{\rm X}$  & (1.7$\pm$0.3)$\times 10^9$      &  2.6$\pm$0.5 & 0.63$\pm$0.10 \\
 $f_{\rm gas}$-$T_{\rm X}$    &  0.005$\pm$0.001                      & 1.6$\pm$0.5 & 0.48$\pm$0.13 \\
 $Y_{\rm X}$-$T_{\rm X}$       &  (1.7$\pm$0.3)$\times 10^9$     & 3.6$\pm$0.5 & 0.75$\pm$0.07 \\

\hline
\end{tabular}}
\end{table}

\section{Discussion}
\label{sec:discussion}

\subsection{Cooling instabilities and the thermodynamic properties of galactic atmospheres}
\label{sec:bimodality}

\subsubsection{Correlation with average X-ray properties}
The distributions of X-ray luminosities ($L_{\rm X}$), gas masses ($M_{\rm gas}$), gas mass fractions ($f_{\rm gas}$) 
and the $Y_{\rm X}= M_{\rm gas} T_{\rm X}$ values, within $r<10$~kpc plotted vs. 
the average X-ray temperatures within the same region (obtained in \S\ref{sec:global_properties}), for the galaxies 
without and with different extents of cool gas are shown in Fig.~\ref{fig:comb_Lx_Mgas_fgas_10kpc_with_cg_morphs}. We 
obtained the gas mass estimates ($M_{\rm gas}$) and the gas mass fractions ($f_{\rm gas}$) for all the galaxies within the same 10 kpc radius circular regions. 
The gas masses were obtained by integrating the densities obtained in \S\ref{sec:depr_profs} ($M_{\rm gas} (r)\,=\,\int 4 \pi r^2\, \mu\, m_{\rm H}\, n\, dr$), 
and the gas mass fractions were obtained as $f_{\rm gas}(r)=M_{\rm gas}(r)/M_{\rm tot}(r)$. The total masses of the galaxies, $M_{\rm tot}(r)$ within a radius $r$, 
were obtained from the gas pressure gradients assuming hydrostatic equilibrium. The pressure gradients were determined using smooth empirical fits\footnote{We 
tried three different models: a powerlaw, a beta model and a 4-parameter model, described in \cite{Lakhchaura2016}.} to the pressure profiles 
obtained in \S\ref{sec:depr_profs}. 
We do not see any trends in $L_{\rm X}$, $M_{\rm gas}$ and $f_{\rm gas}$ with the presence or morphology/extent of cool gas. 

The linear correlation coefficients in log-space between $L_{\rm X}-T_{\rm X}$, $M_{\rm gas}-T_{\rm X}$, 
$f_{\rm gas}-T_{\rm X}$ and $Y_{\rm X}-T_{\rm X}$ were found to be 0.67$\pm$0.09, 0.63$\pm$0.10, 0.48$\pm$0.13 and 0.75$\pm$0.07 (obtained using the python {\it linmix} 
package), and the best-fit relations were found to be $L_{\rm X}\propto T_{\rm X}^{3.1\pm0.5}$, $M_{\rm gas}\propto T_{\rm X}^{2.6\pm0.5}$, $f_{\rm gas}\propto T_{\rm X}^{1.6\pm0.5}$, 
and $Y_{\rm X}\propto T_{\rm X}^{3.6\pm0.5}$, respectively. Our best-fit $L_{\rm X}-T_{\rm X}$ relation is shallower than that found in the group-cluster combined
 studies \citep[e.g., see][]{KimFabbiano2015,Goulding2016,Babyk2018a}. 
However, the relation is fully consistent with the ones obtained using group only samples \citep{Sun2012,Bharadwaj2015}. 
The results for all these linear correlations ($Y=A X^B$; in log space) viz., the intercepts (A), slopes (B) and correlation coefficients, 
are given in Table \ref{tab:correlns}. The (weak) positive correlation between the gas mass fractions and the X-ray 
temperatures suggests that the cores of hotter, more massive systems are able to hold on to a larger fraction of their X-ray emitting gas.

\subsubsection{Correlation with thermodynamic profiles}
The deprojected thermodynamic profiles shown in Figs.~\ref{fig:comb_profs_with_cg_morphs} 
and \ref{fig:comb_band_profs_with_cg_morphs} show higher densities for the extended cool gas galaxies (blue) as compared to the cool gas free galaxies (red). 
The galaxies with extended cool gas are also found to have lower entropies and cooling times than their cool gas free counterparts, outside of their innermost 
regions ($\sim$2 kpc). Using a subsample of 10 galaxies which are also included in this study, \cite{Werner2014} 
found a clear separation in the 
entropy profiles of galaxies with extended emission line filaments and cool gas free galaxies. In our study of a much larger sample, however, the separation 
is less clear and becomes more pronounced outside of the innermost regions at $r\gtrsim2$~kpc.

The large scatter seen in the thermodynamic profiles 
is consistent with the short duty cycles \citep[proportional to the cooling time at $r<0.1\;R_{500}$;][]{Gaspari2017} predicted by the CCA-regulated feedback in 
early type galaxies. As suggested by NGC 6868, the few low-density (high-entropy) outliers with cool gas can be understood by the fact that they might possess 
significant rotation. This was also observed in the massive lenticular galaxy NGC 7049 which has a high central entropy, despite having a cool H$\alpha$+[NII] disk \citep[see][]{Juranova2018}. 
Rotation can strongly reduce the SMBH accretion rate \citep{Gaspari2015}, inducing a long-term accumulation of cold/warm gas in the 
central region or in an extended disc, thus making the multiphase state uncorrelated with the current hot halo properties. 
The presence of rotational support can decrease the gravitational potential depth and hence, 
the X-ray surface-brightness. This has also been observed in simulations. Based on 2D high-resolution 
hydrodynamic simulations of early type galaxies, \cite{Negri2014} found that the hot X-ray emitting gas 
in fast-rotating galaxies has a systemically lower surface brightness than the hot gas in the non-rotating 
systems of similar masses. The effect has also been found in some other studies 
\citep{BrighentiMathews1996,Gaspari2015,Gaspari2017}.

Using numerical simulations it has been found that thermal instability is only significant when the cooling time of the gas is less than $\sim10$ free-fall times 
\citep[see][]{Sharma2012,McCourt2012,Gaspari2012a,Gaspari2013,Meece2015}. We calculated the profiles of free fall time for all our galaxies as $t_{\rm ff}=\sqrt{2r/g}$; 
where the acceleration due to gravity $g=d\phi/dr=2\sigma_{\rm c}^2/r$ (assuming an isothermal sphere potential $\phi=2 \sigma_c^2\; log(r) + const.$), 
leading to $t_{\rm ff}=r/ \sigma_{\rm c}$; where $\sigma_{\rm c}$ is the mean central 
velocity dispersion obtained from the Hyperleda database \citep{Hyperleda}.\footnote{We also tried calculating $g$ assuming hydrostatic equilibrium, 
$g=-\rho^{-1} dP/dr$ ($\rho=$ gas mass density $=\mu n m_p$; $\mu=$0.62; $m_p$=proton mass), 
obtained using smooth empirical fits to the pressure profiles. The 
contribution of non-thermal pressure in a small sub-sample of galaxies was checked by implementing the approach 
used in \cite{Churazov2010}. We found a maximum non-thermal pressure support of $\sim30\%$. Therefore, for the small radial distances 
concerned in this paper where the non-thermal pressure can be really significant, we decided to use $g$ obtained using the isothermal sphere potential. 
We found that using the latter method the $t_{\rm cool}/t_{\rm ff}$ values decrease in the outer regions and increase in the inner regions, 
although the min($t_{\rm cool}/t_{\rm ff}$) values are only slightly affected}. 

The $t_{\rm cool}/t_{\rm ff}$ profiles of the individual galaxies are shown in Fig. \ref{fig:tcoolbytff_profs}. 
The $t_{\rm cool}/t_{\rm ff}$ profiles and the 
minimum values of $t_{\rm cool}/t_{\rm ff}$, for the full sample are shown in the top left and bottom panels of Fig. \ref{fig:comb_tcbtf_profs_mintcbtf_with_cg_morphs}, respectively. 
In the top right panel we also show the median $t_{\rm cool}/t_{\rm ff}$ profiles of the cool gas free (red solid lines), nuclear cool gas (green dashed-dotted lines) and extended cool gas 
(blue dashed lines) groups with the shaded regions showing the median absolute deviation about the median profiles. 
The figure shows that galaxies with cool gas emission (extended$+$nuclear) have in general, lower values of $t_{\rm cool}$/$t_{\rm ff}$ than the cool gas free 
galaxies (red),  especially outside the innermost regions ($\sim$3 kpc). There seems to be a separation in the $t_{\rm cool}/t_{\rm ff}$ 
values of the cool gas rich (nuclear and extended) galaxies and the cool gas free galaxies outside $\sim$3 kpc. Also, the $t_{\rm cool}/t_{\rm ff}$ profiles of the extended 
cool gas galaxies seem to be flatter in the 3-10 kpc range as compared to the nuclear cool gas and cool gas free galaxies for which the values seem to be increasing with radius.

\subsubsection{Distributions of cooling instability criteria}
\label{sec:cool_inst_cri_destn}
\begin{figure*}
\centering
  \includegraphics[width=.47\linewidth]{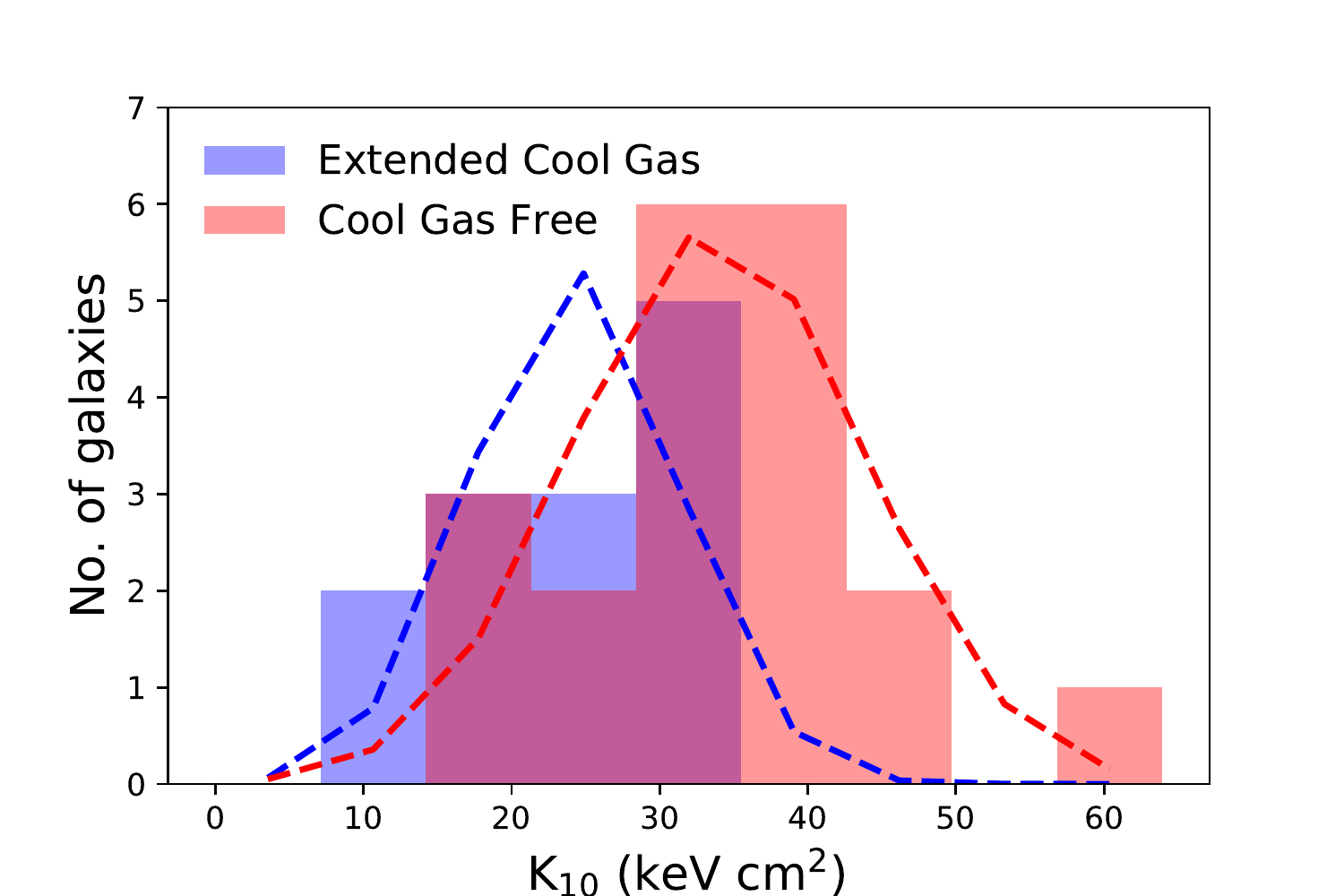}
  \includegraphics[width=.47\linewidth]{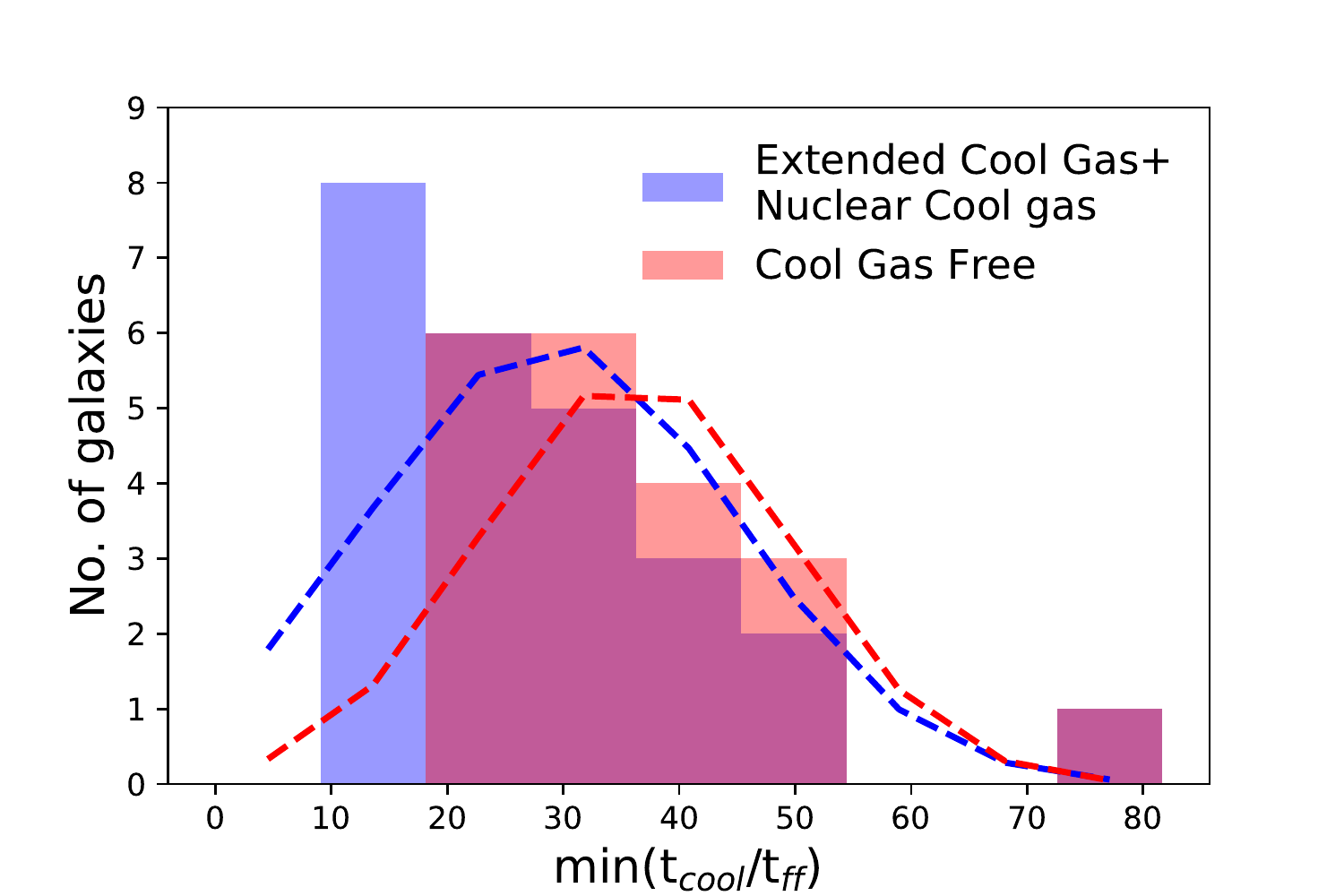}
  \includegraphics[width=.47\linewidth]{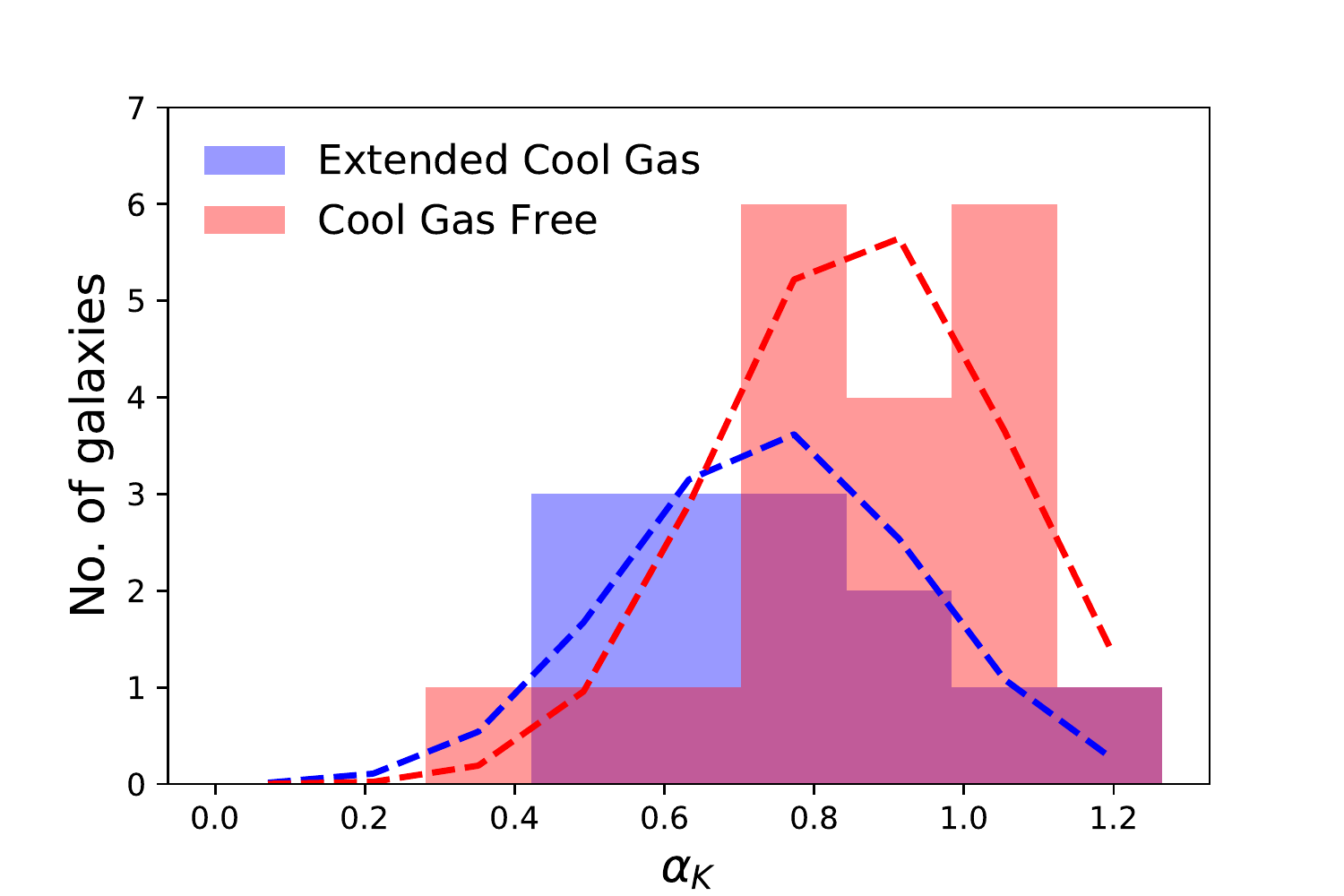} 
  \includegraphics[width=.47\linewidth]{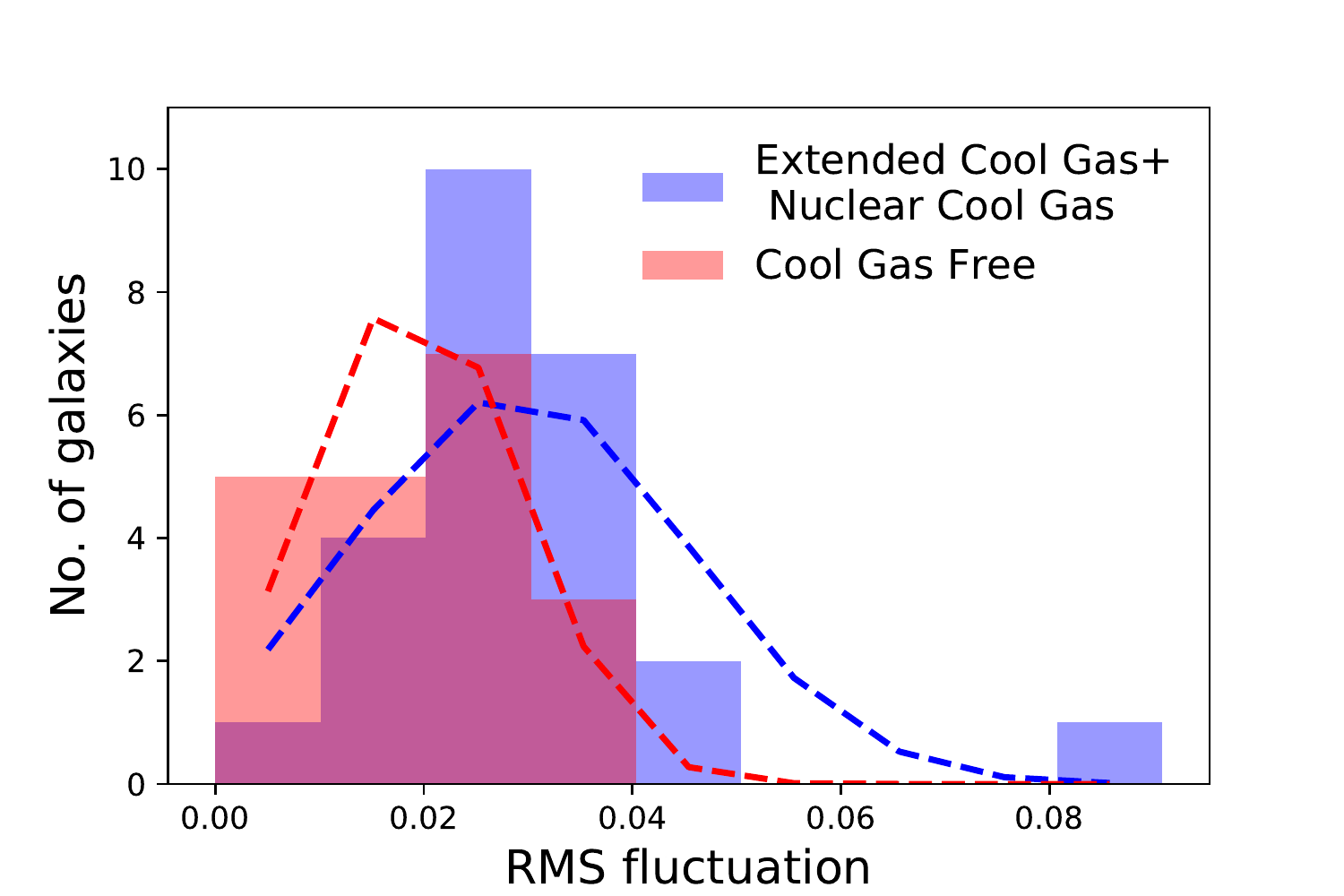} 
\caption{The histograms of $K_{\rm 10}$, min($t_{\rm cool}/t_{\rm ff}$), $\alpha_{\rm K}$ and RMS fluctuation, described in \S\ref{sec:bimodality}. The blue and red 
colors denote the cool gas rich and cool gas poor systems, respectively, and the dotted line shows the best-fitting Gaussians for the two groups. 
The presence of cool gas seems to be preferred in systems with lower values of $K_{\rm 10}$, min($t_{\rm cool}/t_{\rm ff}$), $\alpha_{\rm K}$ and high 
RMS fluctuations.}
\label{fig:comb_K20_mintcoolbytff_alpha_rms_fluc_hist}
\end{figure*}

We fitted powerlaw models to the entropy profiles ($K=K_{10}\,(r/10)^{\alpha_{\rm K}}$) of all the galaxies in the radial range of 1--30 kpc. A histogram of the entropies of the individual 
galaxies at 10 kpc ($K_{\rm 10}$) obtained from the fits, is shown in the top left panel of Fig.~\ref{fig:comb_K20_mintcoolbytff_alpha_rms_fluc_hist}. 
The histogram shows that the galaxies with extended emission line nebulae have lower $K_{\rm 10}$ values than the cool gas free galaxies. However, there 
are outliers and we do not see a clear demarcation in the entropy between the galaxies with and without ongoing cooling, which is also expected because 
of the short duty cycles of these galaxies. The mean$\pm$sigma K$_{10}$ values obtained from a Gaussian fitting of the $K_{\rm 10}$ histograms obtained for 
the cool gas free  and the extended cool 
gas groups were 34$\pm$10 keV cm$^2$  and 24$\pm$7 keV cm$^2$, respectively. The top right panel of 
Fig.~\ref{fig:comb_K20_mintcoolbytff_alpha_rms_fluc_hist} shows histograms of the minimum values of $t_{\rm cool}/t_{\rm ff}$ obtained for the cool gas free 
and cool gas rich (extended$+$nuclear cool gas) groups, which show a similar trend as the entropy. The mean$\pm$sigma  min($t_{\rm cool}/t_{\rm ff}$) values 
obtained from Gaussian fitting of the histograms obtained for the cool gas free  and cool gas rich (extended$+$nuclear) groups were 36$\pm$13  and 
29$\pm$16, respectively.

According to \citet{Voit2017}, the formation of cooling instabilities also depends on the slopes of entropy profiles. The histograms of the slopes $\alpha_{\rm K}$, 
of the best power-law fits to the entropy distributions of the cool gas free and extended cool gas groups are shown in the bottom left panel of 
Fig.~\ref{fig:comb_K20_mintcoolbytff_alpha_rms_fluc_hist}. The separation of the two groups 
appears much weaker here than for the other parameters. The mean$\pm$sigma  $\alpha_{\rm K}$ values obtained from Gaussian fitting of the histograms 
for the cool gas free and extended cool gas groups were 0.86$\pm$0.20  and 0.75$\pm$0.20, respectively.

\citet{McNamara2016}, \citet{Gaspari2017} and \citet{Gaspari2018} argue that uplift and turbulent motions promote nonlinear condensation\footnote{Note that this is a 
more direct and efficient way to produce multiphase gas, since linear thermal instability models (linked to $t_{\rm cool}/t_{\rm ff}$) require
tiny perturbations to grow nonlinear against the restoring buoyancy force. Furthermore, we note that, in a tightly self-regulated loop, the cool gas is also agent of higher
SMBH accretion rates, which will later produce jets and maintain a significant level of
turbulent/RMS fluctuations \citep[cf.,][]{Gaspari2012b,Gaspari2012a}.}. To estimate the disturbedness in our systems, which might be an indication of the level of gas motions, we did the following. We first produced 0.5--7.0 keV exposure-corrected 
images for all the galaxies. Point sources were detected, removed and the empty regions were filled with the average counts from the neighboring pixels. The images were 
then smoothed with Gaussians of  3 pixel ($\sim$1.5$"$) width and were fitted with 2D double $\beta$-models in the \textit{CIAO Sherpa} package. 
As a proxy for the gas motions \citep[e.g.,][]{Hofmann2016,Gaspari2013b,Zhuravleva2014}, we use the root-mean-square (RMS) fluctuations of the residual images within the central 
5 kpc regions. 

The histograms of the RMS fluctuations obtained for the cool gas rich (extended$+$nuclear) and cool gas free galaxies, are shown in the right bottom 
panel of Fig. \ref{fig:comb_K20_mintcoolbytff_alpha_rms_fluc_hist}. Although the plot does not show a clear demarcation value of RMS 
fluctuations between the cool gas rich and free galaxies, it can be seen that in general, the formation of cooling instabilities is preferred in galaxies 
with higher RMS fluctuations. We also tried the scales at 2.5 kpc and 10 kpc. The distinction between cool gas free 
and cool gas rich galaxies seems to get 
better at smaller scales  (2.5 kpc) and almost disappears at larger scales (10 kpc). The mean$\pm$sigma 5 kpc RMS fluctuations obtained from the Gaussian 
fitting to the histograms for the cool gas free and cool gas rich (extended$+$nuclear) groups was found to be 0.02$\pm$0.01 and 
0.03$\pm$0.02, respectively. Note that the value of the RMS fluctuation has some dependence also on the depth of the data and the pixel scale, and the line of sight 
projection effects also complicate these measurements.

We also checked if the $K_{\rm 10}$, min($t_{\rm cool}/t_{\rm ff}$), $\alpha_{\rm K}$  and RMS fluctuations obtained for the cool gas rich and cool gas poor galaxies 
statistically belong to two different populations. For this we used Welch's $t$-test where $t$ is defined as 
\begin{equation}
t=(\overline{X_1}-\overline{X_2})/\sqrt{\frac{S_1^2}{n_1}+\frac{S_2^2}{n_2}}
\end{equation}; 
where $\overline{X_1}$ and $\overline{X_2}$ are the means, $S_1^2$ and $S_2^2$ are the variances and $n_1$ and $n_2$ are the sizes 
of the two samples $X_1$ and $X_2$. 
The test is based on the null-hypothesis that the samples have been taken from the same parent distribution. 
The $t$ values so obtained and the corresponding null hypothesis probabilities ($p$ for $n_1+ n_2 -2$ degrees of freedom), for the $K_{\rm 10}$, min($t_{\rm cool}/t_{\rm ff}$), 
$\alpha_{\rm K}$ and RMS fluctuations obtained for the extended cool gas (ECG), Nuclear cool gas (NCG) and Cool gas free (CGF) groups, using different 
combinations (viz., ECG vs. CGF, ECG$+$NCG vs. CGF and ECG vs. NCG$+$CGF) of the three groups, are given in Table \ref{tab:t_test_results}. The histograms 
of $K_{\rm 10}$, min($t_{\rm cool}/t_{\rm ff}$), 
$\alpha_{\rm K}$ and RMS fluctuations corresponding to only the best combination (highest $t$, lowest $p$) are shown in 
Fig. \ref{fig:comb_K20_mintcoolbytff_alpha_rms_fluc_hist}. We find that the distributions of $K_{10}$, min($t_{\rm cool}/t_{\rm ff}$), 
$\alpha_{\rm K}$ and the RMS fluctuations of the cool gas-rich and cool gas-free galaxies are different at 
$>$99\%, 91\%, 87\% and 98\% confidence levels, respectively.

\begin{table}
 \caption{The results of the Welch's t-test for the $K_{\rm 10}$, min($t_{\rm cool}/t_{\rm ff}$), $\alpha_{\rm K}$  and RMS fluctuation values obtained for the 
 extended cool gas (ECG), Nuclear cool gas (NCG) and Cool gas free (CGF) groups. We used different combinations (viz., ECG vs. CGF, ECG$+$NCG vs. CGF and ECG vs. NCG$+$CGF) 
 for the samples $X_1$ and $X_2$ with means $\overline{X_1}$ and $\overline{X_2}$ and variances $S_1^2$ and $S_2^2$, respectively. 
 The $t$-values and null hypothesis (viz. both samples were taken from the same parent distribution) probabilities ($p$), and the degrees of freedom (DOF) are 
 calculated as defined in \S\ref{sec:cool_inst_cri_destn}.}
\label{tab:t_test_results}
\vskip 0.5cm
\centering
{\scriptsize
\setlength\tabcolsep{4pt} 
\begin{tabular}{c c c c c c c c}
\hline
\hline
\multicolumn{7}{c}{\bf{$K_{\rm 10}$}} \\
\hline
 $X_1$     & $X_2$    &  $\overline{X_1}\pm S_1$  & $\overline{X_2}\pm S_2$ &  $t$  & DOF & $p$ \\
             &            &        (keV cm$^2$)       &        (keV cm$^2$) &       &     &   \\    
\hline
   ECG       &   CGF      &    24$\pm$7           & 34$\pm$10               &  3.20 & 31 & 0.003 \\
  ECG+NCG    &   CGF      &    27$\pm$12          & 34$\pm$10               &  2.01 & 43 & 0.05 \\
   ECG       &  CGF+NCG   &    24$\pm$7           & 33$\pm$12               &  2.78 & 43 & 0.008 \\
\hline
\hline
\multicolumn{7}{c}{\textbf{min($t_{\rm cool}/t_{\rm ff}$)}}\\
\hline
 $X_1$     & $X_2$    &  $\overline{X_1}\pm S_1$  & $\overline{X_2}\pm S_2$ &  $t$  & DOF & $p$ \\
\hline
   ECG       &   CGF      &    28$\pm$15 & 36$\pm$13                        &  1.53 & 31 & 0.14 \\
  ECG+NCG    &   CGF      &    29$\pm$16 & 36$\pm$13                        &  1.75 & 43 & 0.09 \\
   ECG       &  CGF+NCG   &    28$\pm$15 & 34$\pm$15                        &  1.14 & 43 & 0.26 \\
\hline
\hline
\multicolumn{7}{c}{\textbf{$\alpha_{\rm K}$}} \\
\hline
 $X_1$     & $X_2$    &  $\overline{X_1}\pm S_1$  & $\overline{X_2}\pm S_2$ &  $t$  & DOF & $p$ \\
\hline
   ECG       &   CGF      &    0.75$\pm$0.20 & 0.86$\pm$0.20                &  1.56 & 31 & 0.13 \\
  ECG+NCG    &   CGF      &    0.78$\pm$0.23 & 0.86$\pm$0.20                &  1.23 & 43 & 0.20 \\
   ECG       &  CGF+NCG   &    0.75$\pm$0.20 & 0.85$\pm$0.22                &  1.38 & 43 & 0.17 \\
\hline
\hline
\multicolumn{7}{c}{\textbf{RMS fluctuation}} \\
\hline
 $X_1$     & $X_2$    &  $\overline{X_1}\pm S_1$  & $\overline{X_2}\pm S_2$ &  $t$  & DOF &    $p$ \\
\hline
   ECG       &   CGF      &     0.03$\pm$0.01 & 0.02$\pm$0.01               &  2.26  & 31 & 0.03 \\
  ECG+NCG    &   CGF      &     0.03$\pm$0.02 & 0.02$\pm$0.01               &  2.45  & 43 & 0.02 \\
   ECG       &  CGF+NCG   &     0.03$\pm$0.01 & 0.02$\pm$0.22               &  1.22  & 43 & 0.23 \\
\hline
\hline
\end{tabular}}
\end{table}

%\begin{figure*}
%\centering
%  \includegraphics[width=.9\linewidth]{Figures/rms_fluc_histogram_new_classfn.pdf}
%\caption{Histograms of the rms fluctutations for the galxies with and without cold gas (for details, see \S\ref{sec:cool_inst}).}
%\label{fig:rms_fluc_hist}
%\end{figure*}

\begin{figure*}
\centering
  \includegraphics[width=.6\linewidth]{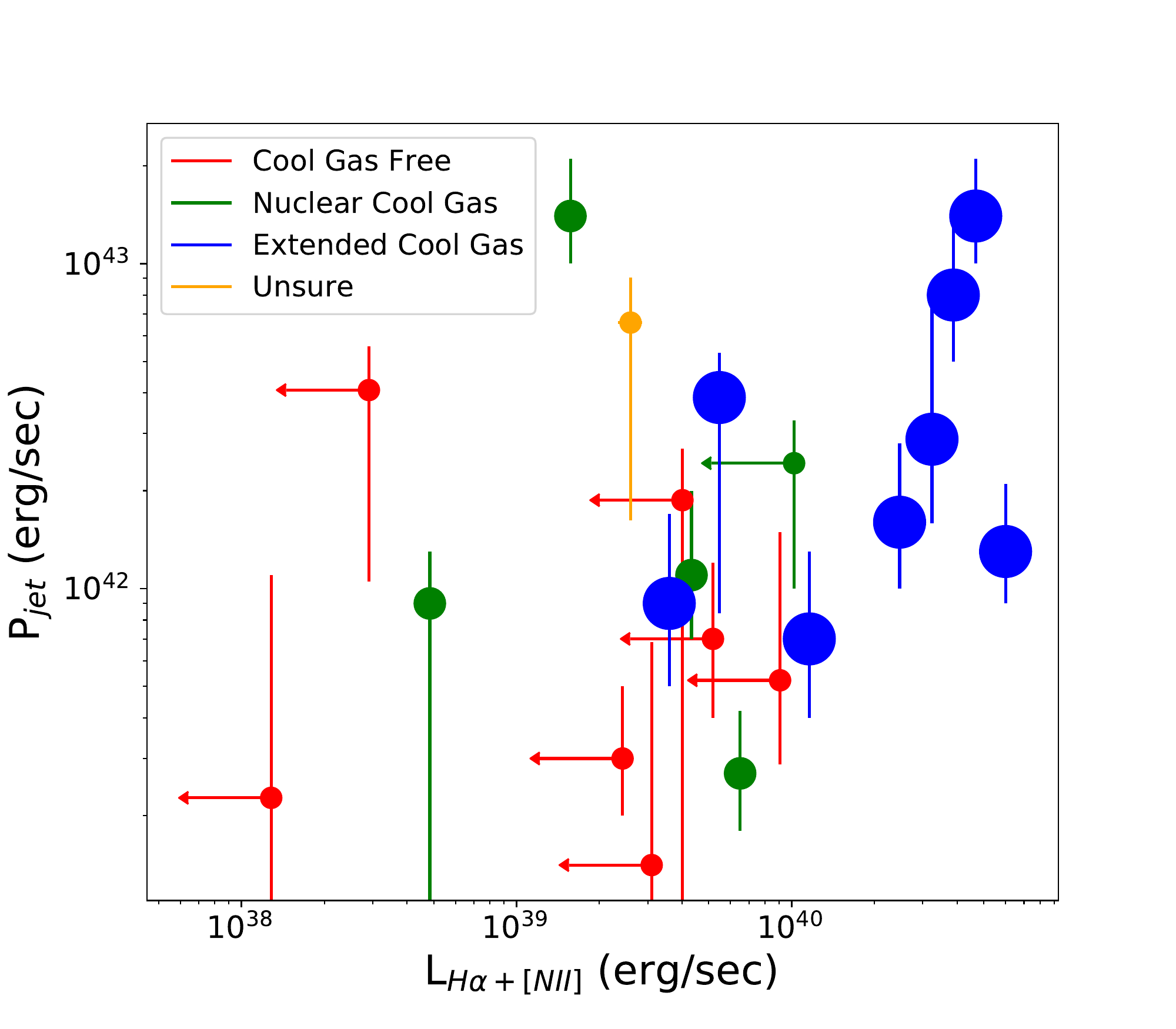}
\caption{Jet powers of the 21 galaxies with X-ray cavities vs. their H$\alpha$+[NII] luminosities (bigger circles represent more extended H$\alpha$+[NII] emission). 
The red, green, blue and orange colors 
denote the cool gas free, nuclear cool gas, extended cool gas and unsure systems, respectively. There seems to be a weak positive correlation between 
the AGN jet powers of the galaxies and their H$\alpha$+[NII] luminosities.}
\label{fig:Pjet_vs_LHa}
\end{figure*}

\subsection{Feedback cycles}
\label{sec:feedback_cycles}
To investigate the connection between the cool gas and AGN activity, we search for a correlation between the AGN 
jet power and the H$\alpha$+[NII] luminosities of the galaxies. The jet powers ($P_{\rm jet}$) are calculated as the 
work required to inflate a cavity with a volume $V$ divided by the age of the cavity, $P_{\rm jet}=4 PV/t_{\rm age}$. 
$P$ is the pressure of the hot gas determined from the X-ray observations. Cavities are usually approximated as ellipsoids 
and their sizes are estimated either from the X-ray images \citep{Cavagnolo2010,Russell2013} or from the radio 
lobes \citep{Allen2006}. The ages of cavities are either assumed to be their buoyancy rise times or sound crossing times 
$r/c_{\rm s}$ (where $r$ is the distance of the cavity from the centre and $c_{\rm s}$ is the sound speed). 

Due to the inconsistencies in the $P_{\rm jet}$ estimates available in the literature, we recalculated these values for 15 of the 21 galaxies 
in our sample that host clear cavities. The sizes for all these cavities (except for NGC~4649) were taken from a single source, \cite{Shin2016}. 
For NGC~4649, we used the X-ray cavity size given in \cite{Paggi2014}. The 
cavity volumes and the associated uncertainties were calculated as described in \cite{Birzan2004}. The ages of the cavities were 
estimated as their buoyancy rise times. For the 
remaining 6 galaxies with cavities, we used the estimates given in 
\cite{Cavagnolo2010}, as their method of calculating jet powers is similar to ours. The jet powers for the 21 galaxies 
are given in Table \ref{tab:jet_power}.

Fig.~\ref{fig:Pjet_vs_LHa} shows the jet powers for the 21 galaxies as a function of their H$\alpha$+[NII] luminosities. For six of the 21 galaxies, the 
H$\alpha$+[NII] luminosities were obtained from \cite{Macchetto1996}; five of these were detected as small disk emission. Also, 
there were three more galaxies for which H$\alpha$+[NII] flux estimates were available in the literature but no H$\alpha$+[NII] emission was detected 
in the SOAR/APO observations. As discussed in \S\ref{sec:sampl_sel}, the H$\alpha$+[NII] luminosities for these eight sources should be interpreted as upper-limits. 
We find a weak positive correlation (Pearson's coefficient $\sim$0.38) between the two quantities, which reduces to $\sim$0.24, if the eight galaxies 
with upper limits are excluded. From Fig. \ref{fig:Pjet_vs_LHa}, in general, the jet power seems to be
increasing with the increase in the cool gas extent and is the
strongest for the galaxies with the most extended H$\alpha$+[NII]
filaments, such as NGC 5044, NGC 4696 and NGC 533. 

Of the 15 galaxies, for which we recalculated the jet powers, the estimate for NGC 4261 was 
found to be too low. The galaxy, however, hosts very powerful jets \citep[see][]{OSullivan2011}. 
Therefore, we also recalculated the correlation coefficient using the lower limit for the jet power 
in this system from \cite{OSullivan2011} (cavity sizes estimated using radio lobes). 
With this, we find a weak positive correlation (Pearson's coefficient $\sim$0.19) between the two quantities which reduces to $\sim$-0.04, if the eight galaxies 
with upper limits are excluded. 

\begin{table}
 \caption{AGN jet power ($P_{\rm jet}$) estimates for 21 galaxies of the sample.}
\label{tab:jet_power}
\vskip 0.5cm
\centering
{\scriptsize
\begin{tabular}{c c c}
\hline
Target & $P_{\rm jet}$ & Ref.$^{*}$\\
Name & (10$^{43}$ erg/sec) & \\
\hline
IC 4296  & 0.39$^{+0.14}_{-0.30}$ & C\\
NGC 315  & 0.66$^{+0.25}_{-0.50}$ & C\\
NGC 533  & 0.29$^{+0.55}_{-0.13}$ & P\\
NGC 777  & 0.41$^{+ 0.15}_{-0.30}$ & C\\
NGC 1316 & 0.05$^{+0.14}_{-0.03}$ & P\\
NGC 1399 & 0.02$^{+0.09}_{-0.02}$ & P\\
NGC 1407 & 0.01$^{+0.05}_{-0.01}$ & P\\
NGC 1600 & 0.19$^{+0.08}_{-0.20}$ & C\\
NGC 4261 & 0.09$^{+0.04}_{-0.08}$ & P\\
  & $>$1.0 & O\\
NGC 4374 & 0.20$^{+0.48}_{-0.11}$ & P\\
NGC 4472 & 0.02$^{+0.03}_{-0.01}$ & P\\
NGC 4486 & 0.44$^{+0.31}_{-0.06}$ & P\\
NGC 4552 & 0.01$^{+0.01}_{-0.001}$ & P\\
NGC 4636 & 0.11$^{+0.11}_{-0.03}$ & P\\
NGC 4649 & 0.05$^{+0.10}_{-0.02}$ & P\\
NGC 4696 & 0.57$^{+0.55}_{-0.08}$ & P\\
NGC 4782 & 0.24$^{+0.09}_{-0.14}$ & C\\
NGC 5044 & 0.51$^{+1.25}_{-0.29}$ & P\\
NGC 5813 & 0.58$^{+1.25}_{-0.27}$ & P\\
NGC 5846 & 0.11$^{+0.13}_{-0.02}$ & P\\
NGC 4778 & 0.70$^{+1.68}_{-0.37}$ & P\\
\hline
\end{tabular}}\\
\footnotesize{* C: \cite{Cavagnolo2010}, O: \cite{OSullivan2011}, P: Present Work.}
\end{table}

The large scatter in the jet powers and the H$\alpha$+[NII] luminosities, the positive correlation between the two, and the increase of jet power with 
the cool gas extent, hint toward the scenario of hysteresis cycles driven by CCA, which has been shown via high-resolution 3D hydrodynamic simulations 
to be the most consistent mechanism for self-regulating AGN feedback in ETGs \citep[cf.,][]{Gaspari2013,Gaspari2015,Gaspari2018} and brightest cluster 
galaxies \citep[cf.,][]{Gaspari2012b,Gaspari2012a,Prasad2015,Voit2017}. Simply put, during CCA, the higher the condensed gas mass (thus higher 
$L_{\rm H\alpha+[NII]}$), the stronger the SMBH accretion rate, and thus feedback power ($P_{\rm feed} \propto \dot{M}_{\rm cool}\,c^2$; more below). 
On top of this trend, the intrinsically chaotic evolution of the colliding clouds/filaments in CCA drives a substantial ($\sim 1$ dex) variability which can hinder 
a strong linear correlation (yet preserving a positive Pearson coefficient). Furthermore, the correlation between P$_{\rm jet}$ and cool gas luminosity (thus 
condensation) is consistent with the turbulent eddy criterion (\S\ref{sec:intro}), as larger jet powers imply larger turbulent velocity dispersions 
(from the turbulent energy flux rate, $\sigma_v \propto P_{\rm jet}^{1/3}$) and hence larger RMS surface brightness fluctuations (as found in \S\ref{sec:cool_inst_cri_destn}). 
We note that alternative models as hot/Bondi accretion would instead have negligible variability and show no correlations with the cool phase (nor turbulence). 
Needless to say, forthcoming investigations should significantly expand the ETG sample and achieve more accurate detections in warm gas, which remains one of 
our main thrusts for our ongoing campaigns.

In more detail, the AGN self-regulation cycle works as follows. In the beginning of the proposed AGN feedback cycle, the galaxies have, in general, 
weak gas motions and smooth and symmetric X-ray morphologies. Galaxies in this phase have neither cold gas 
nor central AGN jets but may have high central entropies as a result of past AGN activity. As the gas in the central regions of the galaxies cools, 
the entropy decreases and the cooling instabilities start forming, giving rise to the cold gas filaments. 
As the cold gas accretion increases, the AGN jet power also increases and the powerful jets start interacting with the surrounding medium, 
driving large scale gas motions and inflating X-ray cavities. The gas motions further increase the formation of cooling instabilities. Eventually, the 
jets start heating the surrounding medium\footnote{We note that the hysteresis cycling mainly occurs within the core region (< 10 kpc, for both initially 
low or high $K_{100}$ systems), while AGN jet feedback rarely affects the large-scale 100 kpc profiles over the Gyr evolution \citep[e.g.,][]{Wang2018}.}, 
preventing further formation of cold gas and might also destroy the existing cold gas filaments. The cold gas fuel further reduces due to the 
AGN jet interaction and the galaxies might then be left with just nuclear cold gas with some AGN activity. Finally due to the lack/absence of cold gas fuel, 
the AGN starves, the jet activity stops and the galaxy returns to its initial phase. 

As also discussed in \S\ref{sec:bimodality}, it is important to note that the 
feedback cycles in early type galaxies are much faster (a few 10s Myr) compared to massive clusters (several 100s Myr). Note also how the 
cooling rate $\dot M_{\rm cool} \propto L_{\rm X}/T_{\rm X}$ is {\it relatively} larger in massive ETGs because of line cooling (< 1 keV) and 
the $T_{\rm x}^{-1}$ dependence. This implies a much
more pronounced hysteresis in the early type galaxies, with high/low feeding and feedback states less separated and
more intertwined, as found in the current observational study.

Based on an analysis of 107 galaxies, groups and clusters, \cite{McDonald2018} found that the correlation between the mass cooling rate 
of the ICM and the star formation rate breaks down at the low mass end, suggesting that the cold gas and star formation are mainly being driven 
by stellar mass loss for the low mass systems. However, in our sample, majority of which includes massive ETGs 
with extended halos, we see clear separations in the density, entropy, and cooling time profiles in the 2-35 kpc radial range, based on the multiphase gas presence. 
These are clear signs of large-scale condensation.

Elliptical galaxies, groups and poor clusters of galaxies are the building blocks of massive clusters and are, therefore, crucial 
for understanding the cosmic structure formation in the Universe. Moreover, X-ray halos are of key importance and appear 
to be ubiquitous, not only for massive ETGs but even for compact or fossil ETGs \citep[e.g.,][]{Werner2018}. As of now, most of the 
studies centred on the non-gravitational processes (cooling, AGN feedback etc.) focus only on the 
bright massive clusters since the current X-ray missions are limited in their capability to study the X-ray emission in the fainter low mass systems 
out to $R_{500}$. The future \textit{Athena} X-ray observatory will allow us to extend the studies of hot halos in giant elliptical galaxies out to redshift $z\sim1$, 
allowing us to investigate the various details (source, effect mass dependence, timescales etc.) of the non-gravitational processes 
\citep{Ettori2013,Roncarelli2018}.

\section{Conclusions}
\label{sec:summary}
We have analysed {\it Chandra} X-ray observations of a sample of 49 nearby X-ray and optically bright giant elliptical galaxies. In particular, we focus on the connection 
between the properties of the hot X-ray emitting gas and the cooler H$\alpha$+[NII] emitting phase, and the possible role of the cool phase in the AGN feedback cycle. 
Our main findings are summarised as follows:
\begin{itemize}
\item We do not find a correlation between the presence of H$\alpha$+[NII] emission and the X-ray luminosity, mass of hot gas, and gas mass fraction.

\item The observed correlation between the gas mass fractions and the X-ray temperatures suggests that the cores of hotter, more massive systems are 
able to hold on to a larger fraction of their X-ray emitting gas.

\item  We find that the presence of H$\alpha$+[NII] emission is more likely in systems with higher densities, lower entropies and cooling times (outside the innermost regions) 
shallower entropy profiles, lower values of min($t_{\rm cool}/t_{\rm ff}$), and more disturbed X-ray morphologies.

\item The distributions of the thermodynamic properties of the nuclear cool gas galaxies are found to be in between the extended cool gas and cool gas free galaxies.

%\item As we see in a few extended cool gas galaxies with low densities, rotation can lead to a de-correlation of the multiphase gas from the hot gas properties.

\item We find that the distributions of entropies at 10 kpc, the min($t_{\rm cool}/t_{\rm ff}$) values, the slope of the entropy profiles ($\alpha_{\rm K}$) 
and the RMS surface-brightness fluctuations within a radius of 5 kpc are statistically 
different between cool gas-rich and cool gas-free galaxies at $>$99\%, 91\%, 87\% and 98\% confidence levels, respectively.

\item The large scatter and the significant overlap between the properties of systems with and without optical emission line nebulae, 
indicate rapid transitions from one group to the other. The continuous distribution might also be a result of the chaotic nature and rapid variability of the feeding
and feedback cycle in these systems.

%\item  We find that while 11 out of 25 H$\alpha$+[NII] emitting galaxies (44\%) host observable central X-ray point sources, only 3 out of 20 cool gas free galaxies appear to contain an X-ray emitting AGN. 

\item The AGN jet power of the galaxies with X-ray cavities hint toward a positive correlation with their H$\alpha$+[NII] luminosity. This feature, the presence of cool gas in more 
disturbed/turbulent halos, and frequent hysteresis cycles in ETGs are consistent with a cold gas nature of AGN feeding and related CCA scenario.

\end{itemize}

\section*{Acknowledgements}
This work was supported by the Lendulet LP2016-11 grant awarded by the Hungarian Academy 
of Sciences. M.S. is supported by NSF grant 1714764, Chandra grant AR7-18016X and NASA grant NNX15AK29A. 
M.G. is supported by NASA through Einstein Postdoctoral Fellowship Award Number PF5-160137
issued by the Chandra X-ray Observatory Center, which is operated by the SAO for and on behalf
of NASA under contract NAS8-03060. Support for this work was also provided by Chandra grant
GO7-18121X. C.S. is supported in part by Chandra grants GO5-16146X, GO7-18122X, and GO8-19106X. 
R.E.A.C. is supported by NASA through Einstein Postdoctoral Fellowship Award Number PF5-160134 issued 
by the Chandra X-ray Observatory Center, which is operated by the SAO for and on behalf of NASA under contract NAS8-03060. 
The scientific results reported in this article are based to a significant degree on data 
obtained from the \textit{Chandra Data Archive}. This research has made use of software provided 
by the \textit{Chandra X-ray Centre} (CXC) in the application packages CIAO, CHIPS and Sherpa. 
This research has made use of data, software and web tools obtained from the High Energy Astrophysics 
Science Archive Research Center (HEASARC), a service of the Astrophysics Science Division at 
NASA/GSFC and of the Smithsonian Astrophysical Observatory's High Energy Astrophysics Division. 
Based on observations obtained at the Southern Astrophysical Research (SOAR) telescope, which is a 
joint project of the Minist\'{e}rio da Ci\^{e}ncia, Tecnologia, Inova\c{c}\~{o}es e Comunica\c{c}\~{o}es 
(MCTIC) do Brasil, the U.S. National Optical Astronomy Observatory (NOAO), the University of North 
Carolina at Chapel Hill (UNC), and Michigan State University (MSU). Based on observations obtained 
with the Apache Point Observatory 3.5-meter telescope, which is owned and operated by the 
Astrophysical Research Consortium.

\bibliography{ref}

\appendix{}

\setcounter{secnumdepth}{-1}
\renewcommand{\thesection}{A}
\renewcommand\thefigure{\thesection.\arabic{figure}}

\section{}

\begin{table}
 \caption{A log of the \textit{Chandra} observations used in the paper.}
\label{tab:obsn_log}
\vskip 0.5cm
\centering
{\scriptsize
\begin{tabular}{c c c c c}
\hline
Target & Obs ID & Instrument & Cleaned Exposure & Date of \\
Name & & & (ks) & Observation\\
\hline
3C 449 & 11737 & ACIS-S & 45.29 & 2010-09-14\\
 & 13123 & ACIS-S & 51.73 & 2010-09-20\\
IC 1860 & 10537 & ACIS-S & 31.12 & 2009-09-12\\
IC 4296 & 2021 & ACIS-S & 14.20 & 2001-09-10\\
 & 3394 & ACIS-S & 16.22 & 2001-12-15\\
IC 4765 & 15637 & ACIS-S & 11.77 & 2013-03-29\\
NGC 57 & 10547 & ACIS-S & 8.89 & 2008-10-29\\
NGC 315 & 4156 & ACIS-S & 37.47 & 2003-02-22\\
NGC 410 & 5897 & ACIS-S & 2.05 & 2004-11-30\\
NGC 499 & 2882 & ACIS-I & 40.13 & 2002-01-08\\
 & 317 & ACIS-S & 19.17 & 2000-10-11\\
 & 10536 & ACIS-S & 17.62 & 2009-02-12\\
 & 10866 & ACIS-S & 8.05 & 2009-02-05\\
 & 10867 & ACIS-S & 6.04 & 2009-02-07\\
NGC 507 & 2882 & ACIS-I & 40.13 & 2002-01-08\\
 & 317 & ACIS-S & 19.17 & 2000-10-11\\
NGC 533 & 2880 & ACIS-S & 29.16 & 2002-07-28\\
NGC 708 & 7921 & ACIS-S & 105.53 & 2006-11-20\\
 & 2215 & ACIS-S & 26.45 & 2001-08-03\\
NGC 741 & 17198 & ACIS-S & 70.94 & 2015-12-04\\
 & 18718 & ACIS-S & 48.36 & 2015-12-06\\
 & 2223 & ACIS-S & 24.20 & 2001-01-28\\
NGC 777 & 5001 & ACIS-I & 7.67 & 2004-12-23\\
NGC 1132 & 801 & ACIS-S & 10.92 & 1999-12-10\\
 & 3576 & ACIS-S & 28.16 & 2003-11-16\\
NGC 1316 & 2022 & ACIS-S & 23.55 & 2001-04-17\\
NGC 1399 & 14529 & ACIS-S & 29.31 & 2015-11-06\\
 & 16639 & ACIS-S & 27.38 & 2014-10-12\\
NGC 1404 & 17549 & ACIS-S & 60.63 & 2015-03-28\\
 & 2942 & ACIS-S & 25.66 & 2003-02-13\\
 & 16231 & ACIS-S & 53.58 & 2014-10-20\\
 & 17540 & ACIS-S & 25.98 & 2014-11-02\\
 & 17541 & ACIS-S & 19.61 & 2014-10-23\\
 & 16232 & ACIS-S & 57.89 & 2014-11-12\\
 & 16233 & ACIS-S & 83.41 & 2014-11-09\\
 & 17548 & ACIS-S & 40.29 & 2014-11-11\\
 & 16234 & ACIS-S & 78.30 & 2014-10-30\\
NGC 1407 & 14033 & ACIS-S & 45.29 & 2012-06-17\\
 & 791 & ACIS-S & 38.85 & 2000-08-16\\
NGC 1521 & 10539 & ACIS-S & 43.51 & 2009-07-04\\
NGC 1550 & 3186 & ACIS-I & 9.22 & 2002-01-08\\
 & 3187 & ACIS-I & 9.13 & 2002-01-08\\
%NGC 1553 & 783 & ACIS-S & 33.73 & 2000-01-02\\
NGC 1600 & 4283 & ACIS-S & 20.90 & 2002-09-18\\
 & 4371 & ACIS-S & 19.07 & 2002-09-20\\
NGC 2300 & 4968 & ACIS-S & 37.12 & 2004-06-23\\
 & 15648 & ACIS-S & 15.53 & 2013-05-24\\
NGC 2305 & 10549 & ACIS-S & 8.64 & 2009-07-19\\
NGC 3091 & 3215 & ACIS-S & 22.51 & 2002-03-26\\
NGC 3923 & 1563 & ACIS-S & 16.24 & 2001-06-14\\
 & 9507 & ACIS-S & 65.37 & 2008-04-11\\
NGC 4073 & 3234 & ACIS-S & 26.10 & 2002-11-24\\
NGC 4125 & 2071 & ACIS-S & 47.34 & 2001-09-09\\
NGC 4261 & 9569 & ACIS-S & 87.63 & 2008-02-12\\
 & 834 & ACIS-S & 17.92 & 2000-05-06\\
NGC 4374 & 803 & ACIS-S & 25.91 & 2000-05-19\\
 & 401 & ACIS-S & 1.25 & 2000-04-20 \\
NGC 4406 & 318 & ACIS-S & 13.31 & 2000-04-07\\
       & 16967 & ACIS-I & 19.04 & 2016-05-02\\
NGC 4472 & 15757 & ACIS-I & 25.07 & 2014-04-18\\
 & 12888 & ACIS-S & 139.77 & 2011-02-21\\
 & 12889 & ACIS-S & 116.73 & 2011-02-14\\
 & 321 & ACIS-S & 32.25 & 2000-06-12\\
 & 322 & ACIS-I &  8.57 & 2000-03-19\\
NGC 4486 & 241 & ACIS-S & 36.76 & 2000-07-17\\
 & 2707 & ACIS-S & 98.69 & 2000-07-17\\
 & 352 & ACIS-S & 37.68 & 2000-07-29\\
 & 5826 & ACIS-I & 126.76 & 2005-03-03\\
 & 5827 & ACIS-I & 156.20 & 2005-05-05\\
 & 6186 & ACIS-I & 51.55 & 2005-01-31\\
 & 7212 & ACIS-I & 65.23 & 2005-11-14\\
\hline
\end{tabular}}
\end{table}

\begin{table}
\ContinuedFloat
 \caption{A log of the \textit{Chandra} observations used in the paper (Continued).}
\label{tab:obsn_log}
\vskip 0.5cm
\centering
{\scriptsize
\begin{tabular}{c c c c c}
\hline
Target & Obs ID & Instrument & Cleaned Exposure & Date of \\
Name & & & (ks) & Observation\\
\hline
NGC 4552 & 14359 & ACIS-S & 44.01 & 2012-04-23\\
 & 14358 & ACIS-S & 41.45 & 2012-08-10\\
 & 2072 & ACIS-S & 44.01 & 2001-04-22\\
 & 13985 & ACIS-S & 41.19 & 2012-04-22\\
NGC 4636 & 323 & ACIS-S & 48.80 & 2000-01-26\\
 & 324 & ACIS-I & 3.17 & 1999-12-04\\
 & 3926 & ACIS-I & 65.24 & 2003-02-14\\
 & 4415 & ACIS-I & 66.43 & 2003-02-15\\
NGC 4649 & 8182 & ACIS-S & 39.83 & 2007-01-30\\
 & 8507 & ACIS-S & 14.96 & 2007-02-01\\
 & 12975 & ACIS-S & 73.44 & 2011-08-08\\
 & 14328 & ACIS-S & 11.92 & 2011-08-12\\
 & 12976 & ACIS-S & 86.94 & 2011-02-24\\
 & 785 & ACIS-S & 33.02 & 2000-04-20\\
NGC 4696 & 1560 & ACIS-S & 45.56 & 2001-04-18\\
 & 16223 & ACIS-S & 175.38 & 2014-05-26\\
 & 16224 & ACIS-S & 40.76 & 2014-04-09\\
 & 16225 & ACIS-S & 29.33 & 2014-04-26\\
 & 16534 & ACIS-S & 54.68 & 2014-06-05\\
 & 16607 & ACIS-S & 44.78 & 2014-04-12\\
 & 16608 & ACIS-S & 33.35 & 2014-04-07\\
 & 16609 & ACIS-S & 80.54 & 2014-05-04\\
 & 16610 & ACIS-S & 16.57 & 2014-04-27\\
 & 4190 & ACIS-S & 34.00 & 2003-04-18\\
 & 4191 & ACIS-S & 32.74 & 2003-04-18\\
 & 4954 & ACIS-S & 87.26 & 2004-04-01\\
 & 4955 & ACIS-S & 44.68 & 2004-04-02\\
 & 504 & ACIS-S & 31.48 & 2000-05-22\\
 & 505 & ACIS-S &  9.96 & 2000-06-08\\
 & 5310 & ACIS-S & 48.56 & 2004-04-04\\
NGC 4778 & 2230 & ACIS-I & 8.70 & 2001-01-08\\
 & 10462 & ACIS-S & 57.57 & 2009-03-02\\
 & 10874 & ACIS-S & 47.59 & 2009-03-03\\
 & 921 & ACIS-S & 40.44 & 2000-01-25\\
 NGC 4782 & 3220 & ACIS-S & 39.16 & 2002-06-16\\
NGC 4936 & 4997 & ACIS-I & 10.23 & 2004-02-09\\
 & 4998 & ACIS-I & 12.54 & 2004-02-15\\
NGC 5044 & 798 & ACIS-S & 17.91 & 2000-03-19\\
 & 9399 & ACIS-S & 73.68 & 2008-03-07\\
 & 17195 & ACIS-S & 66.51 & 2015-06-06\\
 & 17196 & ACIS-S & 75.04 & 2015-05-11\\
 & 17653 & ACIS-S & 30.16 & 2015-05-07\\
 & 17654 & ACIS-S & 21.21 & 2015-05-10\\
 & 17666 & ACIS-S & 73.95 & 2015-08-23\\
 & 3225 & ACIS-S & 63.40 & 2002-06-07\\
 & 3664 & ACIS-S & 47.76 & 2002-06-06\\
NGC 5129 & 7325 & ACIS-S & 21.74 & 2006-05-14\\
 & 6944 & ACIS-S & 17.07 & 2006-04-13\\
NGC 5419 & 5000 & ACIS-I & 12.51 & 2004-06-19\\
 & 4999 & ACIS-I & 14.58 & 2004-06-18\\
NGC5813 & 9517 & ACIS-S & 88.25 & 2008-06-05\\
 & 12951 & ACIS-S & 62.21 & 2011-03-28\\
 & 12952 & ACIS-S & 114.64 & 2011-04-05\\
 & 12953 & ACIS-S & 26.62 & 2011-04-07\\
 & 13246 & ACIS-S & 36.31 & 2011-03-30\\
 & 13247 & ACIS-S & 30.64 & 2011-03-31\\
 & 13253 & ACIS-S & 92.92 & 2011-04-08\\
 & 13255 & ACIS-S & 35.72 & 2011-04-10\\
 & 5907 & ACIS-S & 36.85 & 2005-04-02\\
NGC 5846 & 7923 & ACIS-I & 79.01 & 2007-06-12\\
 & 788 & ACIS-S & 17.58 & 2000-05-24\\
NGC 6407 & 5896 & ACIS-S & 1.90 & 2005-05-24\\
NGC 6861 & 11752 & ACIS-I & 83.01 & 2009-08-13\\
NGC 6868 & 3191 & ACIS-I & 20.65 & 2002-11-01\\
 & 11753 & ACIS-I & 63.90 & 2009-08-19\\
NGC 7619 & 3955 & ACIS-S & 30.31 & 2003-09-24\\
 & 2074 & ACIS-I & 22.14 & 2001-08-20\\
NGC 7796 & 7061 & ACIS-S & 44.51 & 2006-08-28\\
 & 7401 & ACIS-S & 17.65 & 2006-09-03\\
\hline
\end{tabular}}
\end{table}

\begin{figure*}
\centering
  \includegraphics[width=0.9\linewidth]{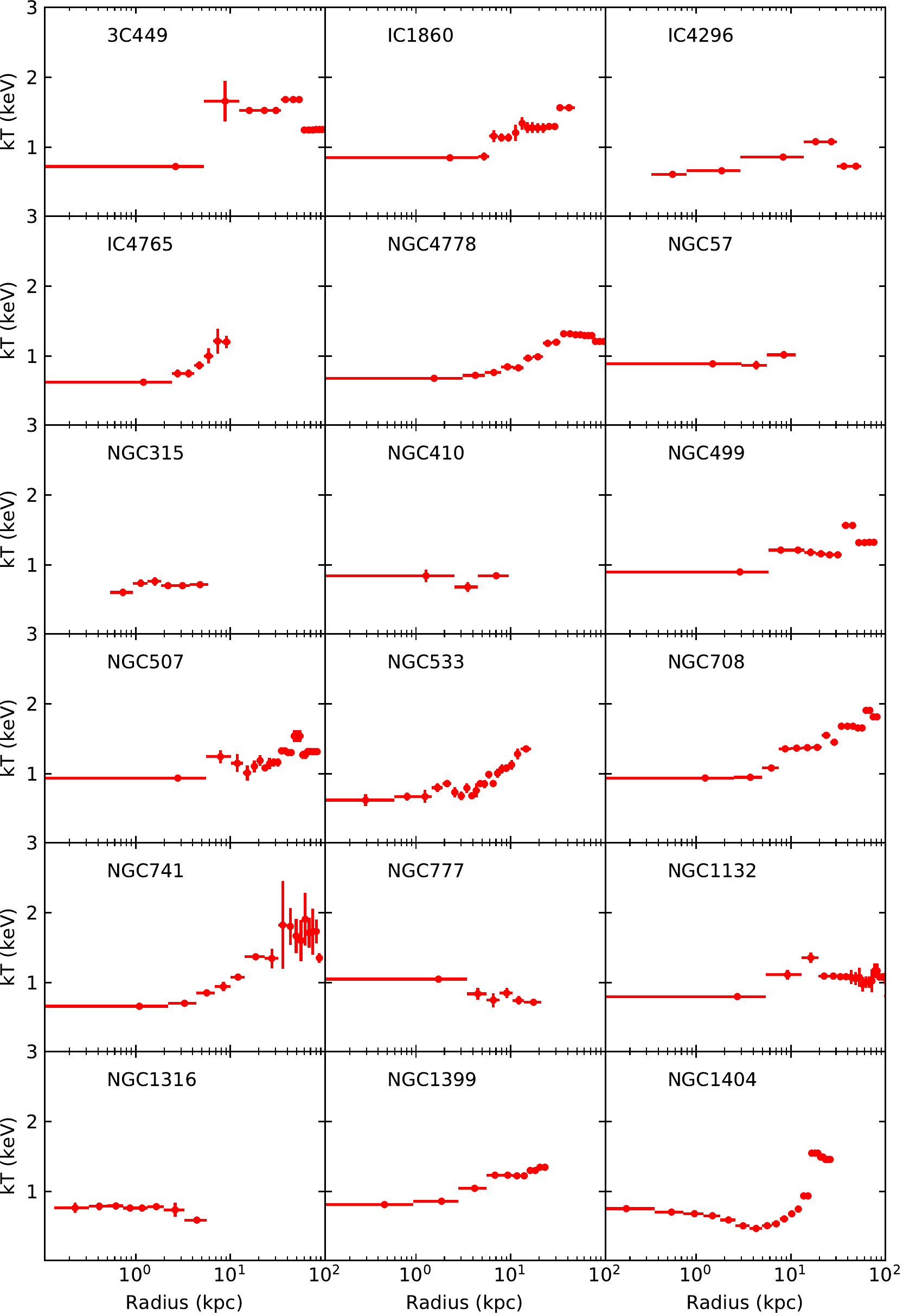}
\caption{Deprojected temperature profiles of the individual galaxies.}
\label{fig:kT_profs}
\end{figure*}

\begin{figure*}
\ContinuedFloat
\centering
  \includegraphics[width=0.9\linewidth]{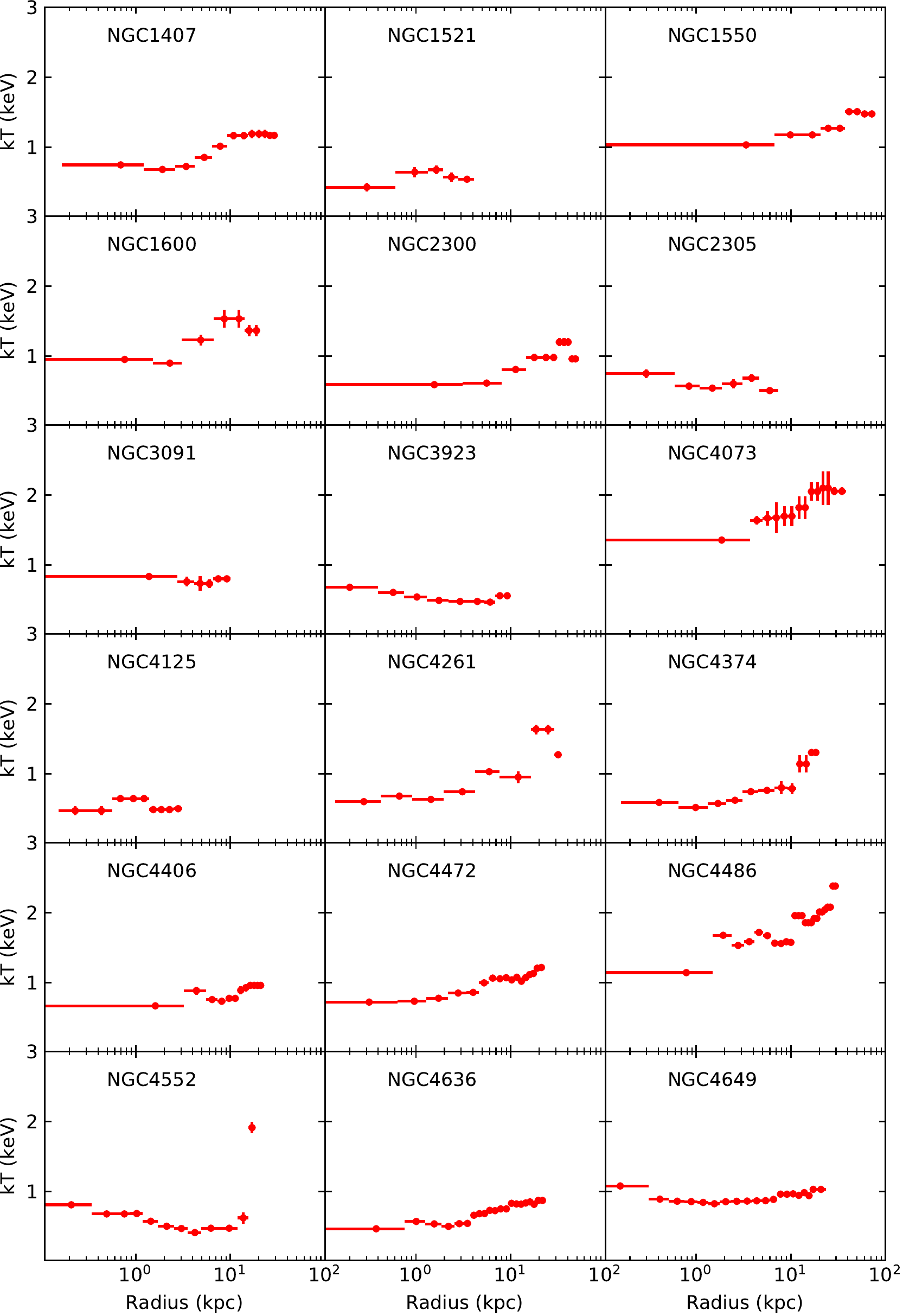}
\caption{Deprojected temperature profiles of the individual galaxies (continued).}
\end{figure*}

\begin{figure*}
\ContinuedFloat
\centering
  \includegraphics[width=0.9\linewidth]{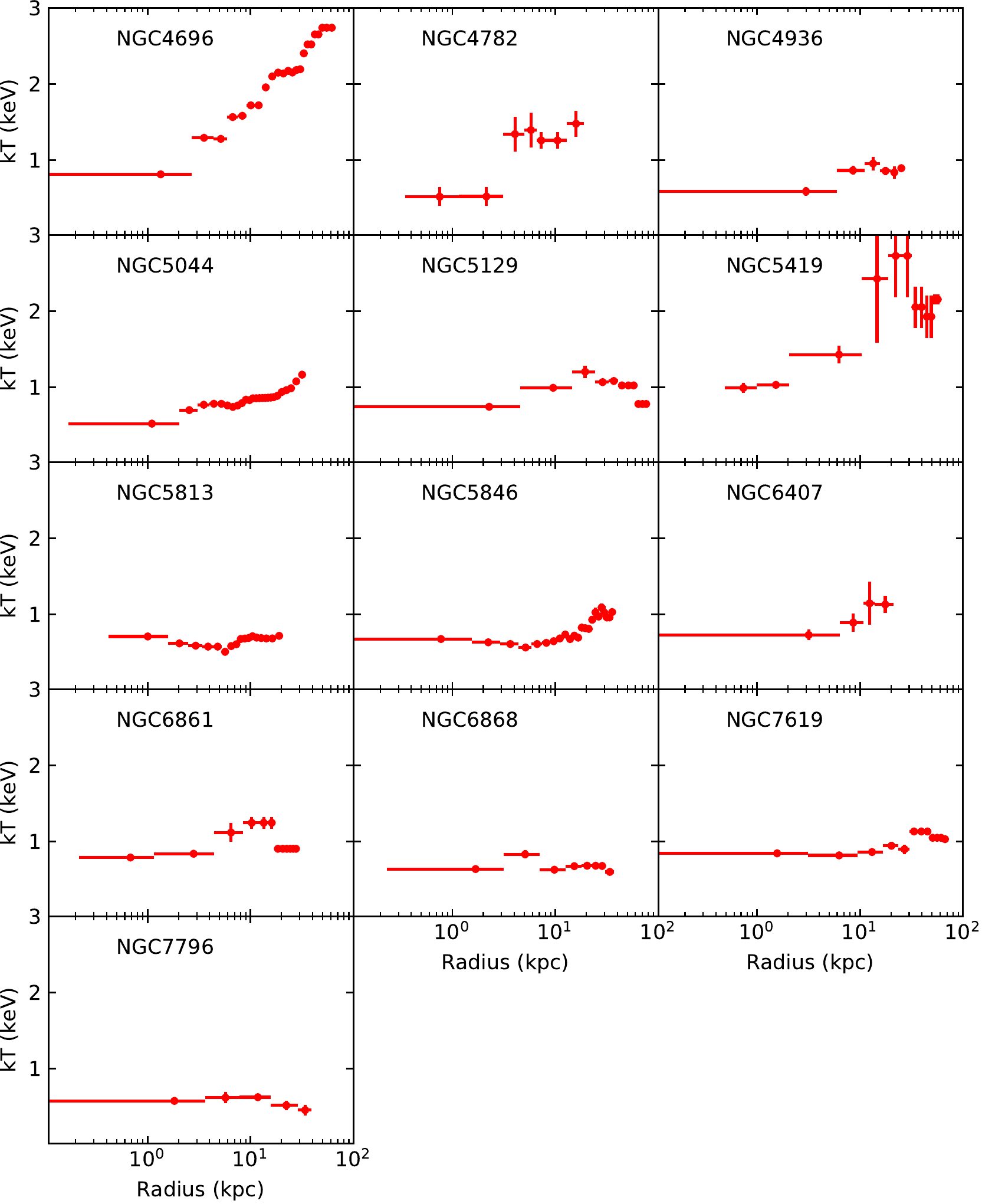}
\caption{Deprojected temperature profiles of the individual galaxies (continued).}
\end{figure*}

\begin{figure*}
\centering
  \includegraphics[width=0.9\linewidth]{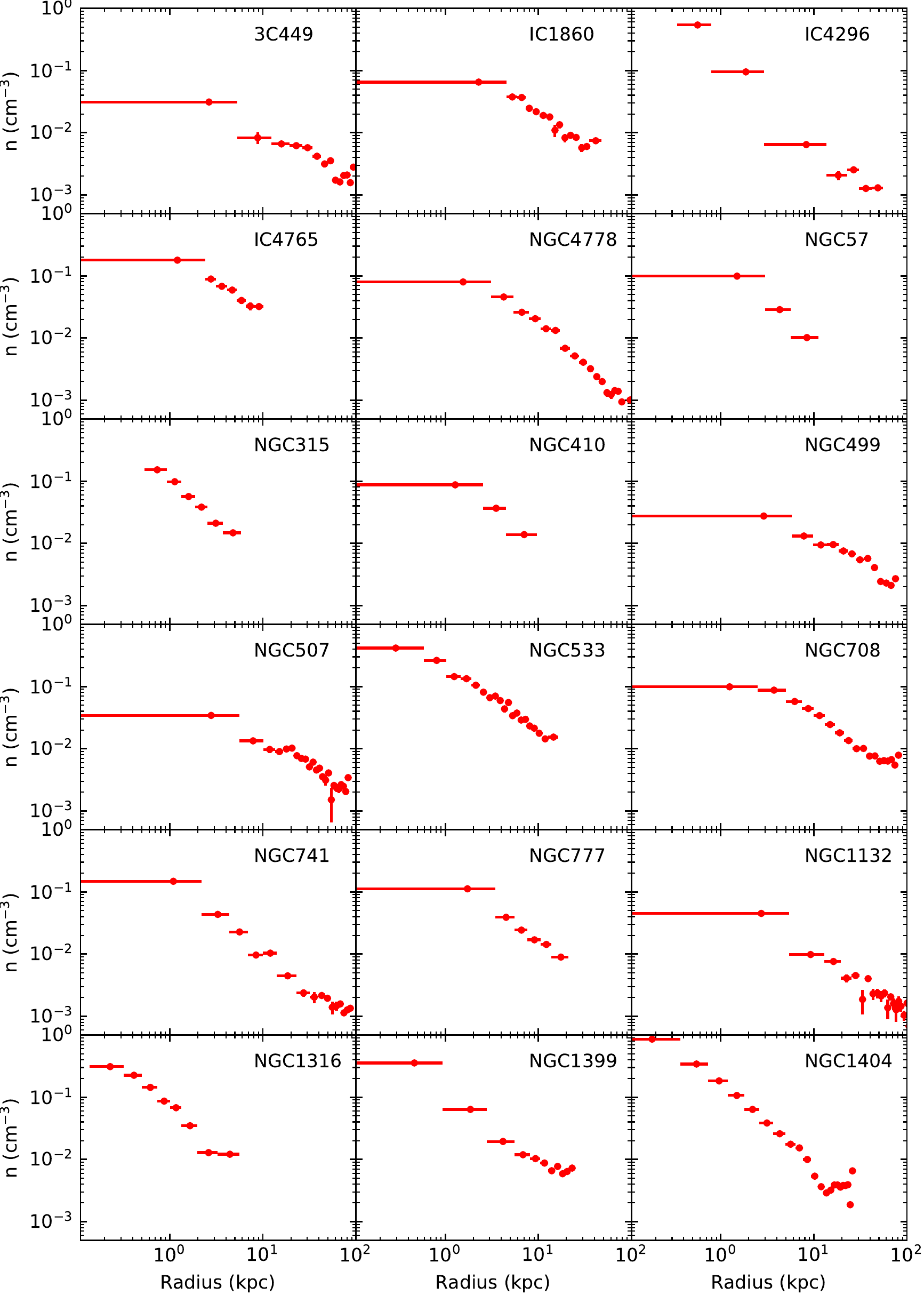}
\caption{Deprojected density profiles of the individual galaxies.}
\label{fig:n_profs}
\end{figure*}

\begin{figure*}
\ContinuedFloat
\centering
  \includegraphics[width=0.9\linewidth]{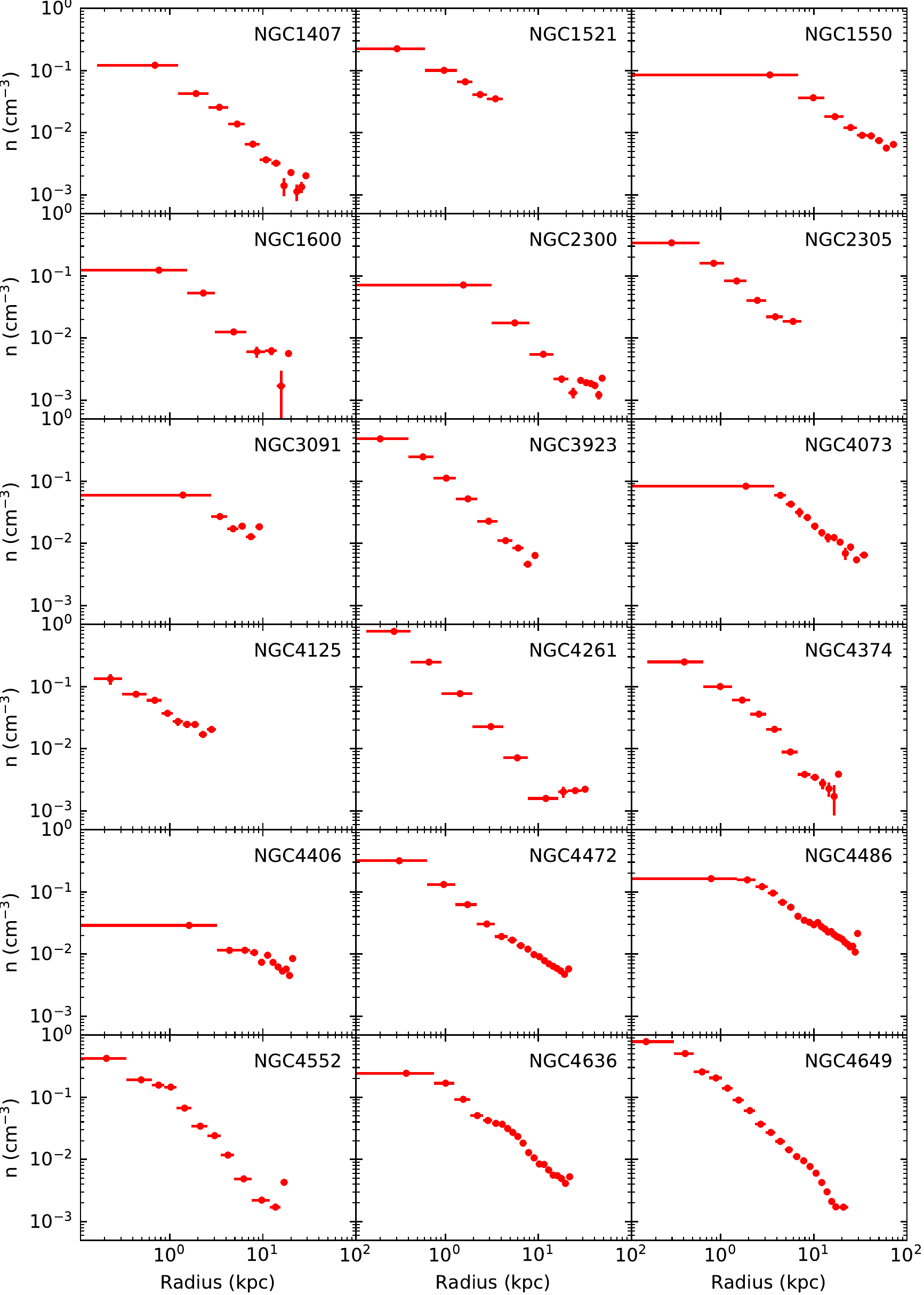}
\caption{Deprojected density profiles of the individual galaxies (continued).}
\end{figure*}

\begin{figure*}
\ContinuedFloat
\centering
  \includegraphics[width=0.9\linewidth]{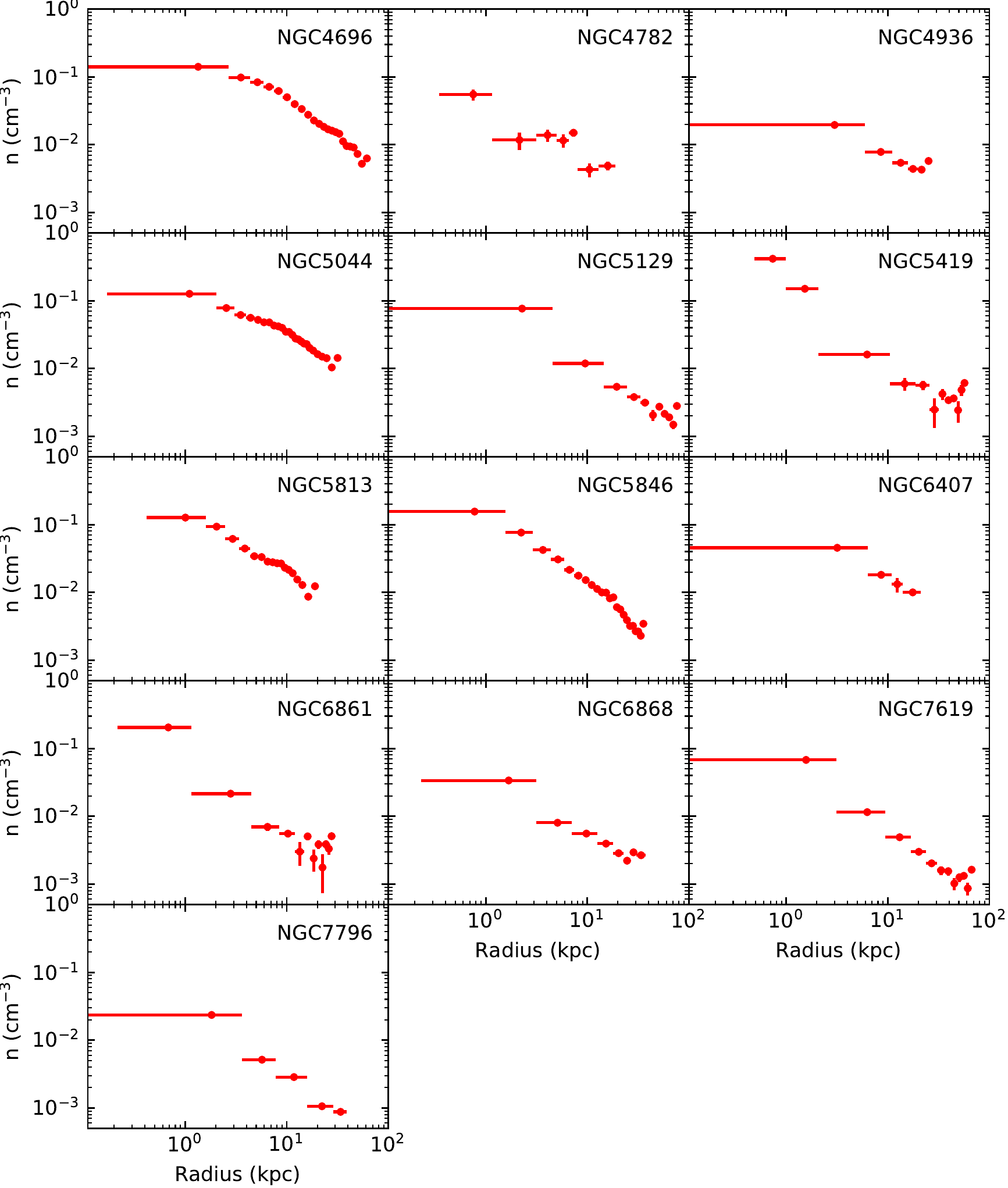}
\caption{Deprojected density profiles of the individual galaxies (continued).}
\end{figure*}

\begin{figure*}
\centering
  \includegraphics[width=0.9\linewidth]{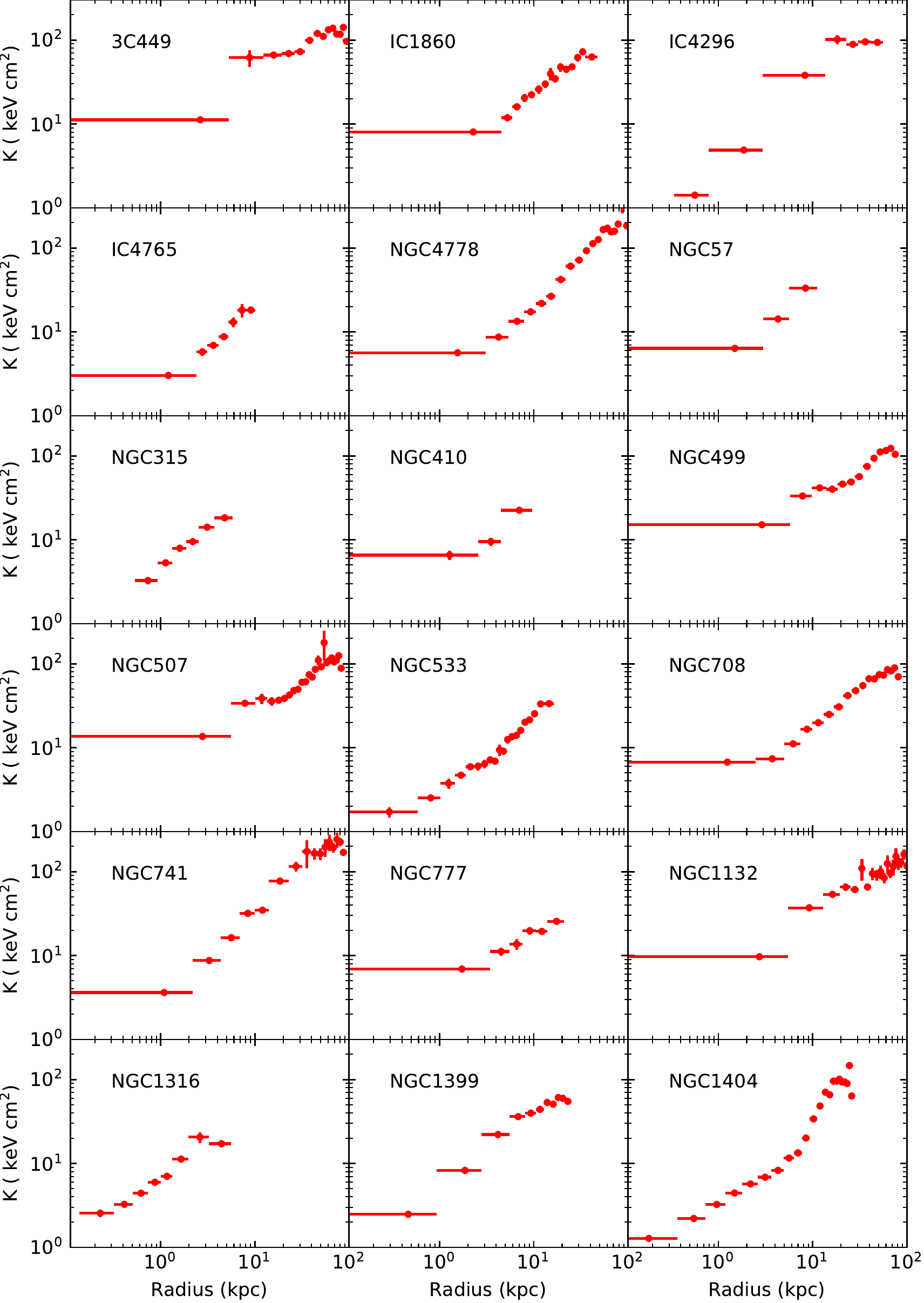}
\caption{Deprojected entropy profiles of the individual galaxies.}
\label{fig:K_profs}
\end{figure*}

\begin{figure*}
\ContinuedFloat
\centering
  \includegraphics[width=0.9\linewidth]{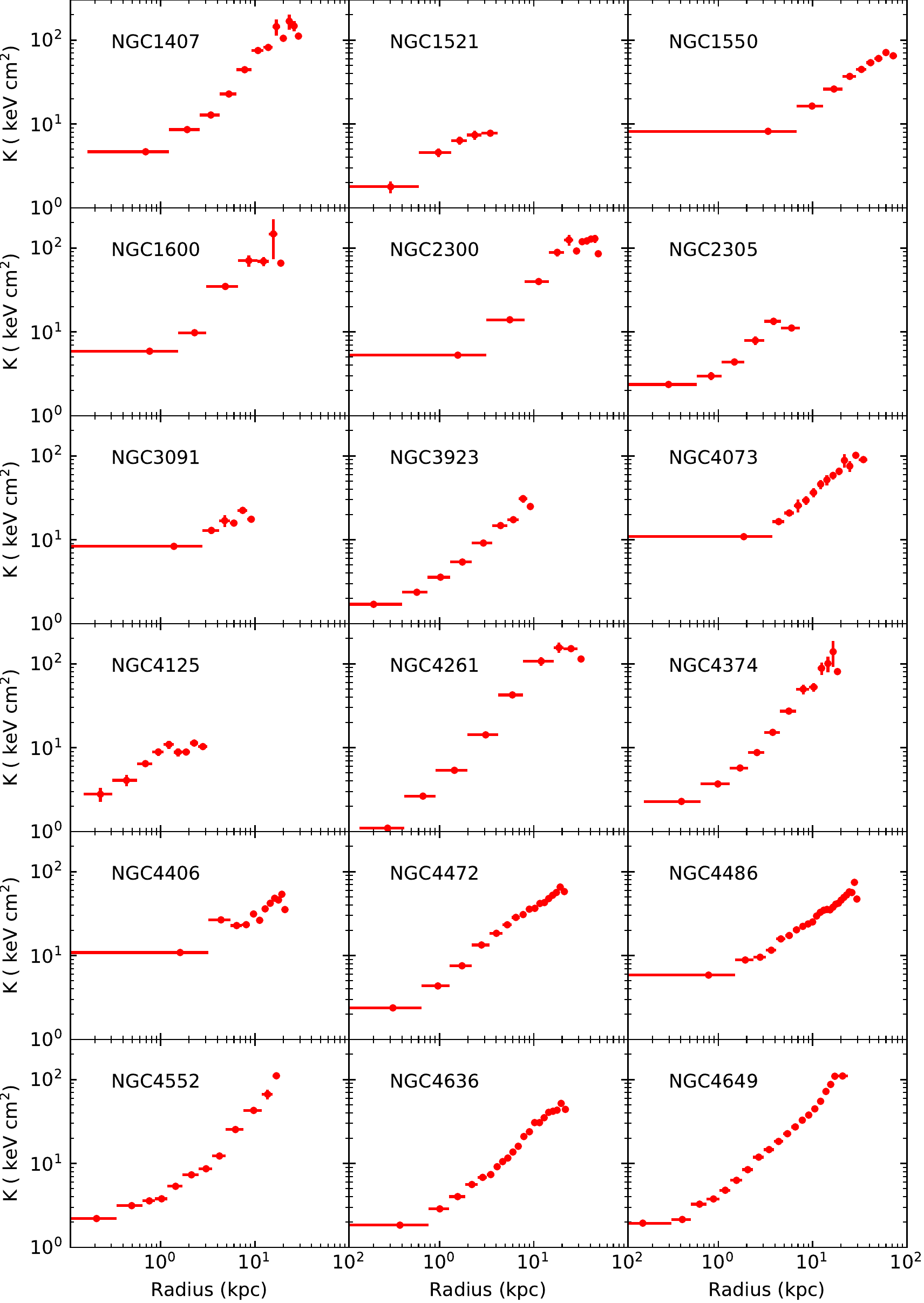}
\caption{Deprojected entropy profiles of the individual galaxies (continued).}
\end{figure*}

\begin{figure*}
\ContinuedFloat
\centering
  \includegraphics[width=0.9\linewidth]{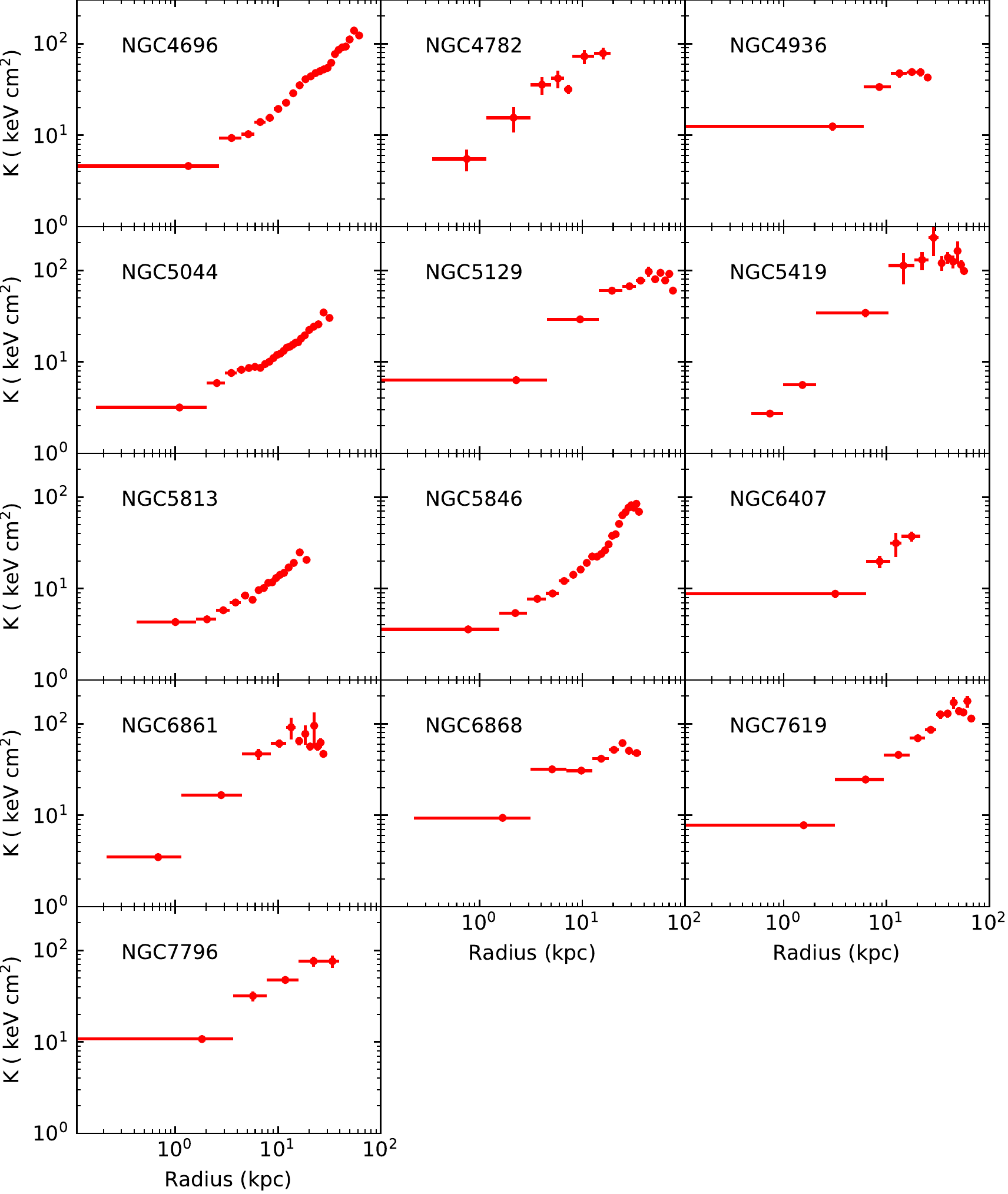}
\caption{Deprojected entropy profiles of the individual galaxies (continued).}
\end{figure*}

\begin{figure*}
\centering
  \includegraphics[width=0.9\linewidth]{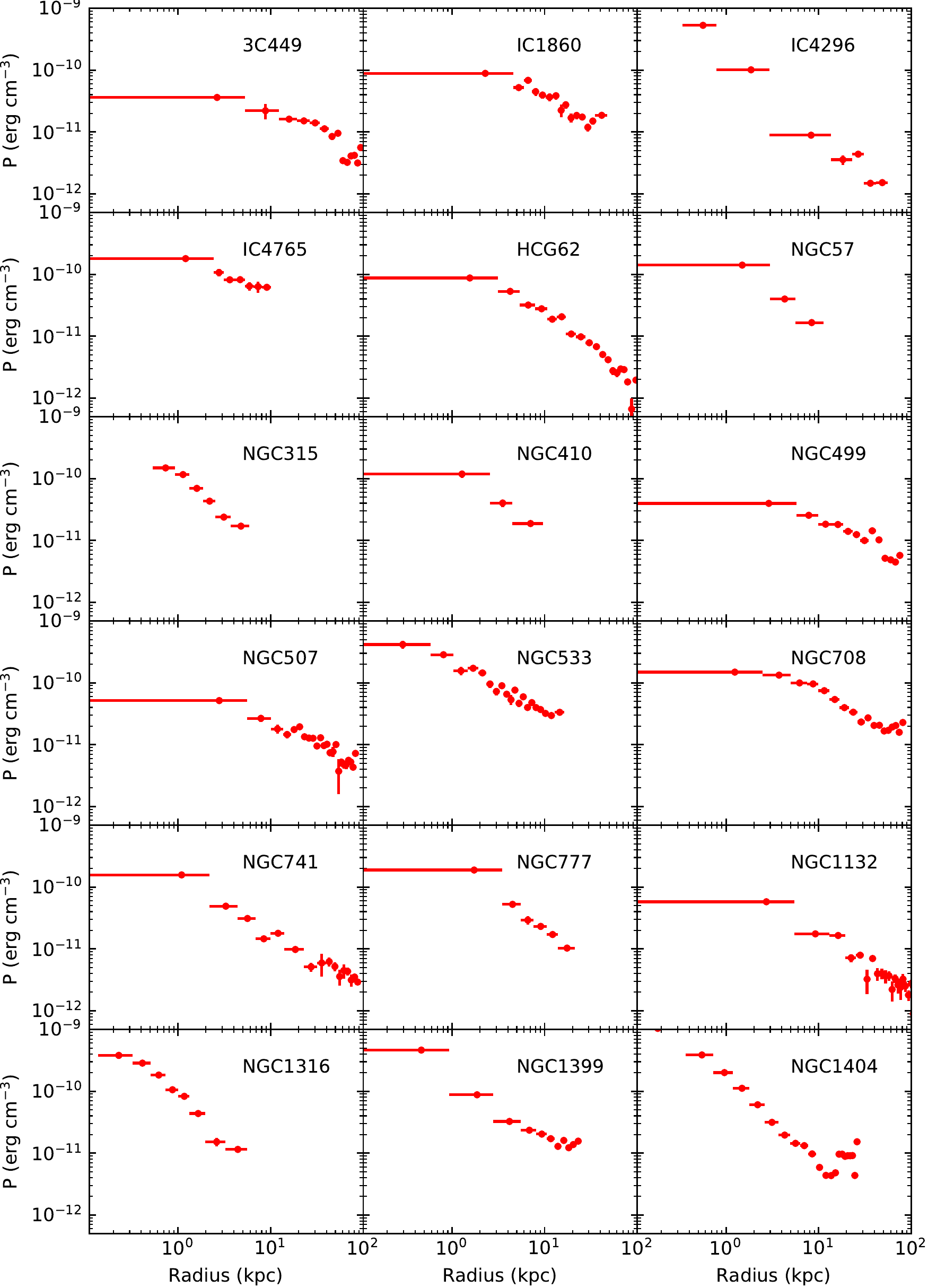}
\caption{Deprojected pressure profiles of the individual galaxies.}
\label{fig:P_profs}
\end{figure*}

\begin{figure*}
\ContinuedFloat
\centering
  \includegraphics[width=0.9\linewidth]{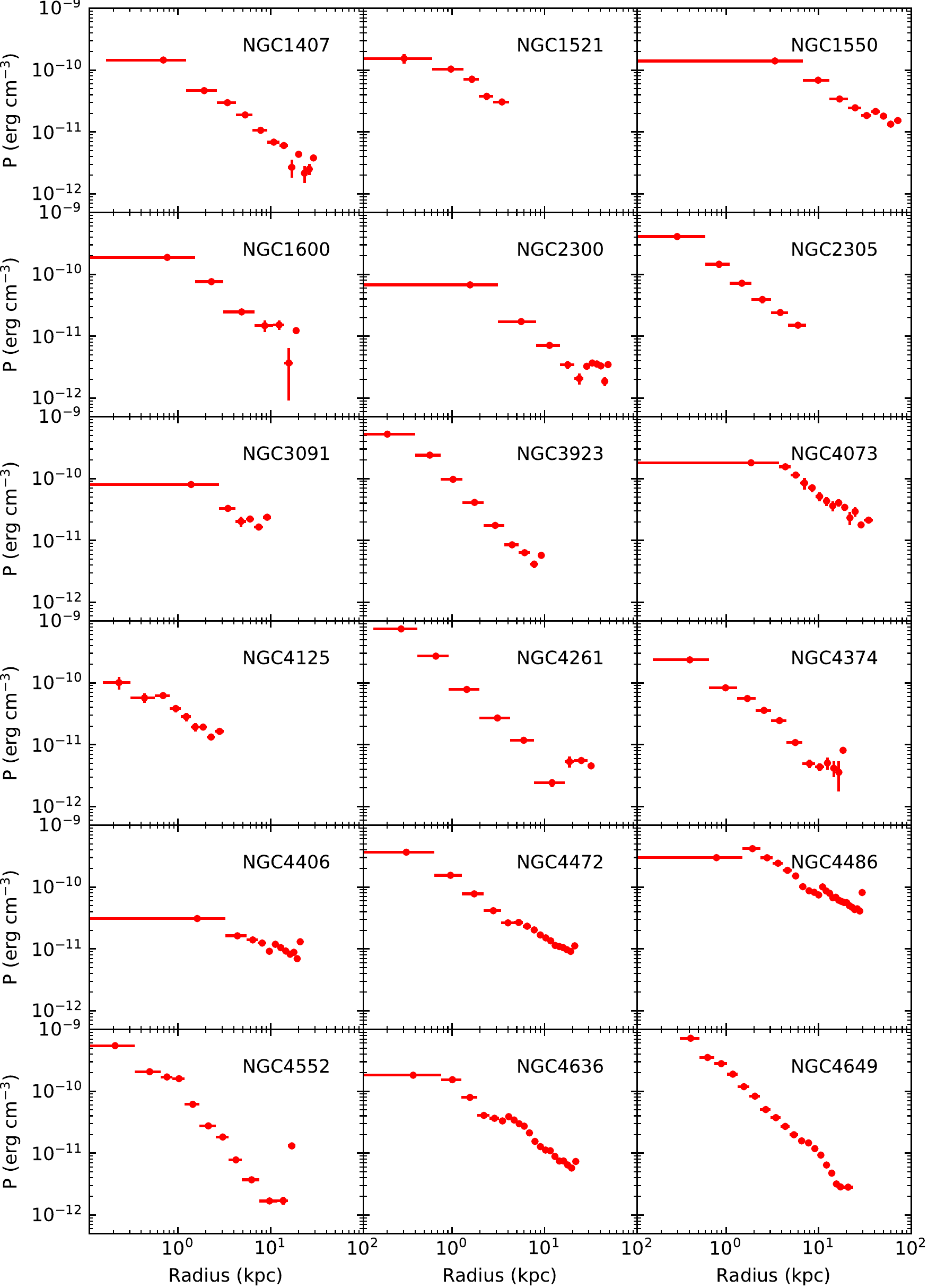}
\caption{Deprojected pressure profiles of the individual galaxies (continued).}
\end{figure*}

\begin{figure*}
\ContinuedFloat
\centering
  \includegraphics[width=0.9\linewidth]{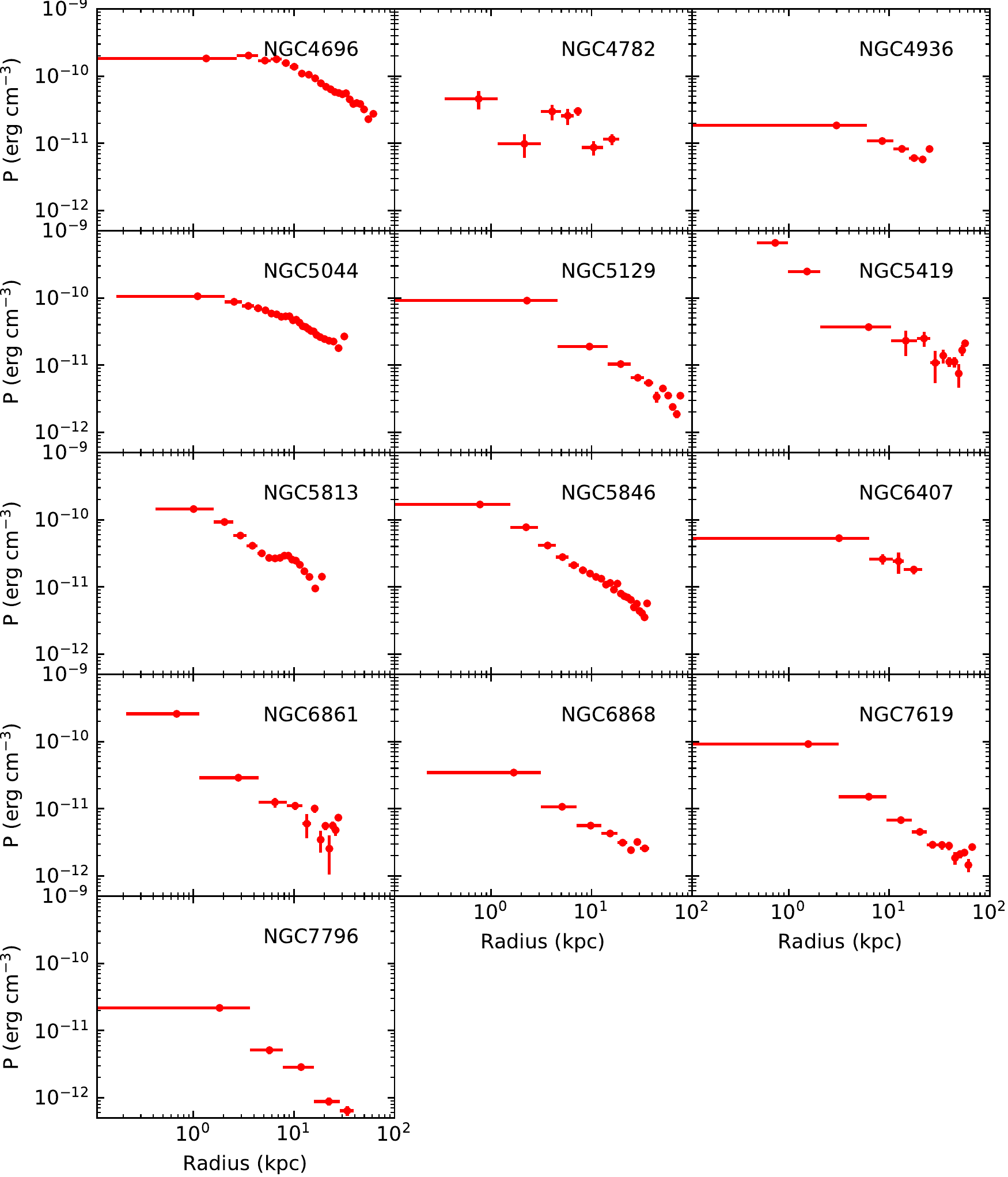}
\caption{Deprojected pressure profiles of the individual galaxies (continued).}
\end{figure*}

\begin{figure*}
\centering
  \includegraphics[width=0.9\linewidth]{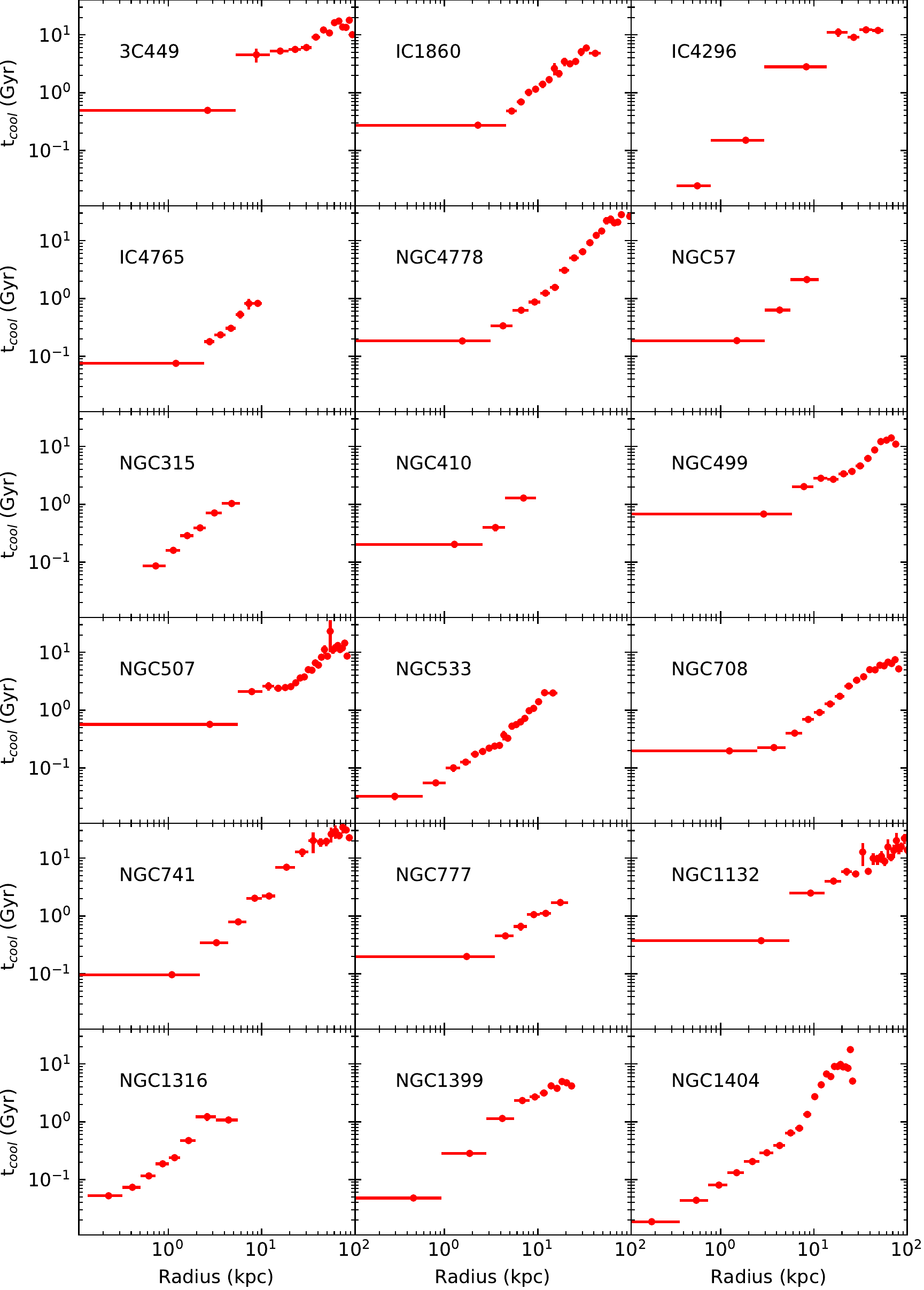}
\caption{Cooling time profiles of the individual galaxies.}
\label{fig:tcool_profs}
\end{figure*}

\begin{figure*}
\ContinuedFloat
\centering
  \includegraphics[width=0.9\linewidth]{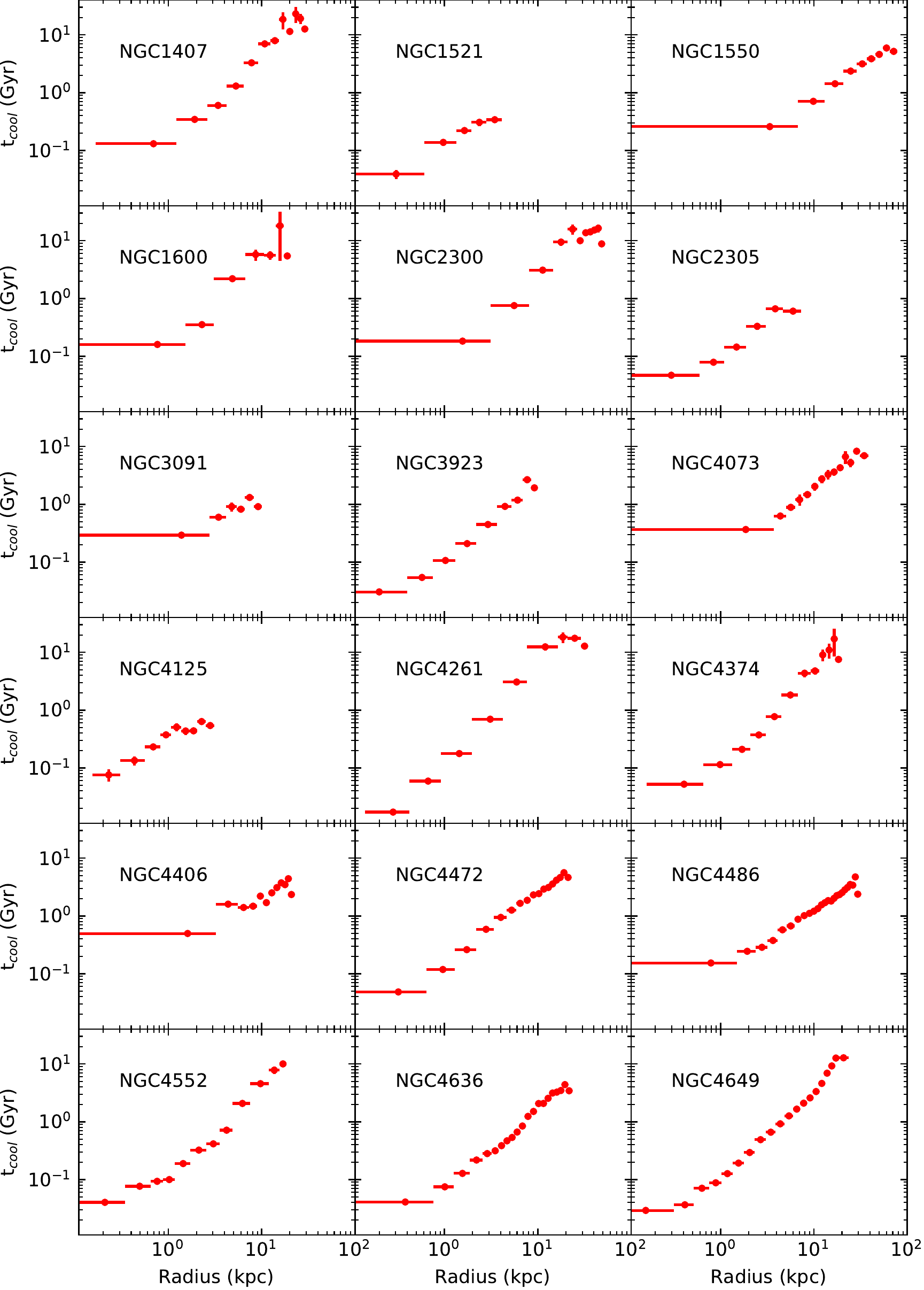}
\caption{Cooling time profiles of the individual galaxies (continued).}
\end{figure*}

\begin{figure*}
\ContinuedFloat
\centering
  \includegraphics[width=0.9\linewidth]{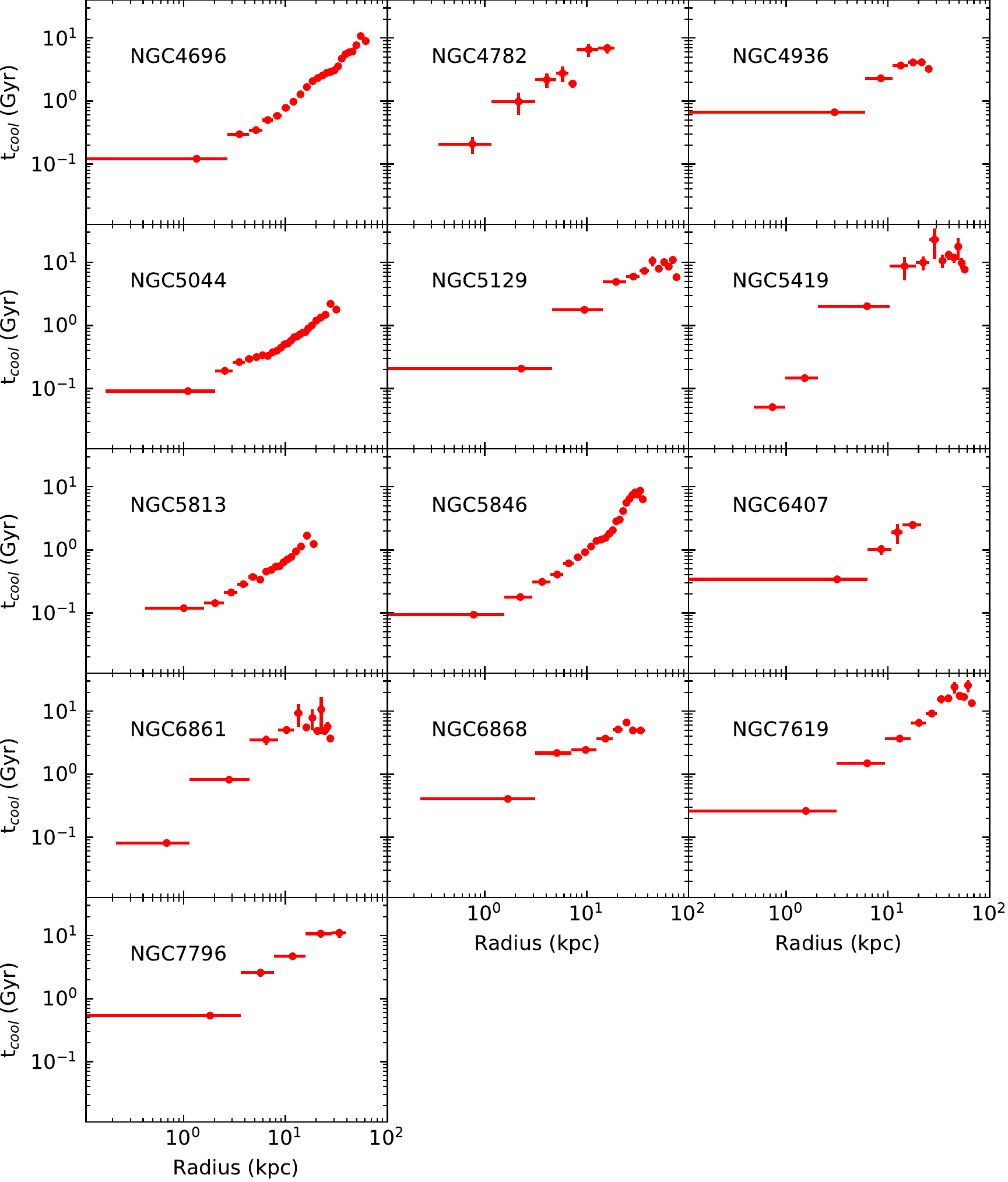}
\caption{Cooling time profiles of the individual galaxies (continued).}
\end{figure*}

\begin{figure*}
\centering
  \includegraphics[width=0.9\linewidth]{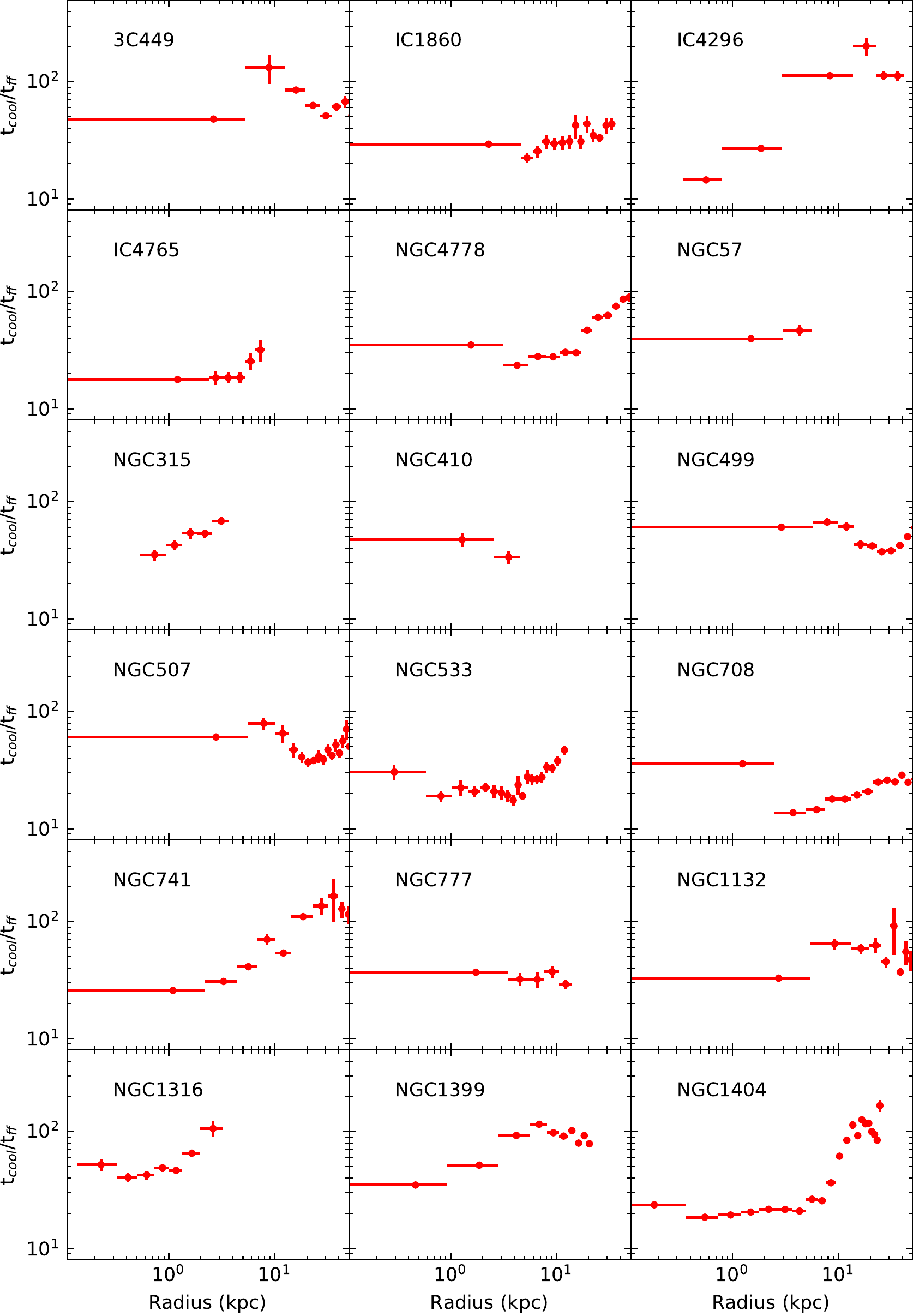}
\caption{$t_{\rm cool}$/$t_{\rm ff}$ profiles of the individual galaxies.}
\label{fig:tcoolbytff_profs}
\end{figure*}

\begin{figure*}
\ContinuedFloat
\centering
  \includegraphics[width=0.9\linewidth]{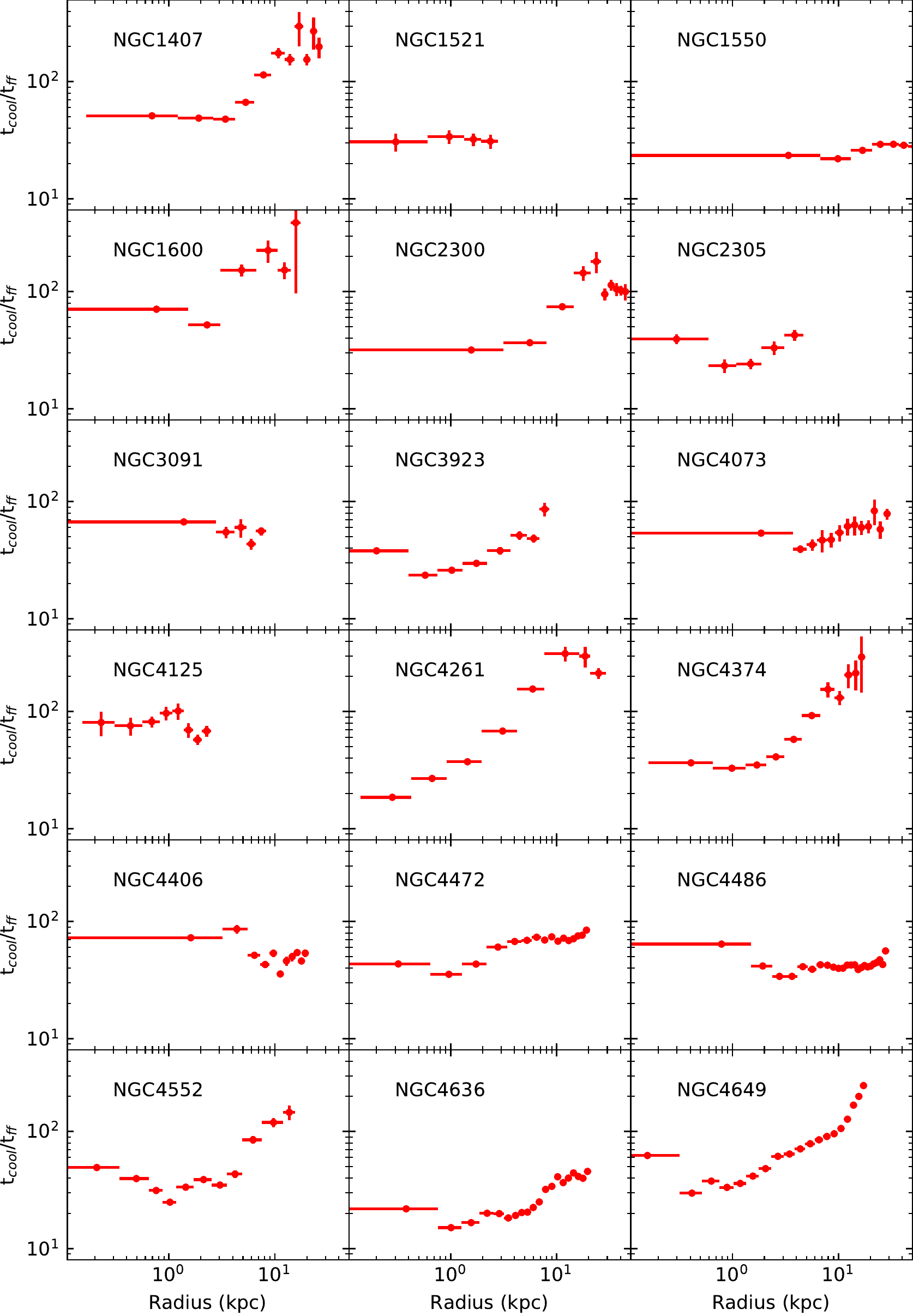}
\caption{$t_{\rm cool}$/$t_{\rm ff}$ profiles of the individual galaxies (continued).}
\end{figure*}

\begin{figure*}
\ContinuedFloat
\centering
  \includegraphics[width=0.9\linewidth]{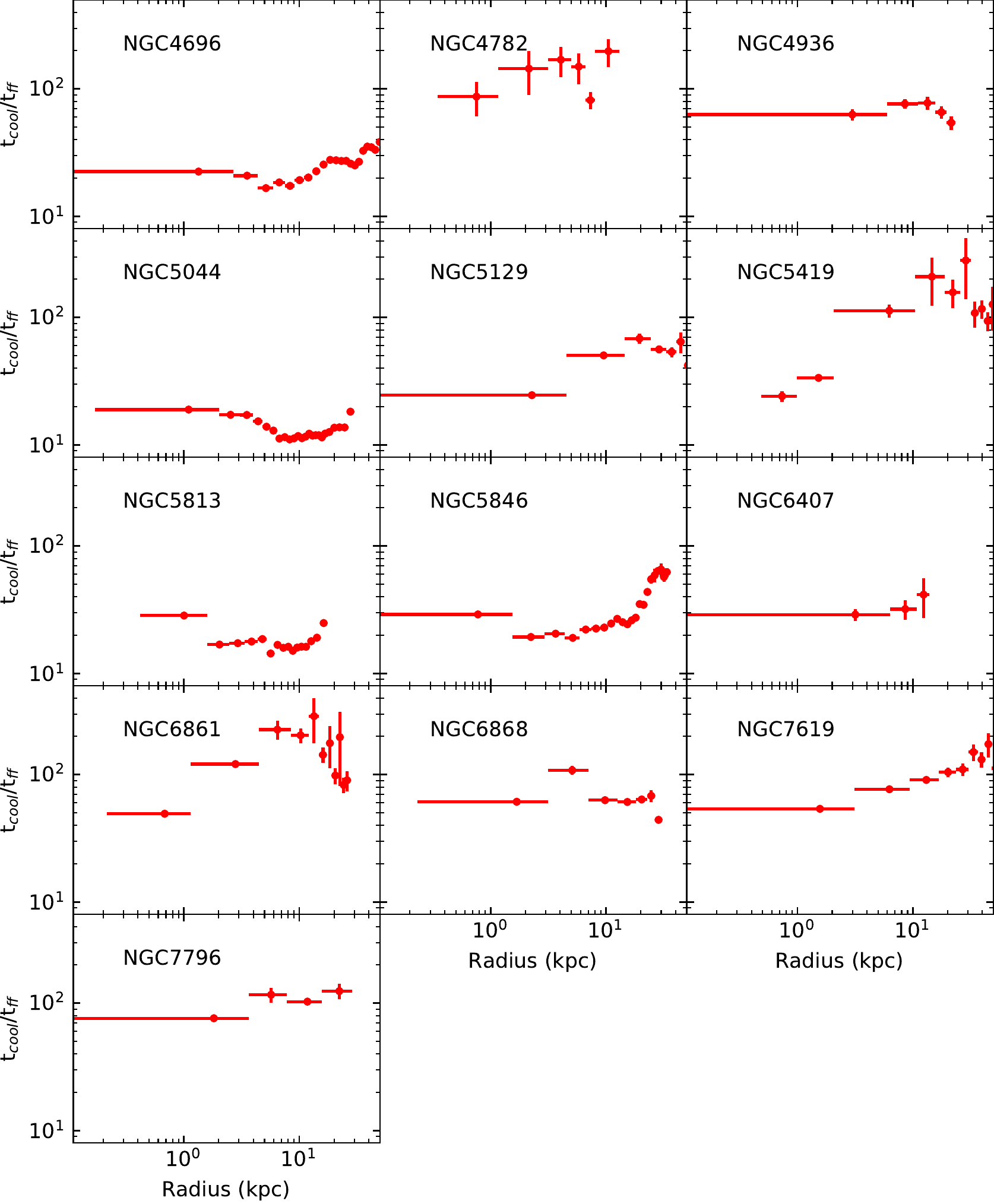}
\caption{$t_{\rm cool}$/$t_{\rm ff}$ profiles of the individual galaxies (continued).}
\end{figure*}

\end{document}